\documentclass[12pt,a4paper]{article}
\usepackage{jheplikemod,ifpdf,tocloft,amsmath,textcomp,mathrsfs,mathtools,braket,rotating,pdflscape,graphicx,multirow,shuffle}
\usepackage[bb=libus]{mathalpha}
\usepackage[T1]{fontenc}
\usepackage{lmodern}
\usepackage{appendix}
\hypersetup{pdftitle={Constructing Concrete Colour Tensors},pdfcreator={},linkcolor=[rgb]{0.15,0.35,0.75},colorlinks=true,citecolor=[rgb]{0.675,0,0.2},urlcolor=[rgb]{0.15,0.35,0.65}}
\usepackage{makeidx}
\usepackage[totoc,itemlayout=singlepar,column sep=50pt]{idxlayout}

\graphicspath{{figures/}}
\let\olditemize\itemize\renewcommand{\itemize}{\vspace{-4.pt}\olditemize\setlength{\itemsep}{0.75pt}\setlength{\parskip}{0pt}\setlength{\parsep}{0pt}}
\let\oldenumerate\enumerate\renewcommand{\enumerate}{\vspace{-4pt}\oldenumerate\setlength{\itemsep}{1pt}\setlength{\parskip}{0pt}\setlength{\parsep}{0pt}}
\makeatletter\renewcommand\section{\addtocontents{toc}{\protect\addvspace{-2.25\p@}}\@startsection {section}{1}{\z@}{-0.0ex \@plus .2ex \@minus 0.2ex}{1ex \@plus.1ex\@minus .5ex}{\normalfont\large\bfseries}}
\renewcommand\subsection{\addtocontents{toc}{\protect\addvspace{-2.25\p@}}\@startsection {subsection}{2}{\z@}{0.5ex \@plus .2ex \@minus 0.2ex}{0.75ex \@plus.1ex\@minus 0.5ex}{\normalfont\bfseries}}
\renewcommand\subsubsection{\addtocontents{toc}{\protect\addvspace{-2.25\p@}}\@startsection {subsubsection}{3}{\z@}{0.5ex \@plus .2ex \@minus 0.2ex}{0.75ex \@plus.1ex\@minus 0.5ex}{\normalfont\bfseries}}
\renewcommand\paragraph{\addtocontents{toc}{\protect\addvspace{-2.25\p@}}\@startsection {paragraph}{4}{\z@}{0.5ex \@plus .2ex \@minus 0.2ex}{0.75ex \@plus.1ex\@minus 0.5ex}{\normalfont\bfseries}}

\newcommand{\sectionAppendix}[2]{\section{\texorpdfstring{#1}{\Alph{section}.\, #1}\label{#2}}}
\newcommand{\subsectionAppendix}[2]{\subsection{\texorpdfstring{#1}{\Alph{section}.\arabic{subsection}\, #1}\label{#2}}}
\newcommand{\subsubsectionAppendix}[2]{\subsubsection{\texorpdfstring{#1}{\Alph{section}.\arabic{subsection}.\arabic{subsubsection}\, #1}\label{#2}}}

\newcommand{\eq}[1]{\vspace{-4.5pt}\begin{equation}#1\vspace{-4.5pt}\end{equation}}

\newcommand{\fwbox}[2]{\text{\makebox[#1][c]{$\hspace{-150pt}\displaystyle#2\hspace{-150pt}$}}}
\newcommand{\fwboxL}[2]{\text{\makebox[#1][l]{$#2$}}}
\newcommand{\fwboxR}[2]{\text{\makebox[#1][r]{$#2$}}}
\newcommand{\equivR}{\fwbox{14.5pt}{\hspace{-0pt}\fwboxR{0pt}{\raisebox{0.47pt}{\hspace{1.25pt}:\hspace{-4pt}}}=\fwboxL{0pt}{}}}
\newcommand{\equivL}{\fwbox{14.5pt}{\fwboxR{0pt}{}=\fwboxL{0pt}{\raisebox{0.47pt}{\hspace{-4pt}:\hspace{1.25pt}}}}}

\newcommand{\bigger}[1]{\raisebox{-0.95pt}{\scalebox{1.25}{$#1$}}}

\renewcommand{\phi}{\varphi}
\renewcommand{\bar}{\overline}
\renewcommand{\hat}{\widehat}
\renewcommand{\tilde}{\widetilde}

\newcommand{\fig}[3][0pt]{\raisebox{#1}{$\begin{array}{@{}c@{}}\includegraphics[scale=#2]{#3}\end{array}$}}

\newcommand{\rpackage}{\textbf{\tt{\textbf{colour}}\rule[-1.05pt]{7.5pt}{.75pt}\tt{\textbf{tensors}}}}

\NewDocumentCommand{\tikzBox}{O{0pt} m m}{\IfFileExists{figures/#2.pdf}{\raisebox{-#1}{\ensuremath{\vcenter{\hbox{\includegraphics[scale=1]{figures/#2}}}}}}{}}

\newcounter{specialfoot}
\newcommand{\footnotespecial}[1]{\addtocounter{specialfoot}{1}\renewcommand*{\thefootnote}{\fnsymbol{specialfoot}}\footnote{#1}\renewcommand*{\thefootnote}{\arabic{footnote}}}

\definecolor{mhvBlue}{rgb}{0.3,0.2,0.75}
\definecolor{varcolor}{rgb}{0.1,0.55,0.25}
\definecolor{functioncolor}{rgb}{0.1,0.35,0.75}
\definecolor{paper_blue}{rgb}{0.3,0.2,0.75}
\definecolor{paper_red}{rgb}{0.65,0.1,0.15}
\definecolor{paper_green}{rgb}{0.05,0.35,0.125}
\definecolor{dim}{rgb}{0.55,0.55,0.55}
\definecolor{hteal}{rgb}{0.0,0.545,0.7451}
\definecolor{rindou1}{rgb}{0.4431,0.2862,0.7960}
\definecolor{rindou2}{rgb}{0.0078,0.1215,0.4392}
\definecolor{lapis}{rgb}{0.0470,0.2941,0.5568}
\definecolor{emerald}{rgb}{0.31, 0.78, 0.47}
\definecolor{pinegreen}{rgb}{0.0, 0.47, 0.44}
\definecolor{jade}{rgb}{0.0, 0.66, 0.42}
\definecolor{teal}{rgb}{0.0, 0.5, 0.5}
\definecolor{hblue}{rgb}{0,0,0.725}
\definecolor{hred}{rgb}{0.575,0.0,0.225}
\definecolor{hgreen}{rgb}{0.0,0.4,0.2}
\definecolor{hteal}{rgb}{0.0,0.545,0.7451}

\renewcommand{\r}[1]{{\color{hred}#1}}
\renewcommand{\b}[1]{{\color{hblue}#1}}

\newcommand{\g}[1]{{\color{paper_green}#1}}
\renewcommand{\t}[1]{{\color{hteal}#1}}

\newcounter{lineNo}\def\inItemQ{0}
\renewcommand{\brace}[1]{\hspace{-2pt}{\texttt{[}}#1{\texttt{]}}\hspace{-1pt}}
\newcommand{\optArg}[1]{\textnormal{\texttt{:}}#1\!}

\newcommand{\var}[1]{{{\color{varcolor}{\sl#1}}}}
\newcommand{\fun}[1]{\text{\texttt{\textbf{\texttt{\color{functioncolor}#1}}}}}

\newcommand{\built}[1]{{\color{black}\textbf{\texttt{#1}}}}

\newcommand{\indices}[2]{{\hspace{-0.0pt}}^{\smash{#1}}_{\phantom{\smash{#1}}\smash{#2}}}

\newcommand{\pattern}{{\color{varcolor}\rule[-1.05pt]{7.5pt}{.65pt}}}
\newcommand{\patternTwo}{{\color{varcolor}\rule[-1.05pt]{12pt}{.65pt}}}

\newcommand{\defnBox}[4][0]{\def\inItemQ{1}\fwboxL{0pt}{\hspace{-22pt}\rule{0pt}{0.pt}}\hspace{\fill}\mbox{~}\\[-12pt]\noindent\rule{0.0pt}{12pt}\fwbox{24pt}{}\fwboxL{425pt}{\begin{minipage}[b]{0.932\textwidth}%
\mbox{%
\ifthenelse{\equal{#1}{31}}{\hspace{-20pt}\hspace{-2pt}{$\bullet$\,\,\ifthenelse{\equal{#3}{}}{\built{#2}}{\fun{#2}}}%
}{%
\ifthenelse{\equal{#1}{21}}{\hspace{-20pt}\hypertarget{#2}{\hspace{-2pt}{$\bullet$\,\,\ifthenelse{\equal{#3}{}}{\built{#2}}{\fun{#2}}}}%
\index{#2@\mbox{\fun{#2}\hspace{-1pt}\texttt{\textbf{[}}\var{algebra}\pattern\texttt{\textbf{]}}\hspace{-2pt}}|textbf}%
}{%
\ifthenelse{\equal{#1}{81}}{\hspace{-20pt}\hypertarget{#2}{\hspace{-2pt}{$\bullet$\,\,\ifthenelse{\equal{#3}{}}{\built{#2}}{\fun{#2}}}}%
\index{#2@\mbox{\fun{#2}\hspace{-1pt}\texttt{\textbf{[}}\var{dynkinLabel}\pattern\texttt{\textbf{]}}\hspace{-2pt}}|textbf}%
}{%
\ifthenelse{\equal{#1}{9}}{\hspace{-20pt}\hypertarget{#2}{\hspace{-2pt}{$\bullet$\,\,\ifthenelse{\equal{#3}{}}{\built{#2}}{\fun{#2}}}}%
\index{#2@\mbox{\fun{#2}\hspace{-1pt}\texttt{\textbf{[}}\var{rep}\pattern\texttt{\textbf{]}}\hspace{-2pt}}|textbf}%
}{%
\ifthenelse{\equal{#1}{8}}{\hspace{-20pt}\hypertarget{d:#2}{\hspace{-2pt}{$\bullet$\,\,\ifthenelse{\equal{#3}{}}{\built{#2}}{\fun{#2}}}}%
\index{#2z@\mbox{\fun{#2}\hspace{-1pt}\texttt{\textbf{[}}\var{dynkinLabel}\pattern\texttt{\textbf{]}}\hspace{-2pt}}|textbf}%
}{%
\ifthenelse{\equal{#1}{7}}{\hspace{-20pt}\hypertarget{amp:#2}{\hspace{-2pt}{$\bullet$\,\,\ifthenelse{\equal{#3}{}}{\built{#2}}{\fun{#2}}}}%
}{%
\ifthenelse{\equal{#1}{6}}{\hspace{-20pt}\hypertarget{irrep:#2}{\hspace{-2pt}{$\bullet$\,\,\ifthenelse{\equal{#3}{}}{\built{#2}}{\fun{#2}}}}%
}{%
\ifthenelse{\equal{#1}{2}}{\hspace{-20pt}\hypertarget{label:#2}{\hspace{-2pt}{$\bullet$\,\,\ifthenelse{\equal{#3}{}}{\built{#2}}{\fun{#2}}}}%
}{%
\ifthenelse{\equal{#1}{3}}{\hspace{-20pt}\hypertarget{operator:#2}{\hspace{-2pt}{$\bullet$\,\,\ifthenelse{\equal{#3}{}}{\built{#2}}{\fun{#2}}}}%
}{%
\ifthenelse{\equal{#1}{4}}{\hspace{-20pt}\hypertarget{tensor:#2}{\hspace{-2pt}{$\bullet$\,\,\ifthenelse{\equal{#3}{}}{\built{#2}}{\fun{#2}}}}%
\index{$\a$b(tensor)#2@\fun{#2} (tensor)|textbf}%
}{%
\ifthenelse{\equal{#1}{5}}{\hspace{-20pt}\hypertarget{algebra:#2}{\hspace{-2pt}{$\bullet$\,\,\ifthenelse{\equal{#3}{}}{\built{#2}}{\fun{#2}}}}%
}{%
\ifthenelse{\equal{#1}{0}}{\hspace{-20pt}\hypertarget{#2}{\hspace{-2pt}{$\bullet$\,\,\ifthenelse{\equal{#3}{}}{\fun{#2}}{\fun{#2}}}}%
\ifthenelse{\equal{#3}{}}{\index{#2@\fun{#2}|textbf}}{\index{#2@\fun{#2}|textbf}}%
}{%
\hspace{-22pt}{$\bullet$\,\,\ifthenelse{\equal{#3}{}}{\built{#2}}{\fun{#2}}}{\ifthenelse{\equal{#3}{}}{\index{#2@\built{#2}|textnormal}}{%
\index{#2@\fun{#2}|textnormal}}}}}}}}}}}}}}}%
\ifthenelse{\equal{#3}{}}{\hspace{-0.05pt}}{\brace{#3}}%
\hspace{-2pt}\textnormal{\texttt{:\!}}}\,\,#4\mbox{~}\hspace{\fill}\mbox{~}\\[-12pt]\end{minipage}\mbox{~}\hspace{\fill}\def\inItemQ{0}\vspace{5pt}\hspace{\fill}}\\[-12pt]~}

\newcommand{\defnBoxTwo}[5][0]{\def\inItemQ{1}\fwboxL{0pt}{\hspace{-22pt}\rule{0pt}{0.pt}}\hspace{\fill}\mbox{~}\\[-12pt]\noindent\rule{0.0pt}{12pt}\fwbox{24pt}{}\fwboxL{425pt}{\begin{minipage}[b]{0.932\textwidth}%
\ifthenelse{\equal{#1}{2}}{\hspace{-20pt}\hypertarget{label:#2}{\hspace{-2pt}{$\bullet$\,\,\ifthenelse{\equal{#3}{}}{\built{#2}}{\fun{#2}}}}%
}{%
\ifthenelse{\equal{#1}{3}}{\hspace{-20pt}\hypertarget{operator:#2}{\hspace{-2pt}{$\bullet$\,\,\ifthenelse{\equal{#3}{}}{\built{#2}}{\fun{#2}}}}%
}{%
\ifthenelse{\equal{#1}{4}}{\hspace{-20pt}\hypertarget{tensor:#2}{\hspace{-2pt}{$\bullet$\,\,\ifthenelse{\equal{#3}{}}{\built{#2}}{\fun{#2}}}}%
}{%
\ifthenelse{\equal{#1}{5}}{\hspace{-20pt}\hypertarget{algebra:#2}{\hspace{-2pt}{$\bullet$\,\,\ifthenelse{\equal{#3}{}}{\built{#2}}{\fun{#2}}}}%
}{%
\ifthenelse{\equal{#1}{0}}{\hspace{-20pt}\hypertarget{#2}{\hspace{-2pt}{$\bullet$\,\,\ifthenelse{\equal{#3}{}}{\built{#2}}{\fun{#2}}}}%
\ifthenelse{\equal{#3}{}}{\index{#2@\built{#2}|textbf}}{\index{#2@\fun{#2}|textbf}}%
}{%
\hspace{-22pt}{$\bullet$\,\,\ifthenelse{\equal{#3}{}}{\built{#2}}{\fun{#2}}}{\ifthenelse{\equal{#3}{}}{\index{#2@\built{#2}|textnormal}}{%
\index{#2@\fun{#2}|textnormal}}}}}}}}%
\ifthenelse{\equal{#3}{}}{\hspace{-1.75pt}}{\brace{#3}\hspace{-2pt}}\ifthenelse{\equal{#4}{}}{\hspace{-1.75pt}}{\hspace{1.5pt}\brace{#4}}%%
\hspace{-2.pt}\textnormal{\texttt{:\!}}\,\,#5\mbox{~}\hspace{\fill}\mbox{~}\\[-12pt]\end{minipage}\mbox{~}\hspace{\fill}\def\inItemQ{0}\vspace{5pt}\hspace{\fill}}\\[-12pt]~}

\newcommand{\mathematicaBox}[1]{\renewcommand{\arraystretch}{0.85}~\\[0pt]~\\[-10pt]\noindent\fwboxL{450pt}{\setcounter{lineNo}{0}\ifthenelse{\inItemQ=1}{\phantom{}\hspace{-24.55pt}}{\phantom{}\hspace{-0.5pt}}\boxed{\ifthenelse{\inItemQ=1}{\begin{minipage}[t]{1.0585\textwidth}}{\begin{minipage}{0.9865\textwidth}}\begin{tabular}{@{}l@{$\;\;\;$}p{11cm}@{}}#1~\\[-30pt]&\end{tabular}\end{minipage}\hspace{\fill}}}\hspace{\fill}\\[-3pt]\renewcommand{\arraystretch}{1}}

\newcommand{\mathematicaSequence}[3][0]{\stepcounter{lineNo}\ifthenelse{\value{lineNo}=1}{~\\[-18pt]\vspace{-0pt}}{}%~\\[-14pt]}
{\color{paper_blue}{\scriptsize{\tt In[}\arabic{lineNo}{\tt]}}\raisebox{-0.65pt}{{\scriptsize{\tt\raisebox{0.75pt}{:}=}}}}&\noindent\raisebox{-1.25pt}{\begin{minipage}[t]{14cm}\rule[-2pt]{0pt}{14pt}{\tt #2}\end{minipage}}~\\[0pt]\rule[0pt]{0.0pt}{12pt}%
\addtocounter{lineNo}{#1}\ifthenelse{\equal{#3}{}}{}{{\color{paper_blue}{\scriptsize{\tt Out[\arabic{lineNo}]}}\raisebox{-0.65pt}{{\scriptsize{\tt=}}}}&\rule[0pt]{0.0pt}{12pt}\raisebox{-1pt}{\rule[0pt]{0pt}{10pt}\begin{minipage}[t]{14cm}{\tt #3}\end{minipage}}\\[2pt]}}

\newcommand{\funL}[2][2]{\ifthenelse{\equal{#1}{2}}{\hyperlink{#2}{\fun{#2}\hspace{0pt}}\index{#2@\fun{#2}}}{\ifthenelse{\equal{#1}{4}}{\hyperlink{#2}{\fun{#2}\hspace{0pt}}}{\ifthenelse{\equal{#1}{1}}{\hyperlink{#2}{\textbf{\texttt{{\color{black}#2}}}}}{\ifthenelse{\equal{#1}{3}}{\hyperlink{#2}{\textbf{\texttt{{\color{black}#2}}}}\index{#2@\built{#2}}}{\ifthenelse{\equal{#1}{4}}{\hyperlink{#2}{\fun{#2}\hspace{0pt}}}{\hyperlink{#2}{\textbf{\texttt{{\color{black}#2}}}}\index{#2@\fun{#2}}}}}}}}

\newcommand{\funDL}[2][2]{\ifthenelse{\equal{#1}{4}}{\hyperlink{d:#2}{\fun{#2}\hspace{0pt}}}{\ifthenelse{\equal{#1}{2}}{\hyperlink{d:#2}{\fun{#2}\hspace{0pt}}\index{#2z@\mbox{\fun{#2}\hspace{-1pt}\texttt{\textbf{[}}\var{dynkinLabel}\pattern\texttt{\textbf{]}}\hspace{-2pt}}}}{\ifthenelse{\equal{#1}{1}}{\hyperlink{d:#2}{\textbf{\texttt{{\color{black}#2}}}}}{\ifthenelse{\equal{#1}{3}}{\hyperlink{d:#2}{\textbf{\texttt{{\color{black}#2}}}}\index{#2z@\built{#2}}}{\ifthenelse{\equal{#1}{4}}{\hyperlink{d:#2}{\fun{#2}\hspace{0pt}}}{\hyperlink{d:#2}{\textbf{\texttt{{\color{black}#2}}}}\index{#2z@\mbox{\fun{#2}\hspace{-1pt}\texttt{\textbf{[}}\var{dynkinLabel}\pattern\texttt{\textbf{]}}\hspace{-2pt}}}}}}}}}

\newcommand{\funAL}[2][2]{\ifthenelse{\equal{#1}{2}}{\hyperlink{#2}{\fun{#2}\hspace{0pt}}\index{#2@\mbox{\fun{#2}\hspace{-1pt}\texttt{\textbf{[}}\var{algebra}\pattern\texttt{\textbf{]}}\hspace{-2pt}}}}{\ifthenelse{\equal{#1}{1}}{\hyperlink{#2}{\textbf{\texttt{{\color{black}#2}}}}}{\ifthenelse{\equal{#1}{3}}{\hyperlink{d:#2}{\textbf{\texttt{{\color{black}#2}}}}\index{#2@\mbox{\fun{#2}\hspace{-1pt}\texttt{\textbf{[}}\var{algebra}\pattern\texttt{\textbf{]}}\hspace{-1pt}}}}{\ifthenelse{\equal{#1}{4}}{\hyperlink{#2}{\fun{#2}\hspace{0pt}}}{\hyperlink{#2}{\textbf{\texttt{{\color{black}#2}}}}\index{#2@\mbox{\fun{#2}\hspace{-1pt}\texttt{\textbf{[}}\var{algebra}\pattern\texttt{\textbf{]}}\hspace{-2pt}}}}}}}}

\newcommand{\funRL}[2][2]{\ifthenelse{\equal{#1}{2}}{\hyperlink{#2}{\fun{#2}\hspace{0pt}}\index{#2@\mbox{\fun{#2}\hspace{-1pt}\texttt{\textbf{[}}\var{rep}\pattern\texttt{\textbf{]}}\hspace{-2pt}}}}{\ifthenelse{\equal{#1}{1}}{\hyperlink{#2}{\textbf{\texttt{{\color{black}#2}}}}}{\ifthenelse{\equal{#1}{3}}{\hyperlink{d:#2}{\textbf{\texttt{{\color{black}#2}}}}\index{#2@\mbox{\fun{#2}\hspace{-1pt}\texttt{\textbf{[}}\var{rep}\pattern\texttt{\textbf{]}}\hspace{-1pt}}}}{\ifthenelse{\equal{#1}{4}}{\hyperlink{#2}{\fun{#2}\hspace{0pt}}}{\hyperlink{#2}{\textbf{\texttt{{\color{black}#2}}}}\index{#2@\mbox{\fun{#2}\hspace{-1pt}\texttt{\textbf{[}}\var{rep}\pattern\texttt{\textbf{]}}\hspace{-2pt}}}}}}}}

\newcommand{\algL}[2][0]{\ifthenelse{\equal{#1}{2}}{\hyperlink{algebra:#2}{\fun{#2}\hspace{0pt}}\index{#2@\fun{#2}}}{\ifthenelse{\equal{#1}{1}}{\hyperlink{algebra:#2}{\textbf{\texttt{{\color{black}#2}}}}}{\ifthenelse{\equal{#1}{3}}{\hyperlink{algebra:#2}{\textbf{\texttt{{\color{black}#2}}}}\index{#2@\built{#2}}}{\ifthenelse{\equal{#1}{0}}{\hyperlink{algebra:#2}{\fun{#2}}%\index{#2@\fun{#2}}
}{\hyperlink{algebra:#2}{\textbf{\texttt{{\color{black}#2}}}}\index{#2@\fun{#2}}}}}}}

\newcommand{\symbolL}[2][0]{\ifthenelse{\equal{#1}{2}}{\hyperlink{operator:#2}{\fun{#2}\hspace{0pt}}\index{#2@\fun{#2}}}{\ifthenelse{\equal{#1}{1}}{\hyperlink{operator:#2}{\textbf{\texttt{{\color{black}#2}}}}}{\ifthenelse{\equal{#1}{3}}{\hyperlink{operator:#2}{\textbf{\texttt{{\color{black}#2}}}}\index{#2@\built{#2}}}{\ifthenelse{\equal{#1}{0}}{\hyperlink{operator:#2}{\fun{#2}}%\index{#2@\fun{#2}}
}{\hyperlink{operator:#2}{\textbf{\texttt{{\color{black}#2}}}}\index{#2@\fun{#2}}}}}}}

\newcommand{\repL}[2][0]{\ifthenelse{\equal{#1}{2}}{\hyperlink{irrep:#2}{\fun{#2}\hspace{0pt}}\index{#2@\fun{#2}}}{\ifthenelse{\equal{#1}{1}}{\hyperlink{label:#2}{\textbf{\texttt{{\color{black}#2}}}}}{\ifthenelse{\equal{#1}{3}}{\hyperlink{irrep:#2}{\textbf{\texttt{{\color{black}#2}}}}\index{#2@\built{#2}}}{\ifthenelse{\equal{#1}{0}}{\hyperlink{irrep:#2}{\fun{#2}}%\index{#2@\fun{#2}}
}{\hyperlink{irrep:#2}{\textbf{\texttt{{\color{black}#2}}}}\index{#2@\fun{#2}}}}}}}

\newcommand{\tensor}[2][0]{\ifthenelse{\equal{#1}{4}}{\hyperlink{tensor:#2}{\fun{#2}}}{\ifthenelse{\equal{#1}{0}}{\hyperlink{tensor:#2}{\fun{#2}}\index{$\a$b(tensor)#2@\fun{#2} (tensor)}}{\hypertarget{tensor:#2}{\fun{#2}}\index{$\a$b(tensor)#2@\fun{#2} (tensor)|textbf}}}}

\newcommand{\dynkinLabel}[3][0]{\mbox{\ifthenelse{\equal{#1}{0}}{\hyperlink{label:w}{\fun{w}{\color{black}\brace{\fun{#2}}\brace{#3}}}}{
\hyperlink{label:w}{\fun{w}{\color{black}\brace{\fun{#2}}\brace{#3}}}\index{$\a$d(zDynkin Label)#2@\fun{w}\brace{\fun{#2}} (Dynkin label)}}}}

\newcommand{\builtL}[2][2]{\ifthenelse{\equal{#1}{2}}{\hyperlink{#2}{\built{#2}}\index{#2@\built{#2}}}{\ifthenelse{\equal{#1}{1}}{\hyperlink{#2}{\textbf{\texttt{{\color{black}#2}}}}}{\ifthenelse{\equal{#1}{3}{{\hyperlink{#2}{\textbf{\texttt{{\color{black}#2}}}\index{#2@\fun{#2}}}}
}{\hyperlink{#2}{\textbf{\texttt{{\color{black}#2}}}\index{#2@\built{#2}}}}}}}}

\makeindex
\renewcommand{\@seccntformat}[1]{\csname the#1\endcsname.\;}%\quad}

\title{\texorpdfstring{{\huge \mbox{Constructing \emph{Concrete} Colour}\\[-5pt]\mbox{Tensors for Simple Lie Algebras}}}{Constructing Concrete Colour Tensors for Simple Lie Algebras}}
\author[\hspace{-4pt}\dagger,\ast]{\vspace{-24pt} Jacob~L.~Bourjaily,}\emailAdd{jbourj@ias.edu}
\author[\hspace{-4pt}\ast]{\vspace{-2pt} Michael~Plesser,}\emailAdd{plesser@psu.edu}
\author[\hspace{0pt}\ast]{\vspace{-2pt} Philip~Velie}\emailAdd{velie@psu.edu}

\affiliation[\dagger]{School of Natural Sciences, Institute for Advanced Study, Princeton, NJ, 08540, USA} 
\affiliation[\ast]{Institute for Gravitation and the Cosmos, Department of Physics,\\Pennsylvania State University, University Park, PA 16802, USA}

\abstract{%~\\[-30pt]%
We describe the implementation and use of \rpackage: a \textsc{Mathematica} package built for the {concrete construction} of explicit colour tensors involving arbitrarily charged particles in any gauge theory. These can be used to \emph{directly} determine linear relations among colour tensors and compute a number of otherwise difficult, representation-theoretic quantities relevant to scattering amplitudes such as colour-interference between partial amplitudes.  
}

\preprint{}

\begin{document}

\maketitle\thispagestyle{empty}
\pagenumbering{roman}\clearpage

\setcounter{section}{0}

\newpage
\pagenumbering{arabic}
\vspace{0pt}%

\addtocontents{toc}{\protect~\\[-48pt]}

\newpage
\section{Introduction}\vspace{-4pt}

The theory of Lie algebras and their representations has long had a close relationship with physics. These structures arise in the description of a wide variety of physical phenomena---spanning the range from classical physics to quantum mechanics, quantum field theory, string theory, and model building. As such, it is no surprise that a number of excellent books and review articles have been written to introduce students to the subject, with varying degrees of mathematical rigor and precision (see, for example, \cite{Slansky:1981yr,Cornwell:1997ke,DiFrancesco:1997nk,birdtracks,hamermesh2012group,fulton2013,georgi2018lie,Bourjaily:2025hvq}), and a wide variety of computational tools now exist to answer many of the representation-theoretic questions that arise (see, for example, \cite{Feger:2019tvk,sage,magma}). 

One subject in which representation theory plays an important role is gauge field theory, which generalizes electrodynamics to a theory of any number of mutually interacting (massless) spin-one particles. It is a general theorem that for any local, unitary quantum field theory, the observable labels (`colours') associated to any set of massless spin-one particles must furnish the adjoint representation of \emph{some} Lie algebra, and that any other particles with which these interact, must be grouped according to `charges' that furnish {representations} of that Lie algebra \cite{Benincasa:2007xk}. In the Standard Model, this Lie algebra is $\mathfrak{su}_3\!\times\!\mathfrak{su}_2\!\times\!\mathfrak{u}_1$, with matter charged under various representations thereof.\\[-10pt]

The Feynman rules for gauge theory make explicit reference to the \emph{generators} of particles' representations, which sew together into \emph{colour tensors} encoding how scattering depends on the colours of particles involved. This colour dependence can be factored-out from the kinematic dependence of any Feynman diagram; grouping diagrams into gauge-invariant subsets \cite{Cvitanovic:1976am} results in a \emph{colour decomposition} (see, \emph{e.g.}~\cite{Mangano:1988kk,Mangano:1990by,Dixon:1996wi,DelDuca:1999rs,Dixon:2013uaa,Melia:2015ika,Johansson:2015oia}):
\vspace{-1pt}\eq{A(\left\{p_a,\epsilon_a,c_a\right\}%_{a\in[n]}
)=\sum_i \mathcal{C}_i(\{c_a\})\,\mathcal{A}^i(\{p_a,\epsilon_a\})\,,\label{colour_decomposition}\vspace{-3pt}}
where the gauge-invariant \emph{partial amplitudes} $\mathcal{A}_i$ depend only on the kinematics (momenta and helicities) of the particles, and the colour tensors $\mathcal{C}_i$ depend only on the non-dynamic quantum numbers labelling their colours. These colour tensors can be viewed as intertwiners connecting the various representations of the incoming particles to those of the outgoing ones.

In many cases, the decomposition (\ref{colour_decomposition}) can be defined without reference to any particular gauge group, or with minimal dependence on the representations of the various charged particles involved. For example, for $n$ particles all charged under the adjoint representation (\emph{e.g.} any process involving gauge bosons), one may write \emph{at tree-level}\footnote{For any simple Lie algebra of `classical' type, there exists a known generalization of the trace decomposition valid to any order of perturbation theory; the details of this decomposition depends on the particular algebra, and may be violated non-perturbatively. It remains to be shown whether any such generalization beyond tree-level exists for gauge theories involving exceptional Lie algebras.}
\vspace{-0pt}\eq{\fwbox{350pt}{A=\frac{1}{T(\mathbf{\b{R}})}\hspace{-0pt}\sum_{\r{\vec{\sigma}}\in\mathfrak{S}([2,n])/\mathbb{Z}_2\hspace{-40pt}}\hspace{-0pt}\big(\mathbf{tr}_{\mathbf{\b{R}}}(1,\r{\sigma_1},\r{\sigma_2},\r{\ldots},\r{\sigma_{\text{-}1}}){+}(\text{-}1)^{n}\mathbf{tr}_{\mathbf{\b{R}}}(1,\r{\sigma_{\text{-}1}},\r{\ldots},\r{\sigma_2},\r{\sigma_{1}})\big)\mathcal{A}(1,\r{\sigma_1},\r{\ldots},\r{\sigma_{\text{-}1}})\,,}\label{trace_expansion_at_tree_level}\vspace{-0pt}}
where $\mathcal{A}$ is an ordered partial amplitude, $\mathbf{\b{R}}$ is \emph{any} representation of the gauge group's Lie algebra and $T(\mathbf{\b{R}})$ its \emph{Dynkin index}, and $\mathbf{tr}_{\mathbf{\b{R}}}$ represents a trace over the \emph{generators} of this representation; the sum is over all $(n{-}1)!/2$ dihedrally-inequivalent permutations of the orderings of the $n$ particles.%\\[-12pt] 

Another possible colour decomposition at tree-level involves the tensors defined by Del Duca, Dixon, and Maltoni (`DDM') in \cite{DelDuca:1999rs}. This re-expresses (\ref{trace_expansion_at_tree_level}) as the sum over $(n{-}2)!$ terms
\vspace{-0pt}\eq{\fwbox{350pt}{A=\hspace{-0pt}\sum_{\r{\vec{\sigma}}\in\mathfrak{S}([2,n{-}1])\hspace{-30pt}}\hspace{-0pt}\mathbf{\r{\textit{\textbf{f}}}}\left[{\smash{\b{1}\,\r{\vec{\sigma}}\,\b{n}}}\right]\mathcal{A}(\b{1},\r{\sigma_1},\r{\ldots},\r{\sigma_{\text{-}1}},\b{n})\,,}\label{ddm_expansion_at_tree_level}\vspace{-2pt}}
where the colour tensors that appear are generated only by the generators of the adjoint representation (or `structure constants'). The equivalence between the two colour decompositions, (\ref{trace_expansion_at_tree_level}) and (\ref{ddm_expansion_at_tree_level}), requires the Jacobi identity and that the partial amplitudes satisfy the linear relations described by Kleiss and Kuijf in \cite{KK}; both of these relations are true for any gauge theory.\\[-5pt]

Admittedly, physicists are rarely interested in colour-dependent observables; and so the impracticality of actually \emph{constructing} colour tensors for particular charged particles in particular gauge theories is rarely appreciated; for the most part, we are content to treat colour tensors appearing in these decompositions as symbolic/abstract objects. But even for colour-independent observables such as colour-summed cross sections, there can be non-trivial interference between the partial amplitudes appearing in (\ref{colour_decomposition}): 
\eq{\sum_{i,j}\bar{\mathcal{A}_i}\langle\mathcal{C}^i|\mathcal{C}_j\rangle\mathcal{A}^j\,.\label{overlap_for_cross_section}}

For the \emph{trace expansion} (\ref{trace_expansion_at_tree_level}), it is a classic result (see \emph{e.g.}~\cite{Mangano:1988kk,Mangano:1990by}) that interference between differently ordered tensors appearing in (\ref{trace_expansion_at_tree_level}) diminishes in the limit of large-rank gauge groups (something that makes sense only for classical Lie algebras); but for any particular case---the Standard Model, say---the precise form of such interference terms must be computed. Interestingly, interference between the colour tensors appearing in the DDM expansion (\ref{ddm_expansion_at_tree_level}) does \emph{not} diminish at large rank.

The `overlap' between conjugated sets of colour tensors is obviously something that is purely representation-theoretic---independent of the bases chosen for generators, or even which linear combinations of generators are chosen. In principle, the methods described by Cvitanovi\'{c} in \cite{birdtracks} (see also \cite{Bourjaily:2025hvq}) do allow for the computation of such numbers; but these ideas have yet to be implemented within any publicly available computational tool, and they can prove extremely laborious when computed by hand.\footnote{In \cite{birdtracks}, Cvitanovi\'{c} writes: ``I fear $\mathfrak{e}_8$ will not yield to pencil and paper.''}

Of course, it should be possible to simply \emph{evaluate} all the contractions appearing in (\ref{overlap_for_cross_section}) provided that one had on hand \emph{concrete} representations of the tensors involved---that is, colour tensors encoded as explicit numerical arrays: after all, modern computers are remarkably good at linear algebra and summing terms (especially when actual numbers are involved). To build these tensors, however, one must first have some \emph{concrete} form of the generators of any relevant representations, and have the tools to effectively sew them together. While some computer algebra systems (such as \texttt{Sage} \cite{sage} and \texttt{Magma} \cite{magma}, but notably \emph{not} including \textsc{Mathematica}) do provide the generators of arbitrary (irreducible) representations of simple Lie algebras, the tools required to construct the colour tensors relevant to amplitudes have not been available.\\[-10pt]

The \textsc{Mathematica} package \rpackage, which we describe here, bridges this gap in our public toolbox. It provides convenient, concrete forms of the generators of arbitrary representations of simple Lie algebras, the tools required to build general colour tensors from them, and the functionality to use these tensors to \emph{directly} compute representation-theoretic quantities such as linear dependences and contractions.\\[-10pt]

The complete source code for the \rpackage~package can be found in the file \rpackage\textbf{\texttt{.m}}, together with a fairly pedagogical walkthrough notebook \textbf{\texttt{demonstrating}\rule[-1.05pt]{7.5pt}{.75pt}\texttt{colour}\rule[-1.05pt]{7.5pt}{.75pt}\texttt{tensor}\rule[-1.05pt]{7.5pt}{.75pt}\texttt{tools}.\texttt{nb}}, which are available through this work's submission to the \texttt{arXiv} via the link under `\b{Ancillary files}' on the work's {abstract} page.

\subsection{Organization and Outline}\vspace{-4pt}

This work is organized as follows. \mbox{Section~\ref{review_section}} aims to {briefly} review the key concepts relevant to the abstract description and concrete realization of the representations of Lie algebras. In particular, \mbox{section~\ref{subsec:reps_via_generators}} reviews the formalism of Lie algebra representations in the way it is encountered by most students of physics; this description makes explicit reference to some set of \emph{generators}, but remains largely agnostic about the bases with respect to which they are defined. 

Given any set of generators, \mbox{section \ref{building_generators_from_generators}} describes how other (possibly distinct) sets of generators can be defined, which are guaranteed to satisfy \emph{identical} commutation relations.

\mbox{Section~\ref{classification_of_irreps}} introduces the notion of \emph{irreducibility} of representations and discusses their classification and specification in terms of \hyperlink{label:w}{Dynkin labels}. This classification is best understood through the connection between representations of simple Lie algebras and their \emph{weight systems}, which is reviewed in \mbox{section~\ref{weight_lattices_for_reps}}. This formalism sharpens and clarifies the notions of irreducibility and decompositions. \mbox{Section~\ref{rep_operations_via_weights}} translates the operations of building representations from representations described in  \mbox{section~\ref{building_generators_from_generators}} in terms of operations on weight systems. 

\newpage
Unfortunately, there is considerable disagreement among reference material as to the \emph{ordering} of weight components (equivalently, the ordering of simple roots), and thus no consensus as to which precise representation is identified by any particular Dynkin label. The choices we have made here and in the package \rpackage~are documented and clarified in \mbox{section~\ref{subsec:conventions_for_weight_systems}}.

In \mbox{section~\ref{section:concrete_reps}} we clarify what is meant by a \emph{concrete} representation and discuss the value of having these on hand as explicit, numeric arrays which encode the components of generators with respect to specific choices of basis. Most physicists are comfortable choosing explicit bases for generators of Lie algebra representations largely because of the role of unitarity in physics, which we review in \mbox{section~\ref{avoiding_complexity}}, but we emphasize the value of \emph{not} adhering to physicists' conventions for the purposes of concrete computations. The broad form of concrete representations provided by \rpackage~are discussed in \mbox{section~\ref{concrete_reps_in_package}}, and the details of how their bases were chosen are described in \mbox{section~\ref{basis_choices_for_concrete_reps}}.

The primary colour tensors that appear in the study of scattering amplitudes are reviewed in \mbox{section~\ref{sec:colourInGaugeTheory}}, highlighting their syntax, usage, and form in \rpackage. Many examples are discussed, illustrating much of the functionality of the package \rpackage. 

We describe how the package \rpackage~may be initialized and installed in \mbox{section~\ref{sec:code}} before briefly concluding in \mbox{section~\ref{sec:conclusion}}.\\[-10pt]

The primary functionality of \rpackage~is documented in the \hyperlink{context_organization_of_appendix}{Appendix}. This describes and illustrates each of the functions provided by \rpackage, as organized according to their context. An alphabetical index of these functions can be found in the \hyperlink{index}{index}, which also indicates where useful examples may be found elsewhere in this work.

\newpage

\section{(Rapid) Review of Representation Theory of Lie Algebras}\label{review_section}\vspace{-4pt}

In this section we briefly review the main description of Lie algebras and their finite-dimensional representations. More complete reviews can be found in, \emph{e.g.},~\mbox{\cite{Slansky:1981yr,Cornwell:1997ke,DiFrancesco:1997nk,birdtracks,hamermesh2012group,fulton2013,georgi2018lie,Bourjaily:2025hvq}.} Specifically, this section deals with representations \emph{in the abstract}---in terms of weight systems and lattices, without reference to any \emph{specific} generators of representations or their actions on particular (bases for) vector spaces. 

The understanding of how Lie algebras and their representations can be classified and enumerated comes largely from this more abstract (less \emph{concrete}) description. It allows for extremely efficient and effective algorithms for computing many features of interest in the study of Lie algebra representations, including enumeration of irreducible representations, the decomposition of representations into those of sub-algebras (`branching rules') or tensor product representations into irreducible representations, and so on. Many of these algorithms have been implemented in the package \rpackage, but these have been effectively implemented in a number of other computational tools \cite{Feger:2019tvk,sage,magma}. The primary of the features of \rpackage~which act at the level of weight-lattices are described in \mbox{appendix~\ref{appendix:abstract_rep_operations}}. 

We review this material here in part because this understanding of Lie algebras and their representations can be seen throughout the syntax and usage of \rpackage; but perhaps more importantly, this structure is \emph{manifest} in our particular choices of \emph{concrete} representations provided by \rpackage, as described in \mbox{section~\ref{section:concrete_reps}}.  This coherence in structure greatly simplifies the analysis of concrete representations and the many various tensors constructed therefrom.\\[-10pt]

The description and classification of Lie algebras and their representations is one of the great triumphs of modern mathematics \cite{lie1888theorie,lie1893vorlesungen,killing1888zusammensetzung,cartan1894structure,brauer1935spinors,weyl1927quantenmechanik,chevalley1946theory,curtis1999pioneers}. Any Lie algebra $\mathfrak{g}$ can be uniquely expressed as an outer product of \emph{simple} Lie algebras (classified by Cartan) and any number of Abelian factors ($\mathfrak{u}_1^{\otimes r}$). And \emph{representations} of Lie algebras can be described in terms of those acting separately on these simple (and Abelian) factors; therefore, and without loss of generality, we may focus our attention to the case of simple Lie algebras and their representations. 

Upon `complexification', the simple Lie algebras come in seven \emph{types} named by Cartan as $\{\mathfrak{a},\mathfrak{b},\mathfrak{c},\mathfrak{d},\mathfrak{e},\mathfrak{f},\mathfrak{g}\}$. The first four types consist of the `classical' Lie algebras---infinite families enumerated by their \emph{rank} $\r{k}$: $\mathfrak{a}_{\r{k}\geq1}\!\simeq\mathfrak{su}_{\r{k}{+}1}$, $\mathfrak{b}_{\r{k}\geq2}\!\simeq\mathfrak{so}_{2\r{k}{+}1}$, $\mathfrak{c}_{\r{k}\geq3}\!\simeq\mathfrak{sp}_{2\r{k}}$, and $\mathfrak{d}_{\r{k}\geq4}\!\simeq\mathfrak{so}_{2\r{k}}$; the last three types include the five `exceptional' algebras: $\mathfrak{e}_{\r{6}}$, $\mathfrak{e}_{\r{7}}$, $\mathfrak{e}_{\r{8}}$, $\mathfrak{f}_{\r{4}}$ and $\mathfrak{g}_{\r{2}}$. 

(Notice that we have limited the ranges of ranks of the classical Lie algebras. This allows us to exclude cases of redundancy such as `$\mathfrak{b}_{\r{1}}$'$\simeq\mathfrak{a}_{\r{1}}$, `$\mathfrak{c}_{\r{2}}$'$\simeq\mathfrak{b}_{\r{2}}$, and `$\mathfrak{d}_{\r{3}}$'$\simeq\mathfrak{a}_{\r{3}}$. To be clear, in the package \rpackage, redundant/alternative nomenclature \emph{is} supported, but is automatically converted (even if only internally) to one of the standard cases enumerated above:
\mathematicaBox{
\mathematicaSequence{\funL[1]{nice}@\{\algL{so}\brace{3},\algL{so}\brace{5},\algL{sp}\brace{4},\algL{so}\brace{6},\algL{e}\brace{4},\algL{su}\brace{5},\algL{so}\brace{9},\algL{sp}\brace{10},\algL{e}\brace{5}\}\\[-10pt]}{\fig[-2.5pt]{1}{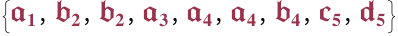}\vspace{-2pt}}
}~\\[-1pt]
This syntax for naming simple Lie algebras is described in more detail in \mbox{appendix~\ref{naming_of_lie_algebras}}.

\subsection{\texorpdfstring{Representations Defined via \emph{Abstract} `{Generators}'}{Representations Defined via Abstract Generators}}\label{subsec:reps_via_generators}\vspace{-4pt}

Let us start from the definition of Lie algebra representations we first learned in school. A \emph{representation} of some Lie algebra may be defined by a collection of \emph{matrices} whose linear span contains the \emph{commutator} of any pair describes some representation of a Lie algebra. These matrices are called \emph{generators} of the representation. Let $\{\mathbf{T}^\r{a}\}_{\r{a}\in\r{[d]}}\!\equivL\mathbf{T}^{\r{[d]}}$ be such a set of generators ($\r{d}$ in number), and let's remain largely agnostic about their size (or action relative to a basis). The statement that these matrices span the commutator of any pair,
\eq{[\mathbf{{T}}^{\r{a}},\mathbf{{T}}^{\r{b}}]\equivR\mathbf{T}^{\r{a}}.\mathbf{T}^{\r{b}}{-}\mathbf{T}^{\r{b}}.\mathbf{T}^{\r{a}}\in\mathrm{span}_{\t{c}}\{\mathbf{T}^{\t{c}}\}\fwboxL{0pt}{\,\qquad\forall\,\r{a},\r{b}\!\in\!\r{[d]}\,,}\label{span_form}}
is equivalent to the requirement that for every pair $\r{a},\r{b}\!\in\!\r{[d]}$ there exists some set of $\r{d}$ numbers $\r{f}\indices{\r{a\,b}}{\t{[d]}}\equivR\{\r{f}\indices{\r{a\,b}}{\t{c}}\}_{\t{c}\in\r{[d]}}$ such that 
\eq{[\mathbf{{T}}^{\r{a}},\mathbf{{T}}^{\r{b}}]\,=\sum_{\t{c}\,\in\r{[d]}}\r{f}\indices{\r{a\,b}}{\t{c}}\,\mathbf{T}^{\t{c}}\,.\label{explicit_form_of_defining_relation}}

Let's pause to note that the statement (\ref{span_form}) is very general: it would be satisfied for any arbitrary linear combination of generators. Suppose we were to consider the generators
\eq{\mathbf{U}^{\r{\smash{{a'}}}}\equivR\sum_{\t{a}\in\r{[d]}}\mathbf{M}\indices{\r{{a'}}}{\t{a}}\mathbf{T}^{\t{a}}\,,\label{reordered_generators}}
defined by the (invertible) $(\r{d}\!\times\!\r{d})$ matrix $\mathbf{M}\indices{\r{[{d}']}}{\r{[d]}}$; it is clear that the statement (\ref{span_form}) would be unchanged. \emph{However}, the precise coefficients appearing in (\ref{explicit_form_of_defining_relation}) would of course change. Specifically, they would change according to
\eq{[\mathbf{{U}}^{\r{{a'}}},\mathbf{{U}}^{\r{{b'}}}]\,=\sum_{\t{{c'}}\,\in\r{[d]}}\r{g}\indices{\r{{a'}\,{b'}}}{\t{{c'}}}\,\mathbf{U}^{\t{{c'}}}\,.\label{explicit_form_of_defining_relation_altered}}
where
\eq{\r{g}\indices{\r{a'\,b'}}{\r{c'}}=\sum_{\t{a},\t{b},\t{c}\in\r{[d]}}\mathbf{M}\indices{\r{a'}}{\t{a}}\mathbf{M}\indices{\r{b'}}{\t{b}}\,\r{f}\indices{\t{a\,b}}{\t{c}}\mathbf{\bar{M}}\indices{\t{c}}{\r{c'}}\,,\label{reshuffling_effect_on_structure_consts}}
with $\mathbf{\bar{M}}\equivR\mathbf{M}^{\text{-}1}$. We may say the sets of generators $\mathbf{U}\indices{\r{[d']}}{}$ and $\mathbf{T}\indices{\r{[d]}}{}$ furnish \emph{isomorphic} but \emph{incompatible} representations, denoted $\mathbf{U}\indices{\r{[d']}}{}\!\sim\!\mathbf{T}\indices{\r{[d]}}{}$. For two representations to be \emph{compatible}, their generators must satisfy \emph{identical} commutation relations as (\ref{explicit_form_of_defining_relation}), with identical coefficients $\r{f}\indices{\r{a\,b}}{\r{[d]}}$ for all $\r{a},\r{b}\!\in\!\r{[d]}$.\\[-10pt]

Describing the generators $\mathbf{T}\indices{\r{[d]}}{}$ as \emph{matrices} may seem unnecessarily prosaic: in common parlance a \emph{matrix} is an array of numbers. It is tempting to take a more abstract (if not rarified) attitude that any such matrix merely describes the action of some automorphism on the \emph{basis} elements of some vector space. Changing this basis would alter the matrices by a \emph{similarity transform}; but it is easy to see that this would not only preserve the general statement of (\ref{span_form}), but also leave unchanged all the coefficients $\r{f}\indices{\r{a\,b}}{\r{[d]}}$. 

Whenever generators $\mathbf{T}\indices{\r{a}}{}$ and $\mathbf{S}\indices{\r{a}}{}$ are related by a change of basis (identical for all $\r{a}\!\in\!\r{[d]}$), then we say that they describe \emph{similar} representations, and write $\mathbf{T}\indices{\r{[d]}}{}\!\simeq\!\mathbf{S}\indices{\r{[d]}}{}$. Of course, they need not be \emph{identical} as arrays of numbers; but they will satisfy \emph{identical} sets of commutation relations with identical coefficients $\r{f}\indices{\r{a\,b}}{\r{[d]}}$. We reserve the symbol for \emph{equality}, `$\mathbf{T}\indices{\r{[d]}}{}\!=\!\mathbf{S}\indices{\r{[d]}}{}$', \emph{exclusively} for when \emph{every component} of \emph{every} generator is identical. Thus, similar sets of generators are automatically \emph{compatible}; this motivates a justifiable degree of abstractness in our present discussion.
%\\[-10pt]

\subsection{Constructing Other Sets of Compatible Generators}\label{building_generators_from_generators}\vspace{-4pt}

Given a set of generators $\mathbf{T}\indices{\r{[d]}}{}$ which satisfy the defining relationship (\ref{explicit_form_of_defining_relation}), it is natural to wonder what \emph{other} sets of generators may exist which satisfy the same commutation relations (\ref{explicit_form_of_defining_relation}) as $\mathbf{T}\indices{\r{[d]}}{}$ with \emph{identical} coefficients $\r{f}\indices{\r{a\,b}}{\r{[d]}}$ for all pairs of indices $\r{a},\r{b}\!\in\!\r{[d]}$. It turns out that given any one representation, an infinitude of others can always be generated.\\[-10pt]

One obvious set of possibly distinct (and possibly \emph{similar}) set of generators can be found by simply transposing both sides of (\ref{explicit_form_of_defining_relation}): 
\eq{\mathbf{\bar{T}}\indices{\r{[d]}}{}\quad\text{with}\quad\mathbf{\bar{T}}\indices{\r{a}}{}\equivR(\text{-}\,\mathbf{T}^{\r{a}})^{T}\,\fwboxL{0pt}{\quad\forall\,\,\r{a}\!\in\!\r{[d]}\,.}\label{conjugate_generators_defined}}
Here, the minus sign is required by the definition of the commutator, and how matrix multiplication reverses under transposition. 

Clearly the generators $\mathbf{T}$ and $\mathbf{\bar{T}}$ have the same size. If they are related by a change of basis (a `similarity transformation'), we say they are \emph{similar}, and that the representation in question is `real' (or, more precisely, \emph{self-conjugate}), and the matrix encoding the similarity transform between them (acting at the level of the components of each generator as a matrix) is called the \emph{self-duality} metric. That is, for each generator labelled by $\r{a}\!\in\!\r{[d]}$, there is some invertible matrix  $\mathbf{S}$ of the same size as each generator such that $\mathbf{T}\indices{\r{a}}{}=\mathbf{S}.\mathbf{\bar{T}}\indices{\r{a}}{}.\mathbf{S}^{\text{-}1}$. We will soon return to the question of how to assess whether or not such a similarity transformation exists.\\[-10pt]

Another (possibly distinct, possibly similar) set of generators which satisfy these same commutation relations are the coefficients $\r{f}\indices{\r{a\,b}}{\r{c}}$ themselves: for each $\r{a}$, define the generator 
\eq{(\mathbf{V}\indices{\r{a}}{})\indices{\r{[d]}}{\r{[d]}}\equivR \r{f}\indices{\r{[d]}\,\r{a}}{\r{[d]}}\,\big(=\text{-}\,\r{f}\indices{\r{a}\,\r{[d]}}{\r{[d]}}\big)\,.\label{induced_ad_generators_defined}}
The fact that these generators always satisfy the same commutation relations with the same commutation relations follows telescopically from applying a commutator to the RHS of (\ref{explicit_form_of_defining_relation}) and applying the Jacobi identity to matrices to re-write the nested commutators on  the LHS (provided none of the $\mathbf{T}\indices{\r{a}}{}$ are identically zero).\\[-10pt]

Notice that these \emph{precise} $(\r{d}\!\times\!\r{d})$ generators agree identically---\emph{component-wise}---for \emph{any} set of generators which satisfy the same commutation relations as those defined by $\mathbf{T}\indices{\r{[d]}}{}$. This is a very special representation, which we call the \emph{adjoint} `$\mathbf{\r{ad}}$'; and it is defined with a very specific set of generators defined with respect to the `basis' of generator indices of $\mathbf{T}\indices{\r{[d]}}{}$---and independent of the basis chosen for these matrices. We describe these particular generators of this \emph{induced} adjoint `$\mathbf{\r{ad}}(\mathbf{T})$'.\\[-10pt]

Because $\mathbf{\r{ad}}(\mathbf{T})$ will be \emph{identical} for \emph{any} set of compatible generators, and because $\mathbf{\r{ad}}(\mathbf{T})$ also satisfies these commutation relations, it is clear that $\mathbf{\r{ad}}(\mathbf{\r{ad}}(\mathbf{T}))=\mathbf{\r{ad}}(\mathbf{T})$; moreover, we see that $\mathbf{\r{ad}}(\mathbf{T})=\mathbf{\r{ad}}(\mathbf{\bar{T}})$ and $\mathbf{\r{ad}}(\mathbf{T})=\mathbf{\r{ad}}(\mathbf{S}.\mathbf{T}.\mathbf{S}^{\text{-}1})$ for any similarity transformation encoded by $\mathbf{S}$. 

To be clear, there will always be an infinitude of \emph{other} sets of generators \emph{similar to} those of (\ref{induced_ad_generators_defined}) defined by changing bases of the generators $\mathbf{V}\indices{\r{[d]}}{}$. These would have components defined by 
\eq{(\mathbf{\tilde{V}}\indices{\r{a}}{})\indices{\r{[{d'}]}}{\r{[{d'}]}}\equivR\sum_{\t{b},\t{c}\in\r{[d]}}\,\mathbf{M}\indices{\r{[d']}}{\t{b}}(\mathbf{V}\indices{\r{a}}{})\indices{\t{b}}{\t{c}}\mathbf{\bar{M}}\indices{\t{c}}{\r{[d']}}\,,}
where $\mathbf{\bar{M}}\equivR\mathbf{M}^{\text{-}1}$ for some invertible matrix $\mathbf{M}$. Notice that such a transformation has no effect on the ordering or basis chosen for the generators themselves; this should be contrasted with (\ref{reshuffling_effect_on_structure_consts}): changing a list of generators by taking some linear combinations has the effect of \emph{both} performing a similarity transformation on the induced adjoint \emph{and} also reshuffling the induced adjoint's generators. For now, we'd like to consider the generator indices as fixed once and for all, and only discuss representations \emph{compatible} with some set of \emph{defining} representation `$\mathbf{T}$'.\\[-10pt]

It turns out that $\mathbf{\r{ad}}(\mathbf{T})$ is necessarily self-conjugate so that $\mathbf{\r{ad}}(\mathbf{T})\!\simeq\mathbf{\r{\bar{ad}}}(\mathbf{T})$. In this case, the matrix encoding the similarity transform between their generators has a special name: the\footnotespecial{\textbf{Note}: matrices encoding similarity are not unique---they are only defined up to a rescaling. This is because $\mathbf{S}.\mathbf{T}.\mathbf{S}^{\text{-}1}=(\b{\gamma}\mathbf{S}).\mathbf{T}.(\b{\gamma}\mathbf{S})^{\text{-}1}$ for arbitrary constants $\b{\gamma}\!\neq\!0$. As such, there is a scaling ambiguity in this definition of `the' Killing metric, which can be fixed only by convention.} \emph{Killing form} or Killing \emph{metric}; this is a $(\r{d}\!\times\!\r{d})$ matrix denoted $\mathcal{\r{K}}\indices{\r{[{d}]}}{\r{[\bar{d}]}}$. Because the generators are transposed under the map defining the conjugate representation, a lower `$\r{[\bar{d}]}$' index is equivalent to an {upper} `$\r{[d]}$' index; and so we more often write this matrix $\mathcal{\r{K}}\indices{\r{[d]\,[d]}}{}=\mathcal{\r{K}}\indices{}{\r{[\bar{d}]\,[\bar{d}]}}$.\\[-10pt]

Before moving on, it is worth mentioning the existence of some other, fairly trivial collections of generators guaranteed to satisfy the same commutation relations as $\mathbf{T}\indices{\r{[d]}}{}$: a representation consisting simply of $\r{d}$ all-\emph{vanishing} matrices. These matrices can be of any size $(\g{q}\!\times\!\g{q})$ with $\g{q}\!>\!0$. We'll call these representations `$\mathbf{1}\indices{\oplus\g{q}}{}$'. Obviously, these generators do not furnish an \emph{interesting} representation; nor can they be used to meaningfully define any particular Lie algebra.

In a similar vein, we can freely add any number of vanishing rows/columns to any set of generators to define new generators of a larger size, which are trivially compatible with the original. Adding $\g{q}$ rows/columns to every generator amounts to defining a new representation we may describe as generated by the generators described as
\eq{\text{`}\mathbf{1}\indices{\oplus\g{q}}{}\oplus\mathbf{T}\text{'}\,.\label{adding_vanishing_rows_columns}}

\newpage
\subsubsection[Building Larger Representations from Representations]{Building Larger Representations from Representations}\vspace{-4pt}

So far, we have seen that given any set of generators $\mathbf{T}\indices{\r{[d]}}{}$ we can construct the generators $\mathbf{\bar{T}}\indices{\r{[d]}}{}$ and those of $\mathbf{\r{ad}}(\mathbf{T})$. These may all be identical, or similar. Nothing about the discussion above depended upon the precise form of the matrices $\mathbf{T}^\r{a}$---not their size, their `reducibility', etc. We'd like to now show that given any set of (non-trivial) generators which satisfy (\ref{explicit_form_of_defining_relation}), we can in fact construct infinitely many \emph{dissimilar} (or \emph{distinct}) other representations. 

To make this discussion somewhat more precise, let us agree that the generators $\mathbf{T}^{\r{a}}$ are of size $(\b{r}\!\times\!\b{r})$; and let's say that they \emph{define} a \emph{representation} we'll call `$\mathbf{\b{R}}$' of \emph{dimension} $\b{r}$. We can gather all the $\r{d}\!\times\!\b{r}\!\times\!\b{r}$ numbers that define the generators into an \emph{array} or `rank-(2,1) tensor' $\mathbf{T}(\mathbf{\b{R}})\indices{\r{[d]}\,\b{[r]}}{\b{[r]}}$ with \emph{dimensions} $\{\r{d},\b{r}|\b{r}\}$. We can associate the generators of the adjoint representation with a similar tensor $\mathbf{T}(\r{\mathbf{ad}})\indices{\r{[d]\,[d]}}{\r{[d]}}$.

One way we can generate new representations is via direct summation. Suppose we have two sets of (not necessarily distinct) generators $\mathbf{T}(\mathbf{\b{R}})$ and $\mathbf{T}(\mathbf{\g{S}})$; we say they are \emph{compatible} if $\mathbf{\r{ad}}(\mathbf{T}(\mathbf{\b{R}}))=\mathbf{\r{ad}}(\mathbf{T}(\mathbf{\g{S}}))$. We can construct a new set of generators defined by taking the block-diagonal `outer' sums of their generators. Somewhat pedantically, we could write,
\eq{\mathbf{T}(\mathbf{\b{R}}\oplus\mathbf{\g{S}})\indices{\r{[d]}\,\t{[(r{+}s)]}}{\t{[(r{+}s)]}}\equivR\left\{\mathbf{T}(\mathbf{\b{R}})\indices{\r{[d]}\,\t{c_1}}{\t{c_2}}\right\}_{\t{c_i}\in\b{[r]}\subset\t{[r{+}s]}}\bigcup\left\{\mathbf{T}(\mathbf{\g{S}})\indices{\r{[d]}\,\t{c_1}}{\t{c_2}}\right\}_{\t{c_i}\in(\t{[r{+}s]}\backslash\b{[r]})}\,.}
To be clear, the inclusion map $i\!:\!\b{[r]}\hookrightarrow\t{[r{+}s]}$ is not unique, and different choices would result in distinct but similar sets of generators for $\mathbf{\b{R}}\oplus\mathbf{\g{S}}$. However, as $\mathrm{dim}(\mathbf{\b{R}}\oplus\mathbf{\g{S}})=\mathrm{dim}(\mathbf{\b{R}}){+}\mathrm{dim}(\mathbf{\g{S}})$, we know at least that this new representation is not similar to either. It is fairly trivial to see that because $\mathbf{\r{ad}}(\mathbf{\b{R}})=\mathbf{\r{ad}}(\mathbf{\g{S}})$, these are both equal to $\mathbf{\r{ad}}(\mathbf{\b{R}}\oplus\mathbf{\g{S}})$, and so the generators $\mathbf{T}(\mathbf{\b{R}}\oplus\mathbf{\g{S}})$ furnish a compatible representation.

We can generalize this construction to include arbitrary summands of representations in the obvious way; for example, we may write
\eq{\mathbf{T}(\mathbf{\b{R}}^{\oplus\,\r{q}})\equivR\mathbf{T}(\underbrace{\mathbf{\b{R}}\oplus\cdots\oplus\mathbf{\b{R}}}_{\r{q}\text{ times}})\,.}

We now understand the notation introduced in the previous section: a set of $\r{d}$ $(\g{q}\!\times\!\g{q})$ vanishing matrices could be constructed as the outer sum of $\g{q}$ copies of the \emph{irreducible} representation `$\mathbf{1}$', defined by vanishing generators of size $(1\!\times\!1)$. Similarly, we see the justification for the notation in (\ref{adding_vanishing_rows_columns}) to describe adding $\g{q}$ copies of $\mathbf{1}$ to any representation's generators.

Given generators of `$\mathbf{1}^{\oplus\,\g{q}}\oplus\mathbf{\b{R}}$' constructed in this way, it would be straight-forward to \emph{project} onto the generators of the representation $\mathbf{\b{R}}$. However, it is worth bearing in mind that \emph{any similarity transformation} performed on the generators would result in a similar representation, from which it may not be so trivial to project out the generators of $\mathbf{\b{R}}$. We'll return to this issue in the following subsection.\\[-10pt]

More interestingly, we can define generators for the \emph{tensor product representation} \mbox{`$(\mathbf{\b{R}}\otimes\mathbf{\g{S}})$'} by starting with the rank-(3,2) tensor
\eq{\mathbf{T}(\mathbf{\b{R}}\otimes\mathbf{\g{S}})\indices{\r{[d]}\,\b{[r]}\,\g{[s]}}{\b{[r]}\,\g{[s]}}\equivR\mathbf{T}(\mathbf{\b{R}})\indices{\r{[d]}\,\b{[r]}}{\b{[r]}}\delta\indices{\g{[s]}}{\g{[s]}}{+}\delta\indices{\b{[r]}}{\b{[r]}}\mathbf{T}(\mathbf{\g{S}})\indices{\r{[d]}\,\g{[s]}}{\g{[s]}}}
and we may use any association $c\!:\!\b{[r]}\otimes\g{[s]}\hookrightarrow\t{[r\times s]}$ to define 
\eq{\mathbf{T}(\mathbf{\b{R}}\otimes\mathbf{\g{S}})\indices{\r{[d]}\,\t{[r\times s]}}{\t{[r\times s]}}\equivR \mathbf{T}(\mathbf{\b{R}}\otimes\mathbf{\g{S}})\indices{\r{[d]}\,{c}(\b{[r]},\g{[s]})}{{c}(\b{[r]},\g{[s]})}\,.\label{tensor_product_defined}}
(This should not be confused with the tensor product \emph{of} representations!) Of course $\mathrm{dim}(\mathbf{\b{R}}\otimes\mathbf{\g{S}})=\mathrm{dim}(\mathbf{\b{R}})\times\!\mathrm{dim}(\mathbf{\g{S}})$.

As with (outer) sums of representations, we can expand our notation for tensor products to define
\eq{\mathbf{T}(\mathbf{\b{R}}^{\otimes\,\r{q}})\equivR\mathbf{T}(\underbrace{\mathbf{\b{R}}\otimes\cdots\otimes\mathbf{\b{R}}}_{\r{q}\text{ times}})\,.}

When taking a tensor product between a representation and itself, we can use the fact that $\b{[r]}\otimes\b{[r]}\simeq(\b{[r]}\wedge\b{[r]})\oplus(\b{[r]}\odot\b{[r]})$ to define generators for the \emph{antisymmetric} tensor product $\mathbf{\b{R}}\wedge\mathbf{\b{R}}$ and \emph{symmetric} tensor product $\mathbf{\b{R}}\odot\mathbf{\b{R}}$. These generalize in the natural way, allowing us to construct
\eq{\mathbf{T}(\mathbf{\b{R}}^{\wedge\r{q}})\equivR\mathbf{T}(\wedge^{\r{q}}\mathbf{\b{R}})=\mathbf{T}(\underbrace{\mathbf{\b{R}}\wedge\cdots\wedge\mathbf{\b{R}}}_{\r{q}\text{ times}})\,,\quad\mathbf{T}(\mathbf{\b{R}}^{\odot\r{q}})\equivR\mathbf{T}(\odot^{\r{q}}\mathbf{\b{R}})=\mathbf{T}(\underbrace{\mathbf{\b{R}}\odot\cdots\odot\mathbf{\b{R}}}_{\r{q}\text{ times}})\,.}

Given the operators $\oplus,\otimes,\wedge,\odot$ it is clear that we can build an unbounded variety of new, \emph{always compatible} representations starting from any one of them (provided it is non-trivial). The vast majority of these, however, will be highly \emph{reducible}. A representation $\mathbf{\b{R}}$ is said to be \emph{reducible} if there exists any pair of (possibly trivial) representations $\mathbf{\g{S}},\mathbf{\r{T}}$ such that
\eq{\mathbf{\b{R}}\simeq\mathbf{\g{S}}\oplus\mathbf{\r{T}}\,.}

\subsubsection[Projecting Representations into Smaller Representations]{Projecting Reducible Representations into Smaller Representations}\vspace{-4pt}

Suppose we know that the representation $\mathbf{\b{R}}$ is reducible and that  $\mathbf{\b{R}}\simeq\mathbf{\g{S}}\oplus\mathbf{\r{T}}$. Then there must exist \emph{orthogonal projectors} $\mathbf{P}_{\mathbf{\g{S}}}\indices{\b{[r]}}{\b{[r]}}$ and $\mathbf{P}_{\mathbf{\r{T}}}\indices{\b{[r]}}{\b{[r]}}$, with ranks $\g{s}$ and $\r{t}$, respectively, such that $\mathbf{P}_{\mathbf{\g{S}}}.\mathbf{P}_{\mathbf{\r{T}}}=\mathbf{P}_{\mathbf{\r{T}}}.\mathbf{P}_{\mathbf{\g{S}}}{=}\mathbf{0}\indices{\b{[r]}}{\b{[r]}}$ and $\mathbf{P}_{\mathbf{\g{S}}}{+}\mathbf{P}_{\mathbf{\r{T}}}=\delta\indices{\b{[r]}}{\b{[r]}}$. Orthogonality and completeness imply that $\mathbf{P}_{\mathbf{\r{T}}}.\mathbf{P}_{\mathbf{\r{T}}}=\mathbf{P}_{\mathbf{\r{T}}}$ and $\mathbf{P}_{\mathbf{\g{S}}}.\mathbf{P}_{\mathbf{\g{S}}}=\mathbf{P}_{\mathbf{\g{S}}}$, and it is easy to see therefore that 
\eq{\mathbf{P}_{\mathbf{\g{S}}}.\mathbf{\b{R}}.\mathbf{P}_{\mathbf{\g{S}}}\quad\text{and}\quad\mathbf{P}_{\mathbf{\r{T}}}.\mathbf{\b{R}}.\mathbf{P}_{\mathbf{\r{T}}}\label{big_projections}}
will necessarily satisfy identical commutation relations as $\mathbf{\b{R}}$ and therefore, represent representations compatible with $\mathbf{\b{R}}$. Of course, the generators of (\ref{big_projections}) are of the same size as those of $\mathbf{\b{R}}$. But we'd like to claim that 
\eq{\mathbf{P}_{\mathbf{\g{S}}}.\mathbf{\b{R}}.\mathbf{P}_{\mathbf{\g{S}}}\simeq\mathbf{\g{S}}\oplus\mathbf{1}^{\oplus \mathrm{dim}(\mathbf{\r{T}})}\quad\text{and}\quad\mathbf{P}_{\mathbf{\r{T}}}.\mathbf{\b{R}}.\mathbf{P}_{\mathbf{\r{T}}}\simeq\mathbf{1}^{\oplus \mathrm{dim}(\mathbf{\g{S}})}\oplus\mathbf{\r{T}}\,.}
Moreover, from these projectors, we'd like to find the $(\g{s}\!\times\!\g{s})$ and $(\r{t}\!\times\!\r{t})$ generators that will result in representations are similar to those of $\mathbf{\g{S}}$ and $\mathbf{\r{T}}$, respectively. 

Consider one of the projectors $\mathbf{P}_{\mathbf{\r{T}}}$ of rank $\r{t}$. Knowing its rank, it should have $\r{t}$ linearly independent rows, so that 
\eq{\mathbf{P}_{\mathbf{\r{T}}}\equivL\sum_{\t{t}\in\r{[t]}}\mathbf{p}\indices{\b{[r]}}{\t{t}}\mathbf{\hat{p}}\indices{\t{t}}{\b{[r]}}\,,}
where $\mathbf{p}\indices{\b{[r]}}{\t{[t]}}$ is defined as a subset of the rows of $\mathbf{P}_{\mathbf{\r{T}}}$, and $\mathbf{\hat{p}}$ is defined as coefficients found for the expansion of all the rows into this linearly-dependent spanning set.

The fact that $\mathbf{P}_{\mathbf{\r{T}}}.\mathbf{P}_{\mathbf{\r{T}}}=\mathbf{P}_{\mathbf{\r{T}}}$ implies that 
\eq{\sum_{\t{r}\in\b{[r]}}\mathbf{\hat{p}}\indices{\r{[t]}}{\t{r}}\mathbf{p}\indices{\t{r}}{\r{[t]}}=\delta\indices{\r{[t]}}{\r{[t]}}\,,}
which allows us to conclude that 
\eq{\mathbf{\hat{p}}.\mathbf{\b{R}}.\mathbf{p}}
is also a representation compatible with $\mathbf{\b{R}}$, but now with generators of size $(\r{t}\!\times\!\r{t})$; moreover, this representation is necessarily similar to $\mathbf{\r{T}}$. The same, of course, applies for the projector $\mathbf{P}_{\mathbf{\g{S}}}$.%\\[-10pt]

Importantly, nothing about the discussion above requires that the representations $\mathbf{\g{S}}$ or $\mathbf{\r{T}}$ are irreducible.\\[-10pt]

One way to construct such projectors is via the construction of \emph{invariants}, such as the quadratic Casimir $\mathcal{C}_2[\mathbf{\b{R}}]$, defined via
\eq{\mathcal{C}_2[\mathbf{\b{R}}]\indices{\b{[r]}}{\b{[r]}}\equivR\sum_{\t{d_i}\in\r{[d]}}(\mathbf{\b{R}}.\mathbf{\b{R}})\indices{\b{[r]}\,\r{d_1\,d_2}}{\b{[r]}}\mathcal{\r{K}}_{\r{d_1\,d_2}}\,.}
It is easy to see that this operator commutes with all the generators of $\mathbf{\b{R}}$, and projectors like those described above can be easily defined for its eigen-subspaces. See \cite{birdtracks} (see also \cite{Bourjaily:2025hvq}) for a more thorough discussion. In principle, it is possible to define and compute a sufficient number of generalized Casimir operators such that their eigenvalues uniquely characterize every irreducible representation of any Lie algebra. In practice, we find that most representations of interest (such as all those of fundamental Dynkin weights) can be constructed using merely the eigen-spaces of the quadratic Casimir.

\newpage
\subsection{Classifying the Irreducible Representations of Simple Lie Algebras}\label{classification_of_irreps}\vspace{-4pt}

Let us now return to the question of how to classify or categorize the possible sets of generators compatible with those of some defining representation. For this discussion, it is useful to introduce the small distinction between a \emph{representation} `$\mathbf{\b{R}}$' and its generators `$\mathbf{T}(\mathbf{\b{R}})$'.\footnote{Later we will define the symbol `$\mathbf{\b{R}}$' as a tensor such that $\mathbf{T}(\mathbf{\b{R}})=\built{Transpose}\brace{\mathbf{\b{R}}}$; but for now, we'd merely like to simplify the notation used in our discussion.} Extending the discussion above, we would now say that `$\mathbf{\b{R}}\simeq\mathbf{\r{S}}$' if $\mathbf{T}(\mathbf{\b{R}})\simeq\mathbf{T}(\mathbf{\r{S}})$.

It is a famous result of representation theory that any representation $\mathbf{\b{R}}$ of a simple Lie algebra may be uniquely decomposed \emph{up to similarity}\footnote{The meaning of `up to similarity' will become sharper when we discuss \emph{concrete} representations in \mbox{section~\ref{section:concrete_reps}}. \textbf{Note}: these similarity transformations are never unique and often involve many free parameters.} as an outer sum of so-called \emph{irreducible} representations (`irreps') $\mathbf{\t{r}}$ with multiplicity $m\indices{\mathbf{\b{R}}}{\mathbf{\t{r}}}$:
\eq{\mathbf{\b{R}}\simeq\bigoplus_{\hspace{-10pt}\text{irreps }\mathbf{\t{r}}\hspace{-20pt}}\,\,m\indices{\mathbf{\b{R}}}{\mathbf{\t{r}}}\,\mathbf{\t{r}}\equivR\bigoplus_{\hspace{-10pt}\text{irreps }\mathbf{\t{r}}\hspace{-20pt}}\,\mathbf{\t{r}}^{\smash{\oplus\,m\indices{\mathbf{\b{R}}}{\mathbf{\t{r}}}}}\,.\label{decompsition_into_irreps}}
For any finite-dimensional representation, the number of terms contributing to (\ref{decompsition_into_irreps}) must be finite; therefore only a finite number of multiplicities can be non-vanishing. Thus, the question of describing general representations of simple Lie algebras can be reduced to one of describing and classifying their irreducible representations.\\[-10pt]

It is a famous result that the \emph{irreducible} representations of any rank-$\r{k}$ simple Lie algebra are uniquely characterized (again, up to isomorphism) by their \emph{Dynkin} (or `highest-weight') \emph{labels} $w(\mathbf{\b{r}})\!\in\!\mathbb{Z}^{\r{k}}_{\geq0}$ \cite{Dynkin:1957um}; and each $\r{k}$-tuple of non-negative integers labels some particular and distinct irreducible representation. 

When the (Cartan-)type of a simple Lie algebra is understood, we write Dynkin labels for irreducible representations as a sequence of non-negative integers in square brackets `$[w_{\r{1}}\cdots w_{\r{k}}]$' with $w_{\r{i}}\!\in\!\mathbb{Z}_{\geq0}$. In the package \rpackage, these labels are encoded by the syntax `\mbox{\hyperlink{label:w}{\fun{w}{\color{black}\brace{\var{alg}}\brace{$\var{w}_{\r{1}},\ldots,\var{w}_{\r{k}}$}}}}' where \var{alg}$\,\in$\{\algL{a},\algL{b},\algL{c},\algL{d},\algL{e},\algL{f},\algL{g}\} identifies the Cartan-{type} of the algebra in question.\\[-10pt]

We may extend the notion of Dynkin labels to non-irreducible representations according to their decompositions (\ref{decompsition_into_irreps}) in the natural way:
\eq{w(\mathbf{\b{R}})=\bigoplus_{\hspace{-10pt}\text{irreps }\mathbf{\t{r}}\hspace{-20pt}}\,\,m\indices{\mathbf{\b{R}}}{\mathbf{\t{r}}}\,w(\mathbf{\t{r}})\equivL\symbolL{plus}\brace{\symbolL{times}\brace{m\indices{\mathbf{\b{R}}}{[\t{0,\cdots,0}]},\hyperlink{label:w}{\fun{w}{\color{black}\brace{\var{alg}}\brace{\t{0,\cdots,0}}}}},\ldots},\vspace{-4pt}\label{dynkin_labels_of_general_reps}}
where we have included the \rpackage~syntax for these decompositions, as seen in the following example:
\mathematicaBox{
\mathematicaSequence[1]{egRep=\symbolL{power}\brace{\dynkinLabel{g}{1,0},3};\\
\funL[1]{nice}@\%}{\fig[-2pt]{1}{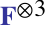}\vspace{-4pt}}
\mathematicaSequence{\funL[1]{tensorPowerDecomposition}@@egRep}{
\symbolL{plus}\texttt{[}\mbox{\symbolL{times}\brace{\hspace{-1pt}1\hspace{-1.02pt},\hspace{-1.02pt}\dynkinLabel{\algL{g}}{0\hspace{-1.17pt},\hspace{-1.17pt}0\hspace{-0.91pt}}\hspace{-0.9pt}}}\hspace{-1.11pt},%
\hspace{-1.11pt}\mbox{\symbolL{times}\brace{\hspace{-1pt}4\hspace{-1.02pt},\hspace{-1.02pt}\dynkinLabel{\algL{g}}{1\hspace{-1.17pt},\hspace{-1.17pt}0\hspace{-0.91pt}}\hspace{-0.9pt}}}\hspace{-1.11pt},%
\hspace{-1.11pt}\mbox{\symbolL{times}\brace{\hspace{-1pt}2\hspace{-1.02pt},\hspace{-1.02pt}\dynkinLabel{\algL{g}}{0\hspace{-1.17pt},\hspace{-1.17pt}1\hspace{-0.91pt}}\hspace{-0.9pt}}}\hspace{-1.11pt},%
\hspace{-1.11pt}\mbox{\symbolL{times}\brace{\hspace{-1pt}3\hspace{-1.02pt},\hspace{-1.02pt}\dynkinLabel{\algL{g}}{2\hspace{-1.17pt},\hspace{-1.17pt}0\hspace{-0.91pt}}\hspace{-0.9pt}}}\hspace{-1.11pt},%
\hspace{-1.11pt}\mbox{\symbolL{times}\brace{\hspace{-1pt}2\hspace{-1.02pt},\hspace{-1.02pt}\dynkinLabel{\algL{g}}{1\hspace{-1.17pt},\hspace{-1.17pt}1\hspace{-0.91pt}}\hspace{-0.9pt}}}\hspace{-1.11pt},%
\hspace{-1.11pt}\mbox{\symbolL{times}\brace{\hspace{-1pt}1\hspace{-1.02pt},\hspace{-1.02pt}\dynkinLabel{\algL{g}}{3\hspace{-1.17pt},\hspace{-1.17pt}0\hspace{-0.91pt}}\hspace{-0.9pt}}}\texttt{]}}
\mathematicaSequence{\funL[1]{nice}@\%}{\fig[-2pt]{1}{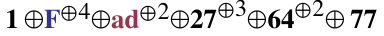}\vspace{-2pt}}
}~\\[-2pt]
(Although not easily translated here, when evaluated in any \textsc{Mathematica} \built{Notebook}, the output of \funL[4]{nice} will show Dynkin labels upon mouse-over.)

Notice the \emph{equality} appearing in (\ref{dynkin_labels_of_general_reps}): the Dynkin label of two similar representations will be \emph{identical}.\\[-10pt]

The best way to understand how irreducible representations can be labelled uniquely by Dynkin labels, and how to extract from this information decompositions such as (\ref{dynkin_labels_of_general_reps}), is through the connection between representations of simple Lie algebras and their associated \emph{weight systems}.

\subsection{Weight Systems for Representations of Simple Lie Algebras}\label{weight_lattices_for_reps}\vspace{-4pt}

Many representation-theoretic questions can be most conveniently answered in terms of the connection between representations and collections (or `lattices') of \emph{weights}---sets of points in $\mathbb{Z}^{\r{k}}$ with multiplicity. We call such a collection a \emph{weight system} for the representation.

For any (not necessarily irreducible) representation $\mathbf{\b{R}}$ of a simple, rank-$\r{k}$ Lie algebra, one may define a collection of \emph{weights} $\Lambda(\mathbf{\b{R}})\subset\mathbb{Z}^{\r{k}}$ with \emph{multiplicities} $m\indices{\mathbf{\b{r}}}{\lambda}$; that is, 
\eq{\Lambda(\mathbf{\b{R}})\equivR\bigoplus_{\hspace{-10pt}\lambda\in\mathbb{Z}^{\r{k}}\hspace{-10pt}}\lambda^{\smash{\oplus\,m\indices{\mathbf{\b{R}}}{\lambda}}}\equivL\bigoplus_{\hspace{-10pt}\lambda\in\mathbb{Z}^{\r{k}}\hspace{-10pt}}\,\smash{m\indices{\mathbf{\b{R}}}{\lambda}}\,\lambda\,.\label{weight_lattice_with_multiplicity}}
The weight lattices of representations (encoded \emph{abstractly} in terms of their \hyperlink{label:w}{Dynkin labels}) can be shown using \rpackage~as in the following example:
\mathematicaBox{
\mathematicaSequence[1]{egRep=\symbolL{power}\brace{\dynkinLabel{g}{1,0},2};\\
\funDL[1]{showWeightLattice}\brace{egRep}\\[-10pt]}{\fig[-5pt]{1}{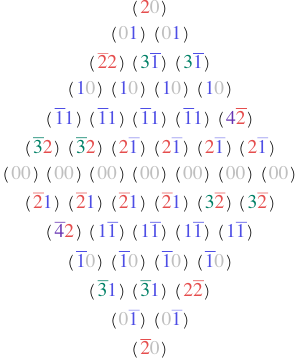}}
}~\\[-4pt]
Notice that the function \funDL[4]{showWeightLattice} shows the list of weights ordered from \emph{highest} to \emph{lowest} \emph{with duplication encoding multiplicity}, and stylizes negative numbers as $\bar{q}\equivR\text{-}q$. (The function \funDL[4]{representationWeights} returns the \emph{unformatted} collection of weights as a \built{SparseArray}.)\\[-8pt]

The \emph{dimension} of the representation is given by a sum over all possible weights, weighted by their multiplicity:
\eq{\mathrm{dim}(\mathbf{\b{R}})\equivL\left|\Lambda(\mathbf{\b{R}})\right|\equivR\sum_{\hspace{-10pt}\lambda\in\mathbb{Z}^{\r{k}}\hspace{-10pt}}\,m\indices{\mathbf{\b{R}}}{\lambda}\,.\vspace{-9pt}}
\mathematicaBox{
\mathematicaSequence[1]{egRep=\symbolL{power}\brace{\dynkinLabel{g}{1,0},2};\\
\funDL[1]{representationWeights}@egRep\\[-10pt]}{\fig[-2pt]{1}{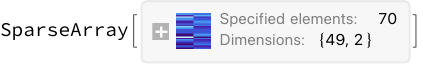}\vspace{-4pt}}
\mathematicaSequence{\funDL[1]{representationDimension}@egRep==\built{Length}@\%}{\texttt{True}}
}~\\[-10pt]

Taking the outer-sum of two representations, weight lattices add according to
\eq{\Lambda(\mathbf{\b{R}}\oplus\mathbf{\r{S}})=\bigoplus_{\lambda\in\mathbb{Z}^{\r{k}}}\,(m\indices{\mathbf{\b{R}}}{\lambda}{+}m\indices{\mathbf{\r{S}}}{\lambda})\lambda\,,\vspace{-4pt}}
which suggests that we allow for a slight abuse of notation to define
\eq{\Lambda(\mathbf{\b{R}}^{\smash{\oplus \r{q}}})=\bigoplus_{\hspace{-10pt}\lambda\in\mathbb{Z}^{\r{k}}\hspace{-10pt}}\left(\r{q}\,m\indices{\mathbf{\b{R}}}{\lambda}\right)\,\lambda\equivL\,\r{q}\,\Lambda(\mathbf{\b{R}})\,,\label{weight_lattice_summands}\vspace{-4pt}}
where multiplication here acts only on \emph{multiplicities} (and \emph{not} as a map $\mathbb{Z}^{\r{k}}\!\mapsto\!\mathbb{Z}^{\r{k}}$ rescaling weights). Given this notation, the decomposition of $\mathbf{\b{R}}$ into \emph{irreducible} representations (\ref{decompsition_into_irreps}) translates into the lattice-level statement that 
\eq{\Lambda(\mathbf{\b{R}})=\bigoplus_{\hspace{-10pt}\text{irreps }\mathbf{\t{r}}\hspace{-20pt}}\,\,m\indices{\mathbf{\b{R}}}{\mathbf{\t{r}}}\,\Lambda(\mathbf{\t{r}})\,.\label{irrep_decomp_via_lattices}\vspace{-4pt}}

In comparison with the example given above, we find that the weight lattice can be decomposed into those of irreducible representations according to:
\mathematicaBox{
\mathematicaSequence[1]{egRep=\symbolL{power}\brace{\dynkinLabel{g}{1,0},2};\\
egRepToIrreps=\funDL[1]{tensorPowerDecomposition}@@egRep}{\symbolL{plus}\texttt{[}\mbox{\symbolL{times}\brace{\hspace{-1pt}1\hspace{-1.02pt},\hspace{-1.02pt}\dynkinLabel{\algL{g}}{0\hspace{-1.17pt},\hspace{-1.17pt}0\hspace{-0.91pt}}\hspace{-0.9pt}}}\hspace{-1.11pt},%
\hspace{-1.11pt}\mbox{\symbolL{times}\brace{\hspace{-1pt}1\hspace{-1.02pt},\hspace{-1.02pt}\dynkinLabel{\algL{g}}{1\hspace{-1.17pt},\hspace{-1.17pt}0\hspace{-0.91pt}}\hspace{-0.9pt}}}\hspace{-1.11pt},%
\hspace{-1.11pt}\mbox{\symbolL{times}\brace{\hspace{-1pt}1\hspace{-1.02pt},\hspace{-1.02pt}\dynkinLabel{\algL{g}}{0\hspace{-1.17pt},\hspace{-1.17pt}1\hspace{-0.91pt}}\hspace{-0.9pt}}}\hspace{-1.11pt},%
\hspace{-1.11pt}\mbox{\symbolL{times}\brace{\hspace{-1pt}1\hspace{-1.02pt},\hspace{-1.02pt}\dynkinLabel{\algL{g}}{2\hspace{-1.17pt},\hspace{-1.17pt}0\hspace{-0.91pt}}\hspace{-0.9pt}}}\texttt{]}}
\mathematicaSequence{\funDL[1]{showWeightLattice}\brace{egRepToIrreps}\\[-10pt]}{\fig[-3pt]{1}{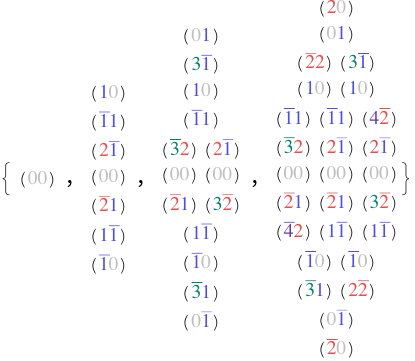}}
}~\\[-14pt]

But in order to perform such a reduction, it is crucial to first know how the lattices of irreducible representations can be constructed.\\[-10pt]

Recall that irreducible representations are uniquely identified by their Dynkin (`highest-weight') labels, which always consist of a collection of non-negative integers. Such a label defines a particular weight lattice as follows. For each simple Lie algebra $\mathfrak{g}$ there exists $\mathrm{rank}(\mathfrak{g})$ \emph{simple} weights (also called `roots') $\alpha_i\!\in\!\mathbb{Z}^{\r{k}}$; these are encoded by the rows of the \emph{Cartan matrix} of $\mathfrak{g}$, which are weights of the \emph{adjoint} (always irreducible) representation. The Dynkin label of the irrep, of course, indicates one particular vector $\lambda\!\in\!\mathbb{Z}^{\r{k}}$: namely, the \emph{highest weight}, which always has multiplicity 1; starting from any weight $\lambda\!\in\!\Lambda(\mathbf{\b{r}})$, one may subtract the $\r{i}$th simple weight $\alpha_{\r{i}}$ up to $\lambda_{\r{i}}$ times, for each $\lambda_{\r{i}}\!>\!0$, resulting in another element in $\Lambda(\mathbf{\b{r}})$. This results in a collection of weights which have non-vanishing multiplicities in $\Lambda(\mathbf{\b{r}})$. The multiplicity each weight that appears can be determined recursively and effectively via Freudenthal's multiplicity formula (see \emph{e.g.}~\cite{MR656202}).\\[-8pt]

Given the weight lattice of some representation $\Lambda(\mathbf{\b{R}})$, it is a straightforward exercise to decompose its lattice---and hence also the representation---into a sum over those of irreducible representations according to (\ref{irrep_decomp_via_lattices}): one starts with the \emph{highest}\footnote{The highest weight is identified by that for which $\sum_{\r{i},\r{j}}\lambda_{\r{i}}.(C[\mathfrak{g}]^{\text{-}1})\indices{\r{i}}{\r{j}}$ is maximal; here, $C[\mathfrak{g}]$ is the \emph{Cartan matrix} for the algebra.} weight appearing in $\Lambda(\mathbf{\b{R}})$, and removes this weight's multiplicity times the weight lattice of the irreducible representation labelled by that as its highest weight, resulting in a new lattice with a necessarily lower highest-weight element; such subtractions are repeated until all weights in $\Lambda(\mathbf{\b{R}})$ have been removed.

\subsection{Actions and Operations on Representations' Weight Systems}\label{rep_operations_via_weights}\vspace{-4pt}

There are a number of useful representation-theoretic constructions that have a natural translation in terms of systems of weights, allowing one to efficiently answer many representation-theoretic questions of interest. We have already seen that the decomposition of any representation into a sum over irreducible ones can be effectively translated into an algorithm acting on weight systems.\\[-10pt]

There is a natural isomorphism between weights and their negations, and it is not hard to show that if there exists a representation lattice $\Lambda(\mathbf{\b{R}})$ with weight multiplicities $m\indices{\mathbf{\b{R}}}{\lambda}$, then a weight system with multiplicities $m\indices{\mathbf{\b{\bar{R}}}}{\lambda}\equivR m\indices{\mathbf{\b{R}}}{\hspace{-5pt}{-}\lambda}$ also describes the weight system of some representation---which we may identify with its (`complex') \emph{conjugate} `$\mathbf{\b{\bar{R}}}$'. Only somewhat glibly, we may write
\eq{\Lambda(\bar{\mathbf{\b{R}}})\equivR{-}\Lambda(\mathbf{\b{R}})\,,}
defining an automorphism among Lie algebra representations. When $\Lambda(\mathbf{\b{R}})=\Lambda(\mathbf{\b{\bar{R}}})$, we say the representation is \emph{real}; otherwise, it is sometimes called `complex'. Many of the simple Lie algebras admit only real representations---namely, $\mathfrak{b}_{\r{k}},\mathfrak{c}_{\r{k}},\mathfrak{e}_{\r{7}},\mathfrak{e}_{\r{8}},\mathfrak{f}_{\r{4}},$ and $\mathfrak{g}_{\r{2}}$.

Given any two (not necessarily irreducible) representations $\mathbf{\b{R}}$ and $\mathbf{\r{S}}$, one may define the tensor product `$\mathbf{\b{R}}\!\otimes\!\mathbf{\r{S}}$' representation by the following action on weight lattices:
\eq{\Lambda(\mathbf{\b{R}}\!\otimes\!\mathbf{\r{S}})=\bigoplus_{\hspace{-10pt}\lambda,{\mu}\in\mathbb{Z}^{\r{k}}\hspace{-10pt}}\,\,\left(m\indices{\mathbf{\b{R}}}{\lambda}m\indices{\mathbf{\r{S}}}{{\mu}}\right)(\lambda{+}{\mu})\,.\label{tensor_product_representations_via_weights}}
The decomposition of the tensor product weight lattice into those of irreducible representations can be effectively accomplished via the Racah-Speiser algorithm, \cite{MR170975,MR171552}.\\[-10pt]

The specialization of (\ref{tensor_product_representations_via_weights}) to the case
\eq{\Lambda(\mathbf{\b{R}}\indices{\otimes\r{n}}{})\equivR\Lambda(\underbrace{\mathbf{\b{R}}\otimes\cdots\otimes\mathbf{\b{R}}}_{\r{n}\text{ times}})\equivL\Lambda(\symbolL{power}\brace{\mathbf{\b{R}},\r{n}})\,\label{tensor_power_defined_0}}
can be refined even further refined to define the symmetric and anti-symmetric tensor powers,
\eq{\Lambda(\mathbf{\b{R}}\indices{\odot\r{n}}{})\equivR\Lambda(\underbrace{\mathbf{\b{R}}\odot\cdots\odot\mathbf{\b{R}}}_{\r{n}\text{ times}})\equivL\Lambda(\symbolL{powers}\brace{\mathbf{\b{R}},\r{n}})\,\label{symmetric_tensor_power_defined_0}}
and 
\eq{\Lambda(\mathbf{\b{R}}\indices{\wedge\r{n}}{})\equivR\Lambda(\underbrace{\mathbf{\b{R}}\wedge\cdots\wedge\mathbf{\b{R}}}_{\r{n}\text{ times}})\equivL\Lambda(\symbolL{powera}\brace{\mathbf{\b{R}},\r{n}})\,,\label{antisymmetric_tensor_power_defined_0}}
respectively. 

Tensor powers and their symmetric and anti-symmetric variants can be effectively implemented at the level of lattices more concretely in terms of \emph{character theory}, which allows one to more directly (functionally) work with weight lattices and their multiplicities as they appear in (\ref{weight_lattice_with_multiplicity}).  We will not have more to say about this construction, other than to note that operationally, characters are no more than a different way to keep track of multiplicities in weight lattices.

\vspace{\fill}
\subsection{Our Conventions for Weight Systems \& Labelling Representations}\label{subsec:conventions_for_weight_systems}\vspace{-4pt}

While many references (and computational tools) describe irreducible representations in terms of their Dynkin labels (and systems of weights), there is broad variability in how authors choose to \emph{order} the indices $w_{\r{i}}(\mathbf{\b{r}})$ appearing in these labels. That is, there is no general agreement as to \emph{which} particular irreducible representation of a given algebra is assigned which particular Dynkin label. It is therefore worthwhile to always check any reference or utility's conventions for such choices before using results. 

Ultimately, this variability stems from the choice made for how the `simple roots' should be ordered, which can be seen in a number of ways. The conventions used by the package \rpackage~can be summarized using \funAL{showLatticeData}:\\[-18pt]
\mathematicaBox{
\mathematicaSequence{\funAL[1]{showLatticeData}\brace{\algL{e}\brace{6}}\\[-10pt]}{\fig[-0pt]{0.72}{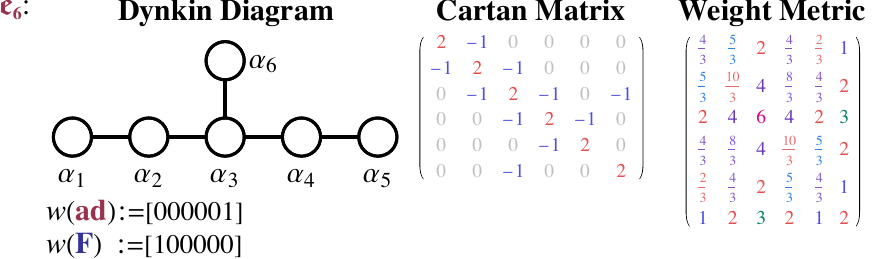}\vspace{-4pt}}
\mathematicaSequence{\funAL[1]{showLatticeData}\brace{\algL{e}\brace{7}}\\[-10pt]}{\fig[-0pt]{0.72}{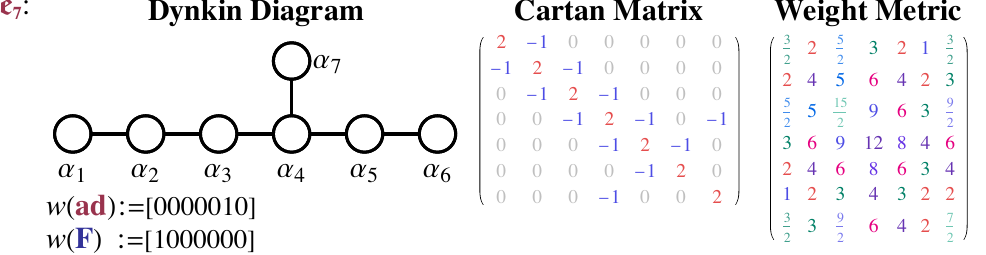}\vspace{-4pt}}
\mathematicaSequence{\funAL[1]{showLatticeData}\brace{\algL{e}\brace{8}}\\[-10pt]}{\fig[-0pt]{0.72}{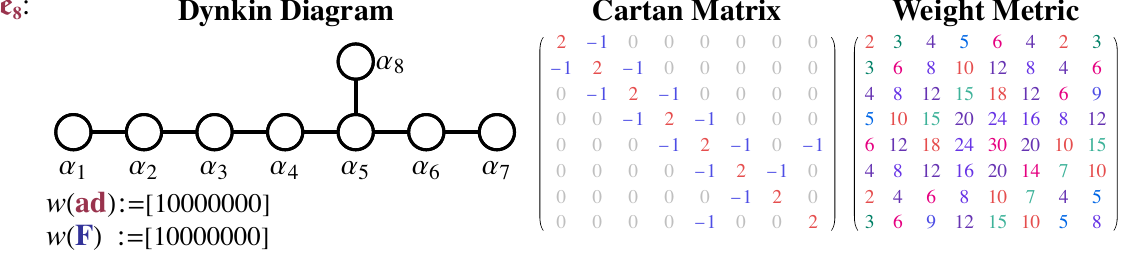}\vspace{-4pt}}
\mathematicaSequence{\funAL[1]{showLatticeData}\brace{\algL{f}\brace{4}}\\[-10pt]}{\fig[-0pt]{0.72}{showLatticeData_f4}\vspace{-4pt}}
\mathematicaSequence{\funAL[1]{showLatticeData}\brace{\algL{g}\brace{2}}\\[-10pt]}{\fig[-0pt]{0.72}{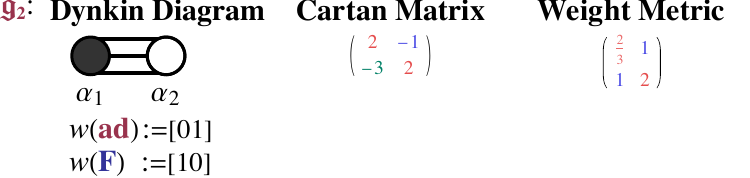}\vspace{-2pt}}
}

These conventions can often be parsed effectively by knowing how Dynkin labels are assigned to a number of irreducible representations of note, such as $\mathbf{\r{ad}}$. Among the most important irreducible representations of any Lie algebra are those of so-called \emph{fundamental} Dynkin weights: irreps $\mathbf{\b{r}}$ for which $\sum_{\r{i}}w_\r{i}(\mathbf{\b{r}}){=}1$. There are always exactly $\r{k}$ such `fundamental' representations denoted $\mathbf{\b{F}}_{\r{i}}$, and these can often be identified simply by their dimension.

The conventions for labelling the irreducible representations of fundamental weight can be summarized via the function \funAL{showFundamentalIrrepLabels}. For the classical algebras, there is relatively less variability in these conventions,
\mathematicaBox{
\mathematicaSequence{\funL{showFundamentalIrrepLabels}/@\{\algL{a}\brace{5},\algL{b}\brace{5},\algL{c}\brace{5},\algL{d}\brace{5}\}\\[-10pt]}
{\{\fig[-3.2pt]{1}{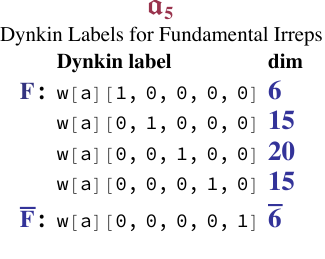},\fig[-0pt]{1}{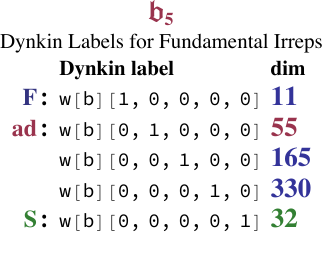},\\[-20pt]\phantom{\{}\fig[-0.9pt]{1}{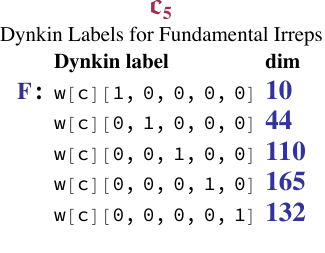},\fig[0.8pt]{1}{showFundamentalIrrepLabels_d5}\}\vspace{-8pt}}
}~\\[-4pt]
with most conventional variation being for the exceptional algebras:
\mathematicaBox{\mathematicaSequence{\funL[1]{showFundamentalIrrepLabels}/@\{\algL{f}\brace{4},\algL{g}\brace{2}\}\\[-18pt]}{\{
\phantom{\texttt{\{}}\fig[-14.pt]{1}{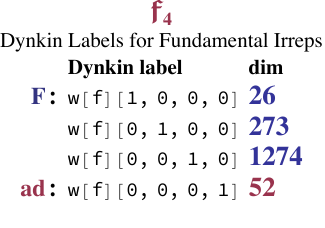},\fig[-0pt]{1}{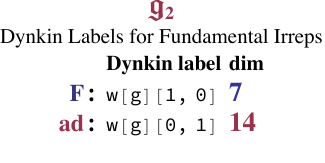}\}\vspace{-13pt}}
}~\\[-4pt]
\newpage
\mathematicaBox{
\mathematicaSequence{\funL[1]{showFundamentalIrrepLabels}/@\{\algL{e}\brace{6},\algL{e}\brace{7},\algL{e}\brace{8}\}\\[-18pt]}{\{\fig[-0pt]{1}{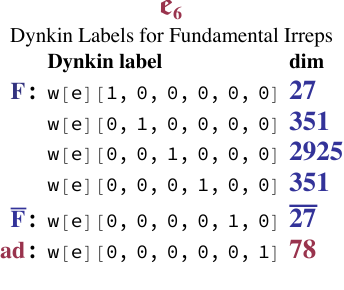},\fig[-5.6pt]{1}{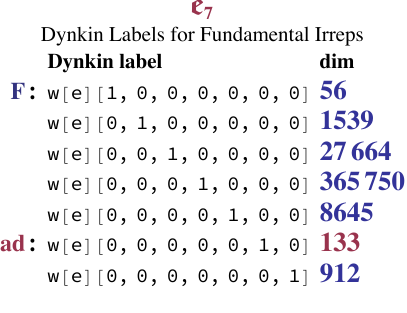},\\[-16pt] \phantom{\{}\fig[-7.4pt]{1}{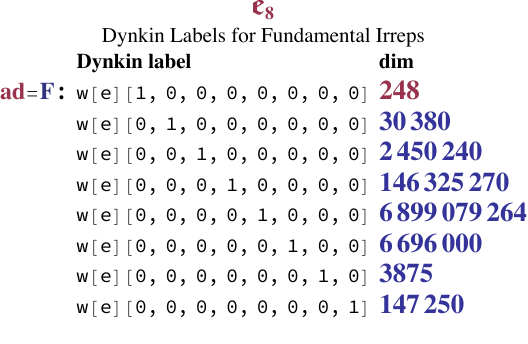}\}\vspace{-13pt}}
}~\\[-4pt]

Notice that the package always identifies $\mathbf{\b{F}}_{\r{1}}\equivL$`$\mathbf{\b{F}}$' as \emph{the} `fundamental' representation; also, the \emph{spinor} representation `$\mathbf{\g{S}}$' of $\mathfrak{b}_{\r{k}}$ is always associated with $\mathbf{\b{F}}_{\r{k}}$, and the two spinor representations `$\mathbf{\g{S}}_{\g{1}}$' and `$\mathbf{\g{S}}_{\g{2}}$' of $\mathfrak{d}_{\r{k}}$ are identified with $\mathbf{\b{F}}_{\r{k}}$ and $\mathbf{\b{F}}_{\r{k\text{-}1}}$, respectively. (For algebras for which the adjoint representation $\mathbf{\r{ad}}$ is of fundamental weight, it too is labelled accordingly.)\\[-10pt]

Beyond their usefulness in disambiguating conventions, the fundamental representations play a particularly important role, as \emph{all} irreducible representations can be \emph{generated} (upon projections) from those of fundamental weight via tensor products, as\\[-12pt]
\eq{\bigotimes_{\r{i}}\left(\mathbf{\b{F}}^{\smash{\odot\,{w_\r{i}(\mathbf{\b{r}})}}}_{\r{i}}\right)\simeq\mathbf{\b{r}}\oplus\ldots\,.\vspace{-2pt}}
By this, we mean that the irrep $\mathbf{\b{r}}$ appears on the right hand side with \emph{unit multiplicity}.\\[-10pt]

In fact, a more powerful statement can be made: there is always at least one fundamental-weight irrep whose tensor products can generate all other representations. Many authors choose to designate this fundamental-weight irrep (or one of them) as \emph{the fundamental}, but we do not follow that choice here; what we have designated by $\mathbf{\b{F}}\equivR\mathbf{\b{F}}_{\r{1}}$ is more often identified as merely a \emph{defining} representation. 

\newpage 
To be clear, in our conventions, what we designate as $\mathbf{\b{F}}$ does generate most other irreps of fundamental weight fairly directly. Specifically, starting from $\mathbf{\b{F}}$, we can generate the other fundamental-weight irreducible representations as follows:
\eq{\mathbf{\b{F}}^{\smash{\wedge\r{i}}}\simeq\mathbf{\b{F}}_{\r{i}}\oplus\ldots\quad\text{for all }\left\{\begin{array}{ll}\r{i}\leq\!\r{k}&\mathfrak{a}_{\r{k}}, \mathfrak{c}_{\r{k}}, \mathfrak{f}_{\r{4}}, \mathfrak{g}_{\r{2}}\\
\r{i}\!<\!\r{k}&\mathfrak{b}_{\r{k}}\\
\r{i}\!<\!\r{k{-}1}&\mathfrak{d}_{\r{k}}\\
\r{i}\!<\!\r{k{-}3}&\mathfrak{e}_{\r{k}}\end{array}\right.\,.\label{generating_fundamental_weight_irreps}}
For the $\mathfrak{e}_{\r{k}}$ algebras, the remaining irreducible representations of fundamental weight can be generated from $\mathbf{\b{F}}$---namely, as appearing (with unit multiplicity) in the following constructions:
\eq{\fwboxR{0pt}{(\mathfrak{e}_{\r{k}})\hspace{57pt}}\mathbf{\b{F}}_{\r{k{-}2}}\!\subset\!\left(\mathbf{\b{F}}\indices{\odot\,2}{}\right)\wedge\left(\mathbf{\b{F}}\indices{\odot\,2}{}\right)\,,\quad\mathbf{\b{F}}_{\r{k{-}1}}\!\subset\!\mathbf{\b{F}}\indices{\odot\,2}\,,\quad\mathbf{\b{F}}_{\r{k}}\!\subset\!\left(\mathbf{\b{F}}\indices{\odot\,2}{}\right)\!\otimes\!\mathbf{\b{F}}\,.
\label{exceptional_fundamental_irrep_constructions}}
But this is not true for the $\mathfrak{so}$-type Lie algebras, for which there exist \emph{spinor} representations which cannot be generated from what we have denoted $\mathbf{\b{F}}$. Specifically, for $\mathfrak{b}_{\r{k}}$ Lie algebras, there exists a single spinor representation denoted `$\mathbf{\g{S}}$' and identified with $\mathbf{\b{F}}_{\r{k}}$ in our conventions. For the Lie algebras $\mathfrak{d}_{\r{k}}$, there are two spinor representations, denoted $\mathbf{\g{S}}_{\g{1}}$ and $\mathbf{\g{S}}_{\g{2}}$, identified with $\mathbf{\b{F}}_{\r{k}}$ and $\mathbf{\b{F}}_{\r{k\text{-}1}}$ in our conventions, respectively.

\newpage
\section{\emph{Concrete} Representations of Lie Algebras}\label{section:concrete_reps}\vspace{-4pt}

The abstract description of Lie algebras and their representations given in the previous section, while sufficient (and powerful enough) to address many questions of interest, there are many reasons why one may want \emph{explicit} sets of generators, encoded by \emph{matrices} of specific \emph{numeric} values (which encode the action of these generators on some choice of basis elements of a vector space).\\[-10pt]

We consider a \emph{concrete} representation $\mathbf{\b{R}}$ to be defined by an \emph{explicit array of numbers} which encode the action of its generators on some choice of basis. There are $\mathrm{dim}(\mathfrak{\r{g}})$ generators, each of some size $(\b{r}\!\times\!\b{r})$; we choose to encode the representation as a rank-(2,1) tensor with dimensions $\{\b{r},\mathrm{dim}(\mathfrak{\r{g}})|\b{r}\}$:
\vspace{-3pt}\eq{\text{`}\mathbf{\b{R}}\text{'}\;\;\;\bigger{\Leftrightarrow}\;\;\;\mathbf{\b{R}}\indices{\b{[r]}\,\r{[\mathfrak{g}]}}{\b{[r]}}\;\;\;\bigger{\Leftrightarrow}\;\;\;\fig[-2pt]{0.8}{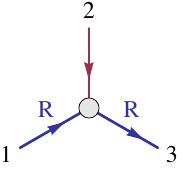}\,.\vspace{-7pt}}
In these figures, the labels on external edges indicate the \emph{slot position} of the corresponding index of the array, and orientations signify whether the index should be understood as raised or lowered.\\[-10pt]

The \emph{generators} of the representation `$\mathbf{T}(\mathbf{\b{R}})$' are most naturally thought of as the rank-(2,1) tensor with components spanning dimensions $\{\mathrm{dim}(\mathfrak{\r{g}}),\!\b{r}|\b{r}\}$; in \textsc{Mathematica}, this corresponds to the association 
\eq{\text{`}\mathbf{T}(\mathbf{\b{R}})\text{'}\;\;\;\bigger{\Leftrightarrow}\;\;\;\mathbf{T}(\mathbf{\b{R}})\indices{\r{[\mathfrak{g}]}\,\b{[r]}}{\b{[r]}}\equivR\built{Transpose}\brace{\mathbf{\b{R}}}\;\;\;\bigger{\Leftrightarrow}\;\;\;\fig[-2pt]{0.8}{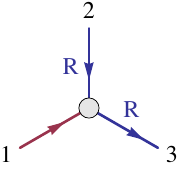}\,}
so that $\mathbf{T}(\mathbf{\b{R}})$ can be understood as a \built{List} (or \built{Array}) of $\mathrm{dim}(\mathfrak{\r{g}})$ generators.\\[-10pt]

Our choice to reorder indices (by a `\built{Transpose}') between `$\mathbf{\b{R}}$' and `$\mathbf{T}(\mathbf{\b{R}})$' is motivated by a desire to more easily contract indices between generators. The function \built{Dot} in \textsc{Mathematica} (or our refinement thereof, `\funL[4]{dot}') sums over indices paired between the \built{Last} \built{Slot} of one argument with the \built{First} \built{Slot} of the next. To construct the tensor `$\mathbf{\b{R}}.\mathbf{\b{R}}$', we can simply take their \built{Dot} (or \funL[4]{dot}) product:
\vspace{-4pt}\eq{\text{`}\mathbf{\b{R}}.\mathbf{\b{R}}\text{'}\;\;\;\bigger{\Leftrightarrow}\;\;\;(\mathbf{\b{R}}.\mathbf{\b{R}})\indices{\b{[r]}\,\r{[\mathfrak{g}]\,[\mathfrak{g}]}}{\b{[r]}}\equivR\funL{dot}\brace{\mathbf{\b{R}},\mathbf{\b{R}}}\;\;\;\bigger{\Leftrightarrow}\;\;\;\fig[-2pt]{0.8}{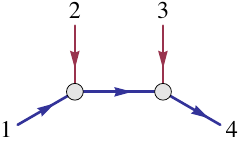}\vspace{-9pt}}
The building of such a tensor from $\mathbf{T}(\mathbf{\b{R}})$ using \built{Dot} (or \funL[4]{dot}) would require additional calls to \built{Transpose} to line up their \built{Sequence} of indices appropriately. 

\newpage
\subsection{Avoiding Unnecessary Complexifications in Concrete Representations}\label{avoiding_complexity}

One reason for an interest in concrete generators is that the bases of vector spaces on which generators act typically \emph{mean something} in physics. For example, the Gell-Mann matrices $\mathbf{\lambda}\indices{\r{[8]}\,\b{[3]}}{\b{[3]}}$, which furnish a 3-dimensional representation of $\mathfrak{su}_{3}$ \cite{Gell-Mann:1962yej} are \emph{Hermitian}, and so encode the \emph{unitary} group elements
\eq{U(\vec{\alpha})\indices{\b{[3]}}{\b{[3]}}\equivR\exp\big(i\!\sum_{\t{a}\in\r{[8]}}\alpha_{\t{a}}\lambda\indices{\t{a}\,\b{[3]}}{\b{[3]}}\big)\in SO(3)\,.}
By this, we mean that for any $\vec{\alpha}\!\in\!\mathbb{R}^{\r{8}}$, $U(\vec{\alpha})$ acts \emph{unitarily} on the space describing the $\b{[3]}$-dimensional space of unitary and normalized \emph{colour-states} of quarks. That is, there is a basis of quarks $q\indices{\b{[3]}}{}\equivR\{q^{\b{1}},q^{\b{2}},q^{\b{3}}\}$ whose labels \emph{mean something} (at least to a physicist): different linear combinations would refer to \emph{different} states.  

Because $\mathrm{exp}(i\,\mathbf{H})$ will be unitary iff $\mathbf{H}$ is Hermitian (equivalently, if $i\,\mathbf{H}$ is anti-Hermitian), physicists are often most interested in \emph{Hermitian} (or anti-Hermitian) generators of Lie algebras. Representations with such generators are sometimes confusingly called `\emph{real}' representations by physicists. We do not follow this vernacular.

Notice that Hermiticity (or anti-Hermiticity) is not preserved by \emph{arbitrary} similarity transformations: requiring that both the representation $\mathbf{\b{R}}$ and $\mathbf{\r{ad}}(\mathbf{\b{R}})$ are (anti-)Hermitian places restrictions on the components of the generators and how they must be normalized. Minimally, these requirements restrict representations' generators to take components over some field extension over $\mathbb{Q}$, necessarily involving $i\equivR\sqrt{\text{-}1}$ and often other algebraic roots.\\[-10pt]

To be clear: this is \emph{not} our motivation for using concrete representations. Indeed, insisting on typical physicists' conventions for choosing bases for generators would be a very bad idea. For example, in the computation of the contraction $\langle\mathrm{tr}_{\mathbf{\b{F}}}(\r{1\,2\,3\,4})|\mathbf{tr}_{\mathbf{\b{F}}}(\r{1\,3\,2\,4})\rangle$, using the Gell-Mann choice of generators would require 7 times more intermediate memory, and 18 times more time than the Chevalley generators provided by \rpackage. Moreover, using Gell-Mann generators to compute this contraction would involve the relatively involved algebraic expression which require considerable efforts to simplify:
\mathematicaBox{
\mathematicaSequence{\funL[0]{contractTensors}\brace{\algL{a}\brace{2}}\brace{\tensor{trF}\brace{\{1,2,3,4\}},\tensor{trF}\brace{\{1,3,2,4\}}}\\[-10pt]}{\texttt{144}}
\mathematicaSequence{\funL[1]{contractTensors}\brace{gellMannLambdas}\brace{\tensor{trF}\brace{\{1,2,3,4\}},\tensor{trF}\brace{\{1,3,2,4\}}}\\[-10pt]}{\fig[-36.5pt]{0.85}{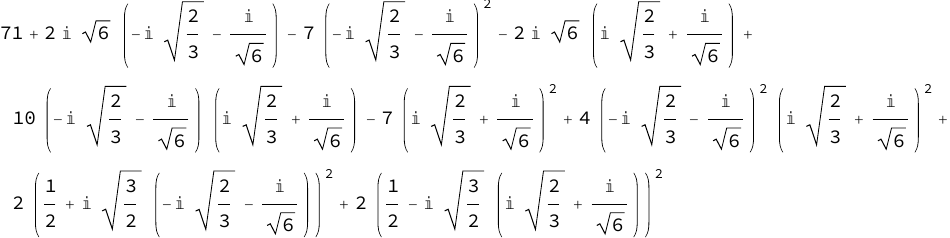}\vspace{-2pt}}
\mathematicaSequence{\built{FullSimplify}@\%}{\texttt{144}}
}
Of course, the two computations agree---as a complete contraction such as that between trace tensors $\langle\mathrm{tr}_{\mathbf{\b{F}}}(\r{1\,2\,3\,4})|\mathbf{tr}_{\mathbf{\b{F}}}(\r{1\,3\,2\,4})\rangle$ is entirely representation-theoretic, independent of the bases chosen for the generators (or even which linear combination of generators are chosen). 

Rather, the reason we find it useful to have \emph{concrete} representations on hand is simply because even representation-theoretic quantities such as $\langle\mathrm{tr}_{\mathbf{\b{F}}}(\r{1\,2\,3\,4})|\mathbf{tr}_{\mathbf{\b{F}}}(\r{1\,3\,2\,4})\rangle$ can be \emph{easily computed} by simply summing arrays of numbers; and this is something that can be done extremely efficiently using modern computers.

\subsection{\texorpdfstring{\emph{Concrete} Representations of Lie Algebras in \rpackage}{Concrete Representations of Lie Algebras in the Package}}\label{concrete_reps_in_package}\vspace{-4pt}

The generators of representations can often be chosen so that each generator acts on small number of basis vectors; that is, they can be chosen to be fairly \emph{sparse}. As such, it is natural to encode the components of $\mathbf{\b{R}}$ as \built{SparseArray}. In \textsc{Mathematica}, this is a built-in, relatively light-weight object type for which a number of sparse-optimized algorithms for linear algebra exist. 
\mathematicaBox{
\mathematicaSequence{egRep=\funL[0]{fundamentalRep}\brace{\algL{f}\brace{4}}\\[-10pt]}{\fig[-2pt]{1}{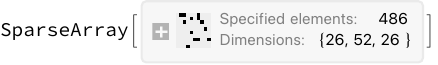}\vspace{-4pt}}
\mathematicaSequence{\{\funL[1]{nice}@\%,\funL[1]{drawTensors}@\%\}}{\fig[-2pt]{1}{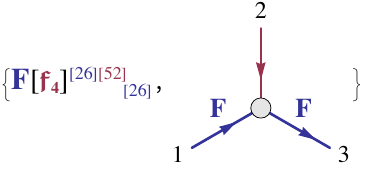}\vspace{-6pt}}
}~\\[-18pt]

Notice that in the example above, the `fundamental' representation of $\mathfrak{f}_{\r{4}}$ provided by \rpackage~requires only 486 non-vanishing components (among its 35,152 components). This number of non-vanishing components is visible in \textsc{Mathematica}'s formatting of \built{SparseArray} objects: it is indicated on the upper right of the output glyph, above the range of each \built{Slot}. We will return to the discussion of the sparsity of the generators when we describe the choices made in their preparation for \rpackage~below.\\[-10pt]

The sparsity of representations allows \built{SparseArray} objects in \textsc{Mathematica} to be stored and worked with more efficiently than \built{Normal} objects (such as a \built{List}). Consider for example the fundamental/adjoint representation of $\mathfrak{e}_{\r{8}}$:
\mathematicaBox{
\mathematicaSequence{egRep=\funL[1]{fundamentalRep}\brace{\algL{e}\brace{8}}\\[-10pt]}{\fig[-2pt]{1}{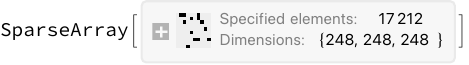}\vspace{-6pt}}
\mathematicaSequence{\funL[0]{byteCount}\brace{egRep}\\[-10pt]}{\fig[-2pt]{1}{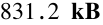}\vspace{-4pt}}
\mathematicaSequence{\funL[1]{byteCount}\brace{\built{Normal}\brace{egRep}}\\[-10pt]}{\fig[-2pt]{1}{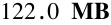}\vspace{-4pt}}
}~\\[-0pt]
As can be seen above, the \built{SparseArray} requires $\sim\!150$ times less memory than ordinary, \built{Normal} \built{Lists}. This has many implications for how fast these representations can be used and manipulated, and for the memory overhead required.\\[-10pt]

Let us now check that the defining relation (\ref{explicit_form_of_defining_relation}) holds for the concrete representations provided by \rpackage. We start by constructing the \funL{dot} product
\mathematicaBox{
\mathematicaSequence[1]{egRep=\funL[1]{fundamentalRep}\brace{\algL{f}\brace{4}};\\
\funL[1]{dot}\brace{egRep,egRep}\\[-10pt]}{\fig[-2pt]{1}{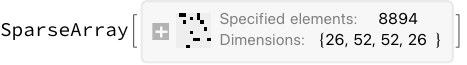}\vspace{-5pt}}
\mathematicaSequence{\funL[1]{drawTensors}@\%}{\fig[-2pt]{1}{dotPower_f4_nice}\vspace{-6pt}}
}~\\[0pt]
%\mbox{}\nonident 
Using \funL[4]{dotPower}, longer sequences $(\mathbf{\b{R}}.\mathbf{\b{R}}.\ldots.\mathbf{\b{R}})$ can be constructed; moreover, the function \funL[4]{dotPower} also allows one to specify how the inner \built{Slot} \built{Sequence} should be ordered. This allows us to rapidly construct, for example, the commutator $`[\mathbf{\b{R}},\mathbf{\b{R}}]$'\\[-16pt]
\mathematicaBox{
\mathematicaSequence[1]{egRep=\funL[1]{fundamentalRep}\brace{\algL{f}\brace{4}};\\
commTensor=\funL[0]{dotPower}\brace{egRep,\{1,2\}}-\funL[1]{dotPower}\brace{egRep,\{2,1\}}\\[-10pt]}{\fig[-4pt]{1}{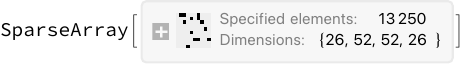}\vspace{-5pt}}
\mathematicaSequence{\funL[1]{drawTensors}@\%}{\fig[-2pt]{1}{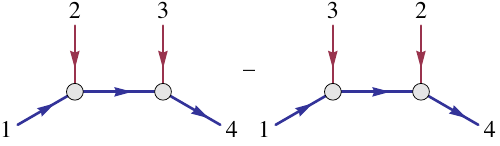}\vspace{-6pt}}
}~\\[-16pt]

Notice that we have now constructed the left hand side of the defining identity (\ref{explicit_form_of_defining_relation}). To construct the right-hand side, we must determine the coefficients $\r{f}\indices{\r{[d]\,[d]}}{\r{[d]}}$, which corresponds to the generators of the \emph{induced} adjoint $\mathbf{\r{ad}}(\mathbf{\b{R}})$. This is generated by \funL[4]{inducedAdjoint}\brace{$\mathbf{\b{R}}$}. Dotting this with the generators of $\mathbf{\b{R}}$ would result in:
\mathematicaBox{
\mathematicaSequence[1]{egAd=\funL[0]{inducedAdjoint}\brace{egRep};\\
fDotRepTensor0=\funL[1]{dot}\brace{egAd,\built{Transpose}\brace{egRep}}\\[-12pt]}{\fig[-4pt]{1}{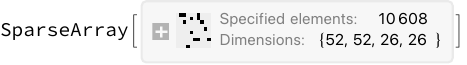}\vspace{-5pt}}
\mathematicaSequence{\funL[1]{drawTensors}@\%}{\fig[-2pt]{1}{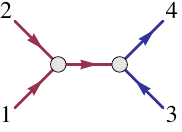}\vspace{-7.5pt}}
}~\\[-2pt]

\noindent But this tensor has the wrong sequence of indices to be compared directly with the \texttt{commTensor}; thus, we need to \built{Transpose} its indices.

Putting all these steps together, we may check the commutation relations are satisfied for any concrete representation via
\mathematicaBox{
\mathematicaSequence[1]{egRep=\funL[0]{fundamentalRep}\brace{\algL{f}\brace{4}};\\
commTensor=\funL[0]{dotPower}\brace{egRep,\{1,2\}}-\funL[1]{dotPower}\brace{egRep,\{2,1\}}\\[-10pt]}{\fig[-4pt]{1}{commTensor}\vspace{-4pt}}
\mathematicaSequence{\funL[1]{drawTensors}@\%}{\fig[-2pt]{1}{commTensor_draw}\vspace{-6pt}}
\mathematicaSequence[1]{egAd=\funL[0]{inducedAdjoint}\brace{egRep};\\
fDotRepTensor=\built{Transpose}\brace{\funL[1]{dot}\brace{egAd,\built{Transpose}\brace{egRep}},\{2,3,1,4\}}\\[-10pt]}{\fig[-4pt]{1}{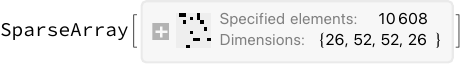}\vspace{-4pt}}
\mathematicaSequence{\funL[1]{drawTensors}@\%}{\fig[-2pt]{1}{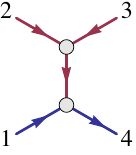}\vspace{-4pt}}
\mathematicaSequence{commTensor==fDotRepTensor}{\text{True}}
}\\[-2pt]
\noindent \textbf{Note}: this check, performed at the level of \emph{entire} rank-(3,1) tensors encoded as \built{SparseArray} objects, \emph{simultaneously} verifies the commutation relations for all pairs of generators.

\subsubsection{\texorpdfstring{Bases Chosen for Concrete Representations of \rpackage}{Bases Chosen for Concrete Representations of the Package}}\label{basis_choices_for_concrete_reps}\vspace{-4pt}

The particular choices made to prepare these representations follow from the desire to maximize the efficiency of their manipulations (both in terms of CPU time and memory). This favours that representations' components involve only (relatively small) integers or rational numbers---without any requirement for algebraic field extensions relative to $\mathbb{Q}$. This can be achieved for representations in the Chevalley basis \cite{chevalley1946theory} (which is closely related to the Cartan-Weyl basis, but with a rescaling of the generators of the Cartan subalgebra). 

In the Chevalley basis, the generators are chosen (and organized) according to the weight system of the adjoint representation. Specifically, the generators are indexed from those of lowest-weight to those of highest-weight, with the generators of the Cartan subalgebra in the middle:
\vspace{-4pt}\eq{\mathbf{T}(\mathbf{\b{R}})=\{\underbrace{\mathbf{F}^{\g{\alpha_1}},\ldots,\mathbf{F}^{\g{\alpha_{{-}1}}}}_{\g{\alpha_i}\in\Delta_-},\underbrace{\mathbf{H}^{\r{1}},\ldots\mathbf{H}^{\r{k}}}_{\mathfrak{h}},\underbrace{\mathbf{E}^{\g{\alpha_1}},\ldots,\mathbf{E}^{\g{\alpha_{{-}1}}}}_{\g{\alpha_i}\in\Delta_+}\}\vspace{-5pt}}
The subsets of generators $\mathbf{F}$ and $\mathbf{E}$ are called `lowering' and `raising', respectively, and are chosen to satisfy the eigenvalue equation
\vspace{-5pt}\eq{[\mathbf{H}^{\r{i}},\mathbf{E}^{\g{\alpha_j}}]=\g{\alpha}\indices{\r{i}}{\g{j}}\,\mathbf{E}^{\g{\alpha_j}}\,\fwboxL{0pt}{\qquad\text{(no sum on $\g{j}$).}}\vspace{-5pt}\label{organization_of_generators}}

The generators of the Cartan subalgebra are diagonalized, resulting in the `\mbox{(co-)}root' lattice of $\mathbf{\b{R}}$ appearing along the diagonal
\eq{\tilde{\Lambda}(\mathbf{\b{R}})=\bigoplus_{\t{r}\in\b{[r]}}(\mathbf{H}\indices{\r{1}}{}(\mathbf{\b{R}})\indices{\t{r}}{\t{r}},\ldots,\mathbf{H}\indices{\r{k}}{}(\mathbf{\b{R}})\indices{\t{r}}{\t{r}})\equivL\bigoplus_{\t{r}\in\b{[r]}}\mathbf{H}\indices{\r{[k]}}{}(\mathbf{\b{R}})\indices{\t{r}}{\t{r}},\vspace{-5pt}}
where elements of the \emph{root} lattice $\tilde{\lambda}\in\tilde{\Lambda}(\mathbf{\b{R}})$ are related to those of the weight lattice by $\lambda_j\equivR\sum_{\t{i}\in\r{[k]}}\tilde{\lambda}_{\t{i}}C[\mathfrak{g}]\indices{\t{i}}{j}$ where $C[\mathfrak{g}]$ is the Cartan matrix encoding the simple roots. In contrast to the weights (which are necessarily \emph{integral}), the `roots' (or more precisely, the coefficients of the positive, simple roots) are sometimes rational.

Not only are the representations efficiently stored by \rpackage, but they have been organized to maximize their simplicity. Components of arbitrary representations are \emph{rational}-valued, and those of adjoints are always (small) integers: For example, the adjoint representations of all classical Lie algebras involve only the numbers $\pm\!\{1,2\}$, and the `worst' case of $\mathfrak{e}_{\r{8}}$ involves only the integers $\pm\!\{1,2,3,4,5,6\}$.\\[-12pt]
\mathematicaBox{
\mathematicaSequence[3]{egAlgebras=\{\algL{a}\brace{16},\algL{b}\brace{16},\algL{c}{16},\algL{d}\brace{16},\algL{e}\brace{6},\algL{e}\brace{7},\algL{e}\brace{8},\algL{f}\brace{4},\algL{g}\brace{2}\};\\
egAdjoints=\funL[0]{adjointRep}/@egAlgebras;\\
\mbox{\built{Sort}@\built{DeleteDuplicates}\brace{\built{Abs}\brace{\var{\texttt{\#\hspace{-1pt}\,}}\brace{\text{{\color[rgb]{0.4,0.4,0.4}``NonzeroValues''}}}}}\&/@egAdjoints};
\mbox{\built{MatrixForm}@\built{Transpose}\brace{\{\funL[1]{nice}/@\,egAlgebras,\%\}}}\\[-10pt]}{\fig[-5pt]{1}{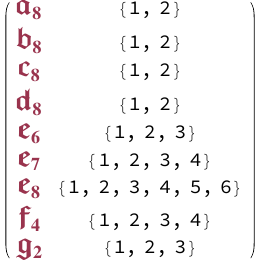}\vspace{-4pt}}
}

In the package \rpackage, the defining `fundamental' representation of each simple Lie algebra $\mathfrak{\r{g}}$ is generated by the function \funL{fundamentalRep}\brace{$\mathfrak{\r{g}}$}, from which \mbox{\funL{adjointRep}\brace{$\mathfrak{\r{g}}$}} and most other representations are defined or may be computed. For the classical Lie algebras, these defining representations are computed internally, and those of the exceptional algebras have been prepared using the computer package \texttt{Magma} \cite{magma}, which uses the algorithm described in \cite{Magma_algorithm} to produce concrete generators of arbitrary irreducible representations.\\[-10pt]

\newpage
As discussed in \mbox{section~\ref{subsec:conventions_for_weight_systems}}, other representations can be generated from $\mathbf{\b{F}}$ via tensor products and projections. A concrete set of generators for $\r{i}$th representation of fundamental weight $\mathbf{\b{F}}_{\r{i}}$ can be obtained using the function \mbox{\funL{fundamentalRep}\brace{$\mathfrak{{g}},\r{i}$}.} For example,~\\[-10pt]\nopagebreak
\mathematicaBox{
\mathematicaSequence[1]{\funL[0]{fundamentalRep}\brace{\algL{d}\brace{5},\var{\texttt{\#\hspace{-1pt}\,}}}\&/@\built{Range}\brace{5};\\
\built{MatrixForm}\brace{\{\funL[0]{dynkinLabel}\brace{\var{\texttt{\#\hspace{-1pt}\,}}},\funL[0]{nice}\brace{\var{\texttt{\#\hspace{-1pt}\,}}},\var{\texttt{\#\hspace{-1pt}\,}},\funL[0]{byteCount}\brace{\var{\texttt{\#\hspace{-1pt}\,}}}\}\&/@\%}\\[-10pt]}{\fig[-2pt]{0.75}{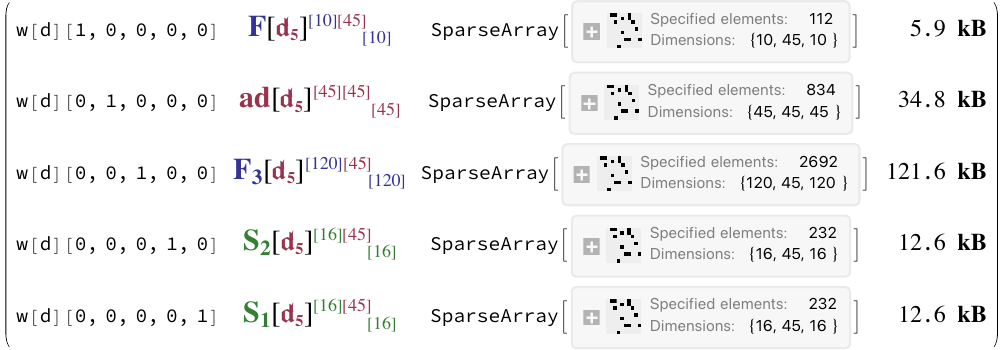}\vspace{-4pt}}
\mathematicaSequence[1]{\funL[1]{fundamentalRep}\brace{\algL{e}\brace{6},\var{\texttt{\#\hspace{-1pt}\,}}}\&/@\built{Range}\brace{6};\\
\built{MatrixForm}\brace{\{\funL[1]{dynkinLabel}\brace{\var{\texttt{\#\hspace{-1pt}\,}}},\funL[0]{nice}\brace{\var{\texttt{\#\hspace{-1pt}\,}}},\var{\texttt{\#\hspace{-1pt}\,}},\funL[1]{byteCount}\brace{\var{\texttt{\#\hspace{-1pt}\,}}}\}\&/@\%}\\[-10pt]}{\fig[-2pt]{0.75}{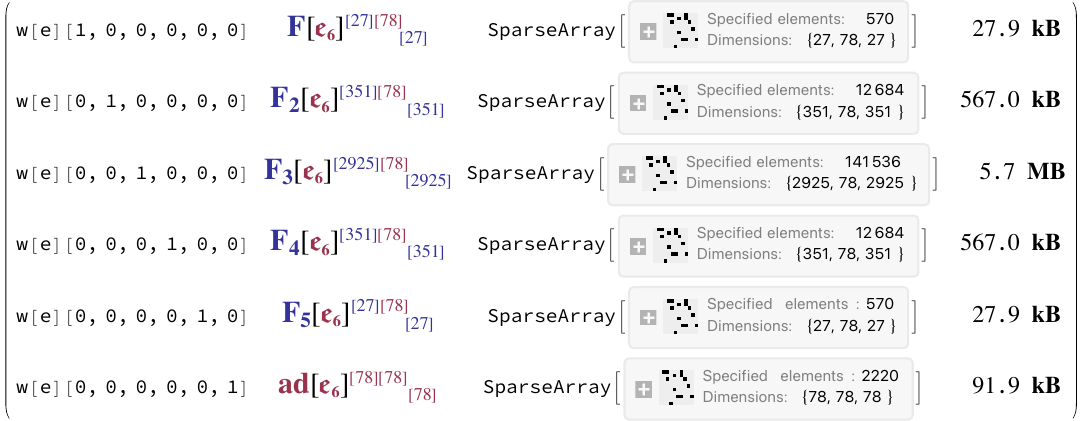}\vspace{-4pt}}
\mathematicaSequence[1]{\funL[1]{fundamentalRep}\brace{\algL{f}\brace{4},\var{\texttt{\#\hspace{-1pt}\,}}}\&/@\built{Range}\brace{4};\\
\built{MatrixForm}\brace{\{\funL[1]{dynkinLabel}\brace{\var{\texttt{\#\hspace{-1pt}\,}}},\funL[0]{nice}\brace{\var{\texttt{\#\hspace{-1pt}\,}}},\var{\texttt{\#\hspace{-1pt}\,}},\funL[1]{byteCount}\brace{\var{\texttt{\#\hspace{-1pt}\,}}}\}\&/@\%}\\[-10pt]}{\fig[-2pt]{0.75}{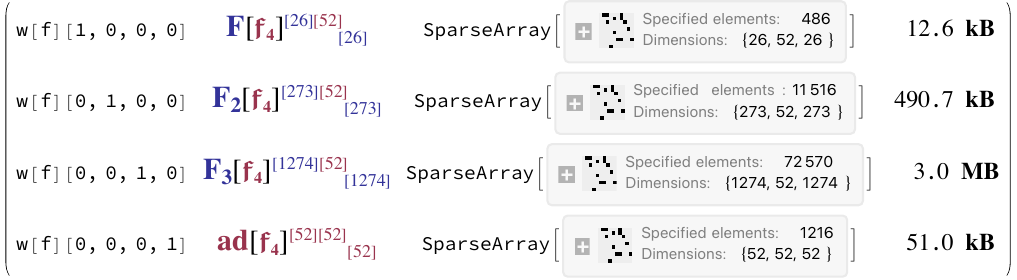}\vspace{-4pt}}
}\nopagebreak~\\[-10pt]

\newpage
\section{Colour Tensors of Scattering Amplitudes in Gauge Theory}\label{sec:colourInGaugeTheory}\vspace{-4pt}

In the Feynman rules of gauge theory, the generators/representations which define the charges of particles are sewn together into colour tensors by summing over the labels of intermediate, virtual particles.

\subsection{Constructing Colour Tensors Relevant to Amplitudes}\label{subsec:constructing_tensors}\vspace{-4pt}

Given the generators of some representation, it is natural to combine them into higher-rank tensors such as the `\funL[4]{dotPower}\brace{$\mathbf{\b{R}}$,\r{$n$}}' $(\mathbf{\b{R}}.\cdots.\mathbf{\b{R}})$,
\mathematicaBox{
\mathematicaSequence[1]{egRep=\funL[0]{fundamentalRep}\brace{\algL{a}\brace{6}}\\%\\[-10pt]}{\fig[-4pt]{1}{fundamentalRep_a6}\vspace{-4pt}}
\{\funL[1]{nice}@\%,\funL[1]{drawTensors}@\%,\funL[1]{byteCount}@\%\}}{\fig[-4pt]{1}{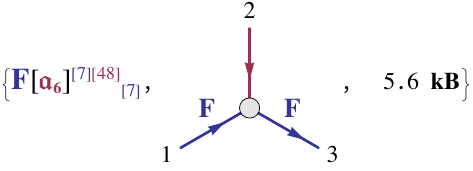}\vspace{-6pt}}
\mathematicaSequence{\funL[0]{dotPower}\brace{egRep,4}\\[-10pt]}{\fig[-4pt]{1}{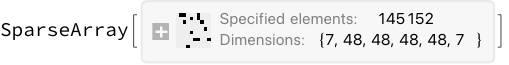}\vspace{-4pt}}
\mathematicaSequence{\{\funL[1]{nice}@\%,\funL[1]{drawTensors}@\%\}}{\fig[-4pt]{1}{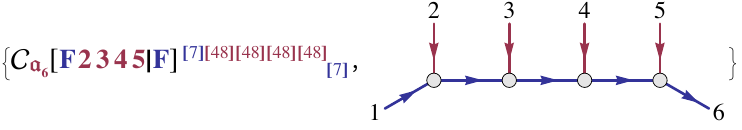}\vspace{-4pt}}
\mathematicaSequence{\funL[1]{byteCount}@\%\%}{\fig[-4pt]{1}{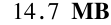}\vspace{-2pt}}
}~\\[-2pt]
or even the `\funL{repTrace}\brace{$\mathbf{\b{R}}$,$\r{n}$}', $\mathbf{tr}(\mathbf{\b{R}}.\cdots.\mathbf{\b{R}})$, 
\mathematicaBox{
\mathematicaSequence{egRep=\funL[0]{fundamentalRep}\brace{\algL{g}\brace{2}}\\[-10pt]}{\fig[-4pt]{1}{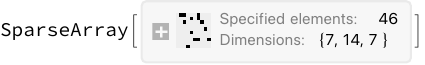}\vspace{-4pt}}
\mathematicaSequence{\funL[1]{repTrace}\brace{egRep,6}\\[-10pt]}{\fig[-4pt]{1}{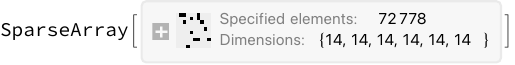}\vspace{-4pt}}
\mathematicaSequence{\{\funL[1]{nice}@\%,\funL[1]{drawTensors}@\%,\funL[1]{byteCount}@\%\}}{\fig[-4pt]{1}{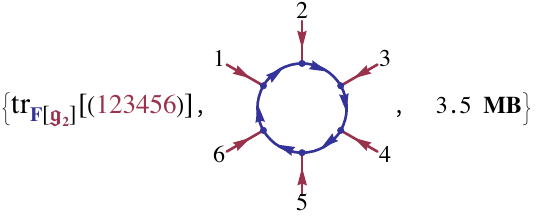}\vspace{-4pt}}
}~\\[4pt]

From these basic operations (combined perhaps with \funL{tensorContract}), a wide variety of colour tensors related to amplitudes can be constructed.

\subsection{Colour Tensors for the Scattering of Adjoints}\label{subsec:tensors_for_adjoints}\vspace{-4pt}

For the scattering of $\r{n}$ adjoint-charged particles, the generators that appear in the Feynman rules are of the adjoint representation; the Jacobi relation
\eq{\fig[-2pt]{0.7}{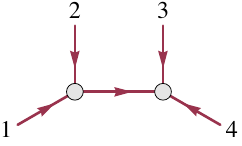}{+}\fig[-2pt]{0.7}{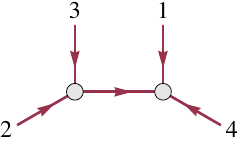}{+}\fig[-2pt]{0.7}{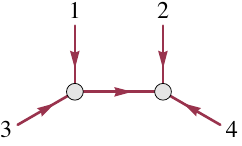}=0,\vspace{-10pt}}
any tree graph constructed from sewing together $\mathbf{\r{ad}}$ tensors can be reduced to a linear combination of those of the form 
\eq{\tensor{ddmTensor}\brace{\b{\alpha},\r{\sigma_1},\ldots,\r{\sigma_{\text{-}1}},\b{\beta}}\;\;\;\bigger{\Leftrightarrow}\;\;\;\fig[-2pt]{0.7}{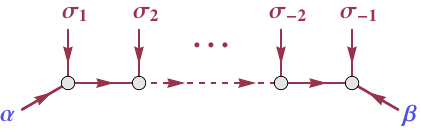}\vspace{-9pt}}
for any fixed choice of $\{\b{\alpha},\b{\beta}\}\in\!\binom{[n]}{2}$ and $\r{\vec{\sigma}}\in\mathfrak{S}([n]\backslash\{\b{\alpha},\b{\beta}\})$. These are the colour tensors appearing in the DDM expansion (\ref{ddm_expansion_at_tree_level}), which is valid at tree-level \cite{DelDuca:1999rs}. In terms of the basic operations,
\eq{\begin{split}\\[-24pt]&\hspace{-20pt}\tensor{ddmTensor}\brace{\b{\alpha},\r{\sigma_1},\ldots,\r{\sigma_{\text{-}1}},\b{\beta}}\,\equivR\\
&\built{Transpose}\texttt{[}\funL{dot}\brace{\funL{dotPower}\brace{\repL{ad}\brace{\mathfrak{g}},\r{n{-}2}},\funAL{killingMetric}\brace{\mathfrak{g}}},\\
&\hspace{62pt}\built{Ordering}\brace{\{\b{\alpha},\r{\sigma_1},\ldots,\r{\sigma_{\text{-}1}},\b{\beta}\}}\texttt{]}.\\[-10pt]
\end{split}}
This basis of colour tensors relevant to all-adjoint scattering was generalized to two loops recently in \cite{Dunbar:2026sle}.\\[-10pt]

The collection of such tensors relevant to tree-level amplitudes is generated by the function \funL{ddmTreeAmpTensors}:
\mathematicaBox{
\mathematicaSequence{egTensors=\funL[1]{ddmTreeAmpTensors}\brace{5}}{\{\mbox{\tensor{ddmTensor}\brace{1\hspace{-1pt},\hspace{-1pt}2\hspace{-1pt},\hspace{-1pt}3\hspace{-1pt},\hspace{-1pt}4\hspace{-1pt},\hspace{-1pt}5}}\hspace{-1pt},\hspace{-1pt}\mbox{\tensor{ddmTensor}\brace{1\hspace{-1pt},\hspace{-1pt}2\hspace{-1pt},\hspace{-1pt}4\hspace{-1pt},\hspace{-1pt}3\hspace{-1pt},\hspace{-1pt}5}}\hspace{-1pt},\hspace{-1pt}\mbox{\tensor{ddmTensor}\brace{1\hspace{-1pt},\hspace{-1pt}3\hspace{-1pt},\hspace{-1pt}2\hspace{-1pt},\hspace{-1pt}4\hspace{-1pt},\hspace{-1pt}5}}\hspace{-1pt},\hspace{-1pt}\\
\phantom{\{}\mbox{\tensor{ddmTensor}\brace{1\hspace{-1pt},\hspace{-1pt}3\hspace{-1pt},\hspace{-1pt}4\hspace{-1pt},\hspace{-1pt}2\hspace{-1pt},\hspace{-1pt}5}}\hspace{-1pt},\hspace{-1pt}\mbox{\tensor{ddmTensor}\brace{1\hspace{-1pt},\hspace{-1pt}4\hspace{-1pt},\hspace{-1pt}2\hspace{-1pt},\hspace{-1pt}3\hspace{-1pt},\hspace{-1pt}5}}\hspace{-1pt},\hspace{-1pt}\mbox{\tensor{ddmTensor}\brace{1\hspace{-1pt},\hspace{-1pt}4\hspace{-1pt},\hspace{-1pt}3\hspace{-1pt},\hspace{-1pt}2\hspace{-1pt},\hspace{-1pt}5}}\}}
\mathematicaSequence{\{\funL[0]{nice}@\var{\texttt{\#\hspace{-1pt}\,}},\funL[0]{drawTensors}@\var{\texttt{\#\hspace{-1pt}\,}}\}\&/@egTensors\\[-10pt]}{\{\fig[-2pt]{0.825}{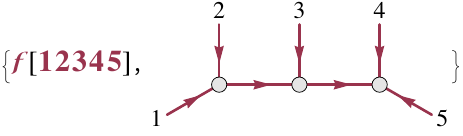},\fig[-2pt]{0.825}{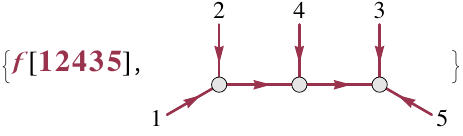},\\
\phantom{\{}\fig[-2pt]{0.825}{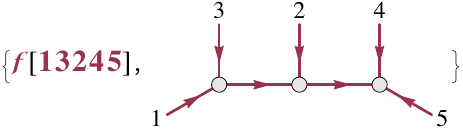},\fig[-2pt]{0.825}{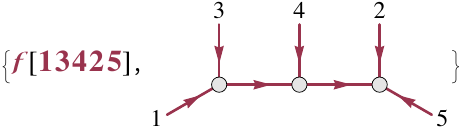},\\
\phantom{\{}\fig[-2pt]{0.825}{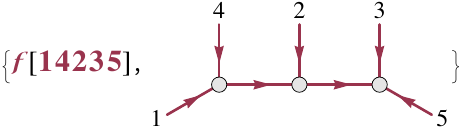},\fig[-2pt]{0.825}{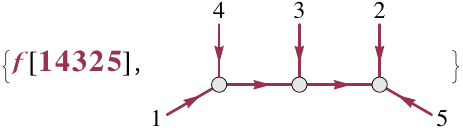}\}}
}~\\[-4pt]
\newpage
\noindent which can be explicitly constructed for any Lie algebra $\mathfrak{\r{g}}$ using $\funL{buildTensors}\brace{\mathfrak{\r{g}}}$:
\mathematicaBox{
\mathematicaSequence[1]{egTensors=\funL[1]{ddmTreeAmpTensors}\brace{5};\\
\funL[0]{buildTensors}\brace{\algL{g}\brace{2}}@egTensors\\[-10pt]}{\{\fig[-2pt]{0.78}{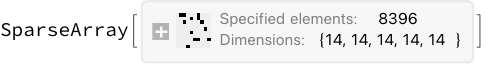},\fig[-2pt]{0.78}{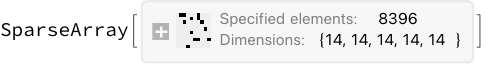},\\
\phantom{\{}\fig[-2pt]{0.78}{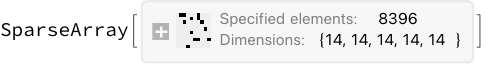},\fig[-2pt]{0.78}{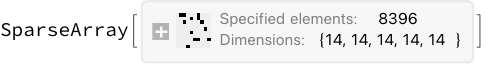},\\
\phantom{\{}\fig[-2pt]{0.78}{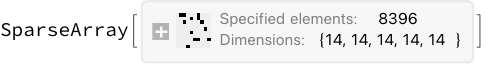},\fig[-2pt]{0.78}{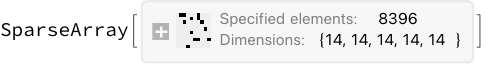}\}}
}~\\[-16pt]

The pairwise overlap between these tensors---contributing to interference in colour-summed, squared amplitudes---can be computed using \funL{tensorOverlap}\brace{$\mathfrak{\r{g}}$}:
\mathematicaBox{
\mathematicaSequence[2]{egTensors=\funL[1]{ddmTreeAmpTensors}\brace{5};\\
\funL[0]{tensorOverlap}\brace{\algL{g}\brace{2}}@egTensors;\\
\funL[1]{nice}@\%\\[-10pt]}{\fig[-6pt]{0.78}{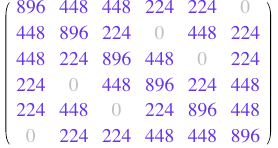}\vspace{-4pt}}
}~\\[-12pt]

Perhaps because the generators of some representation (say, the defining one) are more `available' or accessible than $\mathbf{\r{ad}}$, physicists often make use of the fact that for \emph{any} representation $\mathbf{\b{R}}$, the \funL{structureConstants},
\eq{\mathbf{\r{ad}}\indices{\r{[\mathfrak{g}]\,[\mathfrak{g}]\,\mathfrak{[g]}}}{}\equivR\sum_{\t{a}\in\r{[\mathfrak{g}]}}\mathbf{\r{ad}}\indices{\r{[\mathfrak{g}]\,[\mathfrak{g}]}}{\t{a}}\mathcal{\r{K}}\indices{\t{a}\,\r{[\mathfrak{g}]}}{}}
are related to the single traces via
\eq{\mathbf{\r{ad}}\indices{\r{[\mathfrak{g}]\,[\mathfrak{g}]\,\mathfrak{[g]}}}{}=\frac{1}{T(\mathbf{\b{R}})}\left(\mathbf{tr}_{\mathbf{\b{R}}}(\r{1\,2\,3}){-}\mathbf{tr}_{\mathbf{\b{R}}}(\r{2\,1\,3})\right)\,,\label{structure_constants_via_traces}}
it is always possible to decompose any tree-graph of structure constants into a sum over \emph{single} traces over the chosen representation's generators. Here the constant of proportionality is the \emph{Dynkin index} of the representation $\mathbf{\b{R}}$--which is \emph{conventionally} defined so that 
\eq{T(\mathbf{\b{R}})\,\mathcal{\r{K}}\indices{\r{[\mathfrak{g}]\,[\mathfrak{g}]}}{}\equivR \mathbf{tr}_{\mathbf{\b{R}}}(\r{1\,2})\,.}
It is worth noting that (complex) conjugation of the representation whose generators are being traced amounts to a reversal of arguments; specifically
\eq{\mathbf{tr}_{\mathbf{\b{\bar{R}}}}(\r{\sigma_1}\,\r{\sigma_2}\,\ldots\r{\sigma_{\text{-}1}})=(\text{-}1)^{\r{n}}\mathbf{tr}_{\mathbf{\b{R}}}(\r{\sigma_{\text{-}1}}\cdots\r{\sigma_2}\,\r{\sigma_1})\,.}
For real representations---for which $\mathbf{\b{R}}\simeq\mathbf{\b{\bar{R}}}$---this is a two-term identity, reducing the number of (potentially) independent single trace tensors from $(n{-}1)!$ to $(n{-}1)!/2$. For complex representations, all $(n{-}1)!$ single-trace tensors \emph{may} be distinct; but even in this case, the combinations that appear in the expansion of an adjoint-graph \emph{must} be real---and so only the combinations 
\eq{\mathbf{tr}_{\mathbf{\b{{R}}}}(\r{\sigma_1}\,\r{\sigma_2}\,\ldots\r{\sigma_{\text{-}1}}){-}(\text{-}1)^{\r{n}}\mathbf{tr}_{\mathbf{\b{R}}}(\r{\sigma_{\text{-}1}}\cdots\r{\sigma_2}\,\r{\sigma_1})\,,}
appear in the tree-level expansion of any scattering amplitude involving all adjoint-charged particles.\\[-10pt]

The set of trace cyclically inequivalent tensors appearing in the tree-level colour decomposition (\ref{trace_expansion_at_tree_level}) for $\b{n}$ can be generated by the function \mbox{\funL{singleTraceTensors}\brace{\b{$n$}},} and the set of \emph{dihedrally} inequivalent tensors is given by \mbox{\funL{realSingleTraceTensors}\brace{\b{$n$}}.}

These tensors can be built as concrete \built{SparseArray} objects for any representation $\mathbf{\b{R}}$ using \funL{buildTensors}\brace{$\mathbf{\b{R}}$}, and their pairwise colour contractions determined via \funL{tensorOverlap}\brace{$\mathbf{\b{R}}$}. We can easily verify the diminishment of the overlap between differently ordered traces of $\mathfrak{a}_{\r{k}}$ gauge theory via
\mathematicaBox{
\mathematicaSequence{egTensors=\funL[1]{singleTraceTensors}\brace{4}}{\{\mbox{\tensor{trR}\brace{\{1\hspace{-1pt},\hspace{-1pt}2\hspace{-1pt},\hspace{-1pt}3\hspace{-1pt},\hspace{-1pt}4\}}}\hspace{-1pt},\hspace{-1pt}\mbox{\tensor{trR}\brace{\{1\hspace{-1pt},\hspace{-1pt}2\hspace{-1pt},\hspace{-1pt}4\hspace{-1pt},\hspace{-1pt}3\}}}\hspace{-1pt},\hspace{-1pt}\mbox{\tensor{trR}\brace{\{1\hspace{-1pt},\hspace{-1pt}3\hspace{-1pt},\hspace{-1pt}2\hspace{-1pt},\hspace{-1pt}4\}}}\hspace{-1pt},\hspace{-1pt}\mbox{\tensor{trR}\brace{\{1\hspace{-1pt},\hspace{-1pt}3\hspace{-1pt},\hspace{-1pt}4\hspace{-1pt},\hspace{-1pt}2\}}}\hspace{-1pt},\hspace{-1pt}\\
\phantom{\{}\mbox{\tensor{trR}\brace{\{1\hspace{-1pt},\hspace{-1pt}4\hspace{-1pt},\hspace{-1pt}2\hspace{-1pt},\hspace{-1pt}3\}}}\hspace{-1pt},\hspace{-1pt}\mbox{\tensor{trR}\brace{\{1\hspace{-1pt},\hspace{-1pt}4\hspace{-1pt},\hspace{-1pt}3\hspace{-1pt},\hspace{-1pt}2\}}}\}}
\mathematicaSequence[1]{\funL[1]{tensorOverlap}\brace{\algL{a}\brace{\var{\texttt{\#\hspace{-1pt}\,}}}}\brace{egTensors}\&/@\{2,10\};\\
\funL[0]{heatMap}/@\%\\[-10pt]}{\fig[-2pt]{1}{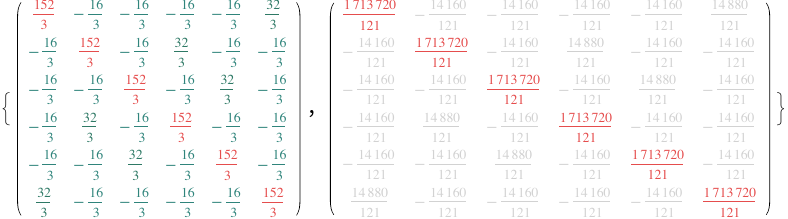}\vspace{-2pt}}
}~\\[-10pt]

Or we can compute the overlaps between 5-particle (dihedral) single-trace tensors built from the fundamental of $\mathfrak{f}_{\r{4}}$ as follows:
\mathematicaBox{
\mathematicaSequence{egTensors=\funL[0]{realSingleTraceTensors}\brace{5}\\
\funL[1]{nice}\brace{\funL[0]{tensorOverlap}\brace{\algL{f}\brace{4}}\brace{egTensors}}\\[-10pt]}{\fig[-2pt]{0.8}{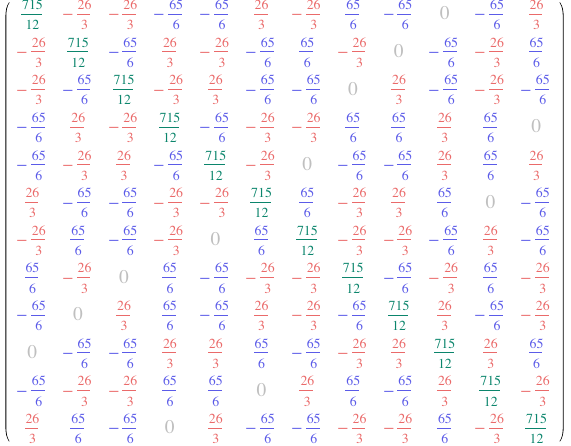}\vspace{-2pt}}
}~\\[-10pt]

\newpage
What about beyond tree-level? We can view any tensor defined by contracting a loop-level graph of structure constants as one defined by contracting some edges of a tree-level graph with additional external edges. Thus, for colour tensors encountered beyond tree-level, the question becomes what is the structure of tensors generated by contracting the legs of single-traces?

For the classical Lie algebras, generalized Fierz identities exist to decompose any such tensor into products of single traces, which we call \emph{multi-}traces (see \emph{e.g.}~\cite{Bourjaily:2025hvq} for more details). These can be listed using \mbox{\funL{multiTraceTensors}\brace{\r{$n$}}} and for dihedrally inequivalent multi-trace tensors,  \mbox{\funL{realMultiTraceTensors}\brace{\r{$n$}}}. 

\subsection{Colour Tensors for Scattering of Charged Matter \& Adjoints}\label{subsec:tensors_for_charged_matter}\vspace{-4pt}

Consider now the colour tensors which appear in the decomposition of amplitudes involving some number $n_G$ of adjoint-charged particles (\emph{e.g.} `gauge bosons') and an arbitrary number of $n_F$ \emph{pairs} of particles which are all charged under the same pair of conjugate representations $\mathbf{\b{R}},\mathbf{\b{\bar{R}}}$. The partial amplitudes relevant to this decomposition were described by Melia in \cite{Melia:2013bta,Melia:2013epa,Melia:2013xok}, and the colour tensors that appear at tree-level were described first by Johansson and Ochirov in \cite{Johansson:2015oia} (see also \cite{Kosower:1988kh,Johansson:2014zca,Melia:2015ika,Ochirov:2019mtf,Bourjaily:2026adf}, and for a discussion at one loop, \cite{Bern1991ColorDO,Kalin:2017oqr}).

We refer the reader to \cite{Bourjaily:2026adf} for a more thorough discussion of the tensors that appear in the colour decomposition of these amplitudes, their precise definition, and the general syntax used to define them. The list of colour tensors in the Melia basis relevant for the scattering of $n_G$ adjoints and $n_F$ fermion-anti-fermion pairs (all assumed to have the same flavour) is obtainable from \mbox{\funL{fermionicTreeAmpTensors}\brace{$n_F$,$n_G$}.}
\mathematicaBox{
\mathematicaSequence{egTensor=\built{RandomChoice}\brace{\funL[1]{fermionicTreeAmpTensors}\brace{3,3}}}{\mbox{\fun{joTensor}\brace{\{\hspace{-1.4pt}\fun{f}\brace{\hspace{-1pt}1\hspace{-1pt}}\hspace{-1.7pt},\hspace{-1.1pt}\fun{g}\brace{\hspace{-1pt}6\hspace{-1pt}}\hspace{-1.7pt},\hspace{-1.1pt}\fun{f}\brace{\hspace{-1pt}3\hspace{-1pt}}\hspace{-1.7pt},\hspace{-1.1pt}\fun{f}\brace{\hspace{-1pt}2\hspace{-1pt}}\hspace{-1.7pt},\hspace{-1.1pt}\fun{g}\brace{\hspace{-1pt}4\hspace{-1pt}}\hspace{-1.7pt},\hspace{-1.1pt}\fun{fb}\brace{\hspace{-1pt}2\hspace{-1pt}}\hspace{-1.7pt},\hspace{-1.1pt}\fun{g}\brace{\hspace{-1pt}5\hspace{-1pt}}\hspace{-1.7pt},\hspace{-1.1pt}\fun{fb}\brace{\hspace{-1pt}3\hspace{-1pt}}\hspace{-1.7pt},\hspace{-1.1pt}\fun{fb}\brace{\hspace{-1pt}1\hspace{-1pt}}\}\hspace{-1.4pt},\hspace{-1.4pt}\!\{\!\{1\hspace{-1.7pt},\hspace{-1.4pt}9\}\hspace{-2.1pt},\hspace{-1.8pt}\{4\hspace{-1.7pt},\hspace{-1.4pt}6\}\hspace{-2.1pt},\hspace{-1.8pt}\{3\hspace{-1.7pt},\hspace{-1.4pt}8\}\!\}\hspace{-1pt}}}}
\mathematicaSequence{\funL[1]{nice}\brace{egTensor}\\[-12pt]}{\fig[-2pt]{1}{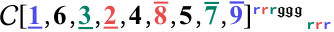}\vspace{-4pt}}
\mathematicaSequence{\funL[1]{drawTensors}\brace{egTensor}\\[-20pt]}{\fig[-2pt]{0.9}{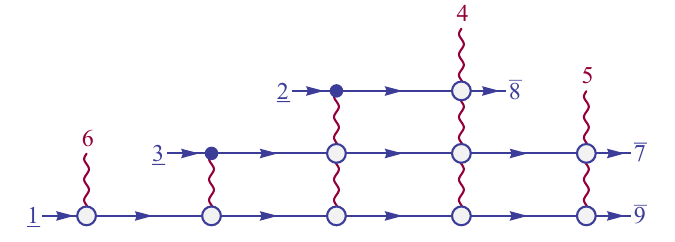}\vspace{-6pt}}
}~\\[-4pt]
Relative to the \textsc{Mathematica} package attached to \cite{Bourjaily:2026adf}, we have merely renamed the \built{Head} of the colour tensors to \tensor{joTensor}. See \cite{Bourjaily:2026adf} for more details.

\newpage
\subsection{\texorpdfstring{Colour Tensor Bases for Arbitrary Charges in $\mathfrak{a}_{\r{1}}$ Gauge Theory}{Colour Tensor Bases for Arbitrary Charges in a1 Gauge Theory}}\label{subsec:a1_clebsch_tensors}\vspace{-4pt}

All of the previous tensors appearing in the colour decomposition of amplitudes have the limitation that they become linearly dependent in the limit of large multiplicity (for any fixed rank) (see \emph{e.g.}~\cite{Bourjaily:2024jbt}), and that they are far from orthogonal in colour space. While for classical Lie algebras, multi-trace tensors are known to become orthogonal in the limit of large rank, at finite-rank there can be substantial interference terms between partial amplitudes appearing in (\ref{overlap_for_cross_section}). 

The problem of orthogonalization has been addressed by a number of authors especially in the context of multi-trace tensors of $\mathfrak{a}$-type gauge theory (see~\emph{e.g.}~\cite{Zeppenfeld:1988bz,Keppeler2012OrthogonalMB,Naculich:2012,Naculich:2024fiy,Sjodahl2015,Sjodahl2024}), but only recently has there been a general construction valid amplitudes involving particles charged under truly \emph{arbitrary} representations of arbitrary gauge groups. 

In the recent work \cite{Bourjaily:2025hvq}, a new construction was proposed that provides colour-orthogonal, linearly independent,and non-perturbatively complete bases of colour tensors suitable for amplitudes involving arbitrary coloured particles. In brief, one can construct such a basis by picking any \emph{fixed}, oriented trivalent tree graph $\Gamma$ with external edges labelled by the representations of particles involved (different assignments of representations to the legs of the graph will result in different bases of colour tensors); each vertex of the graph represents a Clebsch-Gordan coefficient decomposing its incoming edges into irreducible representations of its outgoing edges (all incoming or all outgoing vertices encode Clebsch-Gordans decomposing tensor products into the trivial representation); by scanning over the possible Clebsch-Gordan coefficients and the possible internal representations that appear, one constructs a basis with the properties described. 

The tensors appearing in such a basis therefore are defined by sewing together a number of Clebsch-Gordan coefficients along internal edges. For many cases of interest, it is not too hard to construct the Clebsch-Gordan tensors that appear; \emph{general} strategies for computing them rapidly become computationally costly, but (\emph{e.g.}) in the case of $\mathfrak{a}_{\r{1}}$ gauge theory they can be computed reliably and efficiently. 
% The tensors appearing in such a basis therefore are defined by sewing together a number of Clebsch-Gordan coefficients along internal edges. For many cases of interest, it is not too hard to construct the Clebsch-Gordan tensors that appear; but we know of no general strategy for computing them---outside the case of $\mathfrak{a}_{\r{1}}$ gauge theory, for which the necessary ingredients to construct these tensors are known. 

It may be useful to illustrate this idea with a couple of examples. Consider the case of scattering four particles in representations $\{{\color[rgb]{0.3,0.3,0.9}\mathbf{R}_1},{\color[rgb]{0.4,0.2,0.7}\mathbf{R}_2},{\color[rgb]{0,0.4,0.9}\mathbf{R}_3},{\color[rgb]{0.9,0.3,0.3}\mathbf{R}_4}\}$. One choice of basis would be provided by the set of graphs,
\eq{\mathbf{C}_{{\color[rgb]{0,0.5,0.4}\mathbf{r}}}({\color[rgb]{0.3,0.3,0.9}\mathbf{R}_1}\,{\color[rgb]{0.4,0.2,0.7}\mathbf{R}_2}\,{\color[rgb]{0,0.4,0.9}\mathbf{R}_3}\,{\color[rgb]{0.9,0.3,0.3}\mathbf{R}_4})\indices{{\color[rgb]{0.3,0.3,0.9}[r_1]}\,{\color[rgb]{0.4,0.2,0.7}[r_2]}\,{\color[rgb]{0,0.4,0.9}[r_3]}\,{\color[rgb]{0.9,0.3,0.3}[r_4]}}{}\;\;\;\bigger{\Leftrightarrow}\;\;\;\fig[-2pt]{1}{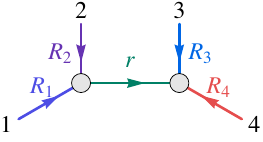}\label{basis_s}\,,}
where the individual tensors are labelled by the irreducible representation ${\color[rgb]{0,0.5,0.4}\mathbf{r}}$ and defined to be
\eq{\mathbf{C}_{{\color[rgb]{0,0.5,0.4}\mathbf{r}}}({\color[rgb]{0.3,0.3,0.9}\mathbf{R}_1}\,{\color[rgb]{0.4,0.2,0.7}\mathbf{R}_2}\,{\color[rgb]{0,0.4,0.9}\mathbf{R}_3}\,{\color[rgb]{0.9,0.3,0.3}\mathbf{R}_4})\indices{{\color[rgb]{0.3,0.3,0.9}[r_1]}\,{\color[rgb]{0.4,0.2,0.7}[r_2]}\,{\color[rgb]{0,0.4,0.9}[r_3]}\,{\color[rgb]{0.9,0.3,0.3}[r_4]}}{}\equivR\sum_{\t{r}\in{\color[rgb]{0,0.5,0.4}[r]}}C({\color[rgb]{0.3,0.3,0.9}\mathbf{R}_1}\,{\color[rgb]{0.4,0.2,0.7}\mathbf{R}_2}|{\color[rgb]{0,0.5,0.4}\mathbf{r}})\indices{{\color[rgb]{0.3,0.3,0.9}[r_1]}\,{\color[rgb]{0.4,0.2,0.7}[r_2]}}{\t{r}}C({\color[rgb]{0,0.5,0.4}\mathbf{r}}\,{\color[rgb]{0,0.4,0.9}\mathbf{R}_3}\,{\color[rgb]{0.9,0.3,0.3}\mathbf{R}_4})\indices{\t{r}\,{\color[rgb]{0,0.4,0.9}[r_3]}\,{\color[rgb]{0.9,0.3,0.3}[r_4]}}{}\label{clebsch_basis_tensors_defined}}
in terms of Clebsch-Gordan coefficients. Because only a finite number of irreps ${\color[rgb]{0,0.5,0.4}\mathbf{r}}$ appear in the decomposition of ${\color[rgb]{0.3,0.3,0.9}\mathbf{R}_1}\!\otimes\!{\color[rgb]{0.4,0.2,0.7}\mathbf{R}_2}$, only a finite number of basis elements exist. 

Alternatively, one could construct the colour tensors
\eq{\mathbf{\tilde{C}}_{{\color[rgb]{0,0,0}\mathbf{s}}}({\color[rgb]{0.3,0.3,0.9}\mathbf{R}_1}\,{\color[rgb]{0,0.4,0.9}\mathbf{R}_3}\,{\color[rgb]{0.4,0.2,0.7}\mathbf{R}_2}\,{\color[rgb]{0.9,0.3,0.3}\mathbf{R}_4})\indices{{\color[rgb]{0.3,0.3,0.9}[r_1]}\,{\color[rgb]{0.4,0.2,0.7}[r_2]}\,{\color[rgb]{0,0.4,0.9}[r_3]}\,{\color[rgb]{0.9,0.3,0.3}[r_4]}}{}\;\;\;\bigger{\Leftrightarrow}\;\;\;\fig[-2pt]{1}{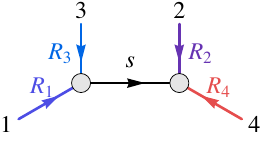}\label{basis_u}\,,}
where
\eq{\mathbf{\tilde{C}}_{{\color[rgb]{0,0,0}\mathbf{s}}}({\color[rgb]{0.3,0.3,0.9}\mathbf{R}_1}\,{\color[rgb]{0,0.4,0.9}\mathbf{R}_3}\,{\color[rgb]{0.4,0.2,0.7}\mathbf{R}_2}\,{\color[rgb]{0.9,0.3,0.3}\mathbf{R}_4})\indices{{\color[rgb]{0.3,0.3,0.9}[r_1]}\,{\color[rgb]{0.4,0.2,0.7}[r_2]}\,{\color[rgb]{0,0.4,0.9}[r_3]}\,{\color[rgb]{0.9,0.3,0.3}[r_4]}}{}\equivR\sum_{\t{r}\in{\color[rgb]{0,0.5,0.4}[r]}}C({\color[rgb]{0.3,0.3,0.9}\mathbf{R}_1}\,{\color[rgb]{0,0.4,0.9}\mathbf{R}_3}|{\color[rgb]{0,0,0}\mathbf{s}})\indices{{\color[rgb]{0.3,0.3,0.9}[r_1]}\,{\color[rgb]{0,0.4,0.9}[r_3]}}{\t{s}}C({\color[rgb]{0,0,0}\mathbf{s}}\,{\color[rgb]{0.4,0.2,0.7}\mathbf{R}_2}\,{\color[rgb]{0.9,0.3,0.3}\mathbf{R}_4})\indices{\t{s}\,{\color[rgb]{0.4,0.2,0.7}[r_2]}\,{\color[rgb]{0.9,0.3,0.3}[r_4]}}{}\,.\label{alt_basis_clebsches_defined}}
Notice that both sets of tensor `bases' in (\ref{clebsch_basis_tensors_defined}) and (\ref{alt_basis_clebsches_defined}) involve rank-(4,0) tensors with \emph{identical} dimensions. 

Let us see how these bases of tensors could be computed. Suppose that the four representations were taken to be $\{\hyperlink{label:w}{\fun{w}{\color{black}\brace{\algL{a}}\brace{{\color[rgb]{0,0.4,0.9}4}}}},\hyperlink{label:w}{\fun{w}{\color{black}\brace{\algL{a}}\brace{{\color[rgb]{0.24,0.7,0.6}14}}}},\hyperlink{label:w}{\fun{w}{\color{black}\brace{\algL{a}}\brace{{\color[rgb]{0.4,0.2,0.9}7}}}},\hyperlink{label:w}{\fun{w}{\color{black}\brace{\algL{a}}\brace{{\color[rgb]{0.9,0,0.5}5}}}}\}$; if we label these by their dimensions $\mathrm{dim}([a])=(a{+}1)$ (which is always unambiguous for $\mathfrak{a}_1$), we'd denote these by `$\{{\color[rgb]{0,0.4,0.9}\mathbf{5}},{\color[rgb]{0.24,0.7,0.6}\mathbf{15}},{\color[rgb]{0.4,0.2,0.9}\mathbf{8}},{\color[rgb]{0.9,0,0.5}\mathbf{6}}\}$'.

Now, 
\eq{\begin{split}~\\[-24pt]{\color[rgb]{0,0.4,0.9}\mathbf{5}}\otimes{\color[rgb]{0.24,0.7,0.6}\mathbf{15}}\simeq&{\color[rgb]{0,0.5,0.4}\mathbf{11}}\oplus{\color[rgb]{0,0.5,0.4}\mathbf{13}}\oplus{\color[rgb]{0,0.5,0.4}\mathbf{15}}\oplus{\color[rgb]{0,0.5,0.4}\mathbf{17}}\oplus{\color[rgb]{0,0.5,0.4}\mathbf{19}}\\
\fwboxR{0pt}{\text{and\hspace{20pt}}}{\color[rgb]{0.4,0.2,0.9}\mathbf{8}}\otimes{\color[rgb]{0.9,0,0.5}\mathbf{6}}\simeq{\color[rgb]{0,0.5,0.4}\mathbf{3}}\oplus{\color[rgb]{0,0.5,0.4}\mathbf{5}}\oplus{\color[rgb]{0,0.5,0.4}\mathbf{7}}\oplus{\color[rgb]{0,0.5,0.4}\mathbf{9}}\,\oplus&{\color[rgb]{0,0.5,0.4}\mathbf{11}}\oplus{\color[rgb]{0,0.5,0.4}\mathbf{13}}\\[-12pt]
\end{split}}
so that the basis tensors $\mathbf{C}_{{\color[rgb]{0,0.5,0.4}\mathbf{r}}}$ are labelled by the representations ${{\color[rgb]{0,0.5,0.4}\mathbf{r}}}\!\in\!\{{\color[rgb]{0,0.5,0.4}\mathbf{11}},{\color[rgb]{0,0.5,0.4}\mathbf{13}}\}$.
while for the basis tensors $\mathbf{\tilde{C}}_{{\color[rgb]{0,0,0}\mathbf{s}}}$, we have 
\eq{\begin{split}~\\[-24pt]{\color[rgb]{0,0.4,0.9}\mathbf{5}}\otimes{\color[rgb]{0.4,0.2,0.9}\mathbf{8}}\simeq{\color[rgb]{0,0,0}\mathbf{4}}\oplus{\color[rgb]{0,0,0}\mathbf{6}}\oplus{\color[rgb]{0,0,0}\mathbf{8}}\,\oplus&{\color[rgb]{0,0,0}\mathbf{10}}\oplus{\color[rgb]{0,0,0}\mathbf{12}}\\
\fwboxR{0pt}{\text{while\hspace{62pt}}}{\color[rgb]{0.24,0.7,0.6}\mathbf{15}}\otimes{\color[rgb]{0.9,0,0.5}\mathbf{6}}\simeq&{\color[rgb]{0,0,0}\mathbf{10}}\oplus{\color[rgb]{0,0,0}\mathbf{12}}\oplus{\color[rgb]{0,0,0}\mathbf{14}}\oplus{\color[rgb]{0,0,0}\mathbf{16}}\oplus{\color[rgb]{0,0,0}\mathbf{18}}\oplus{\color[rgb]{0,0,0}\mathbf{20}}\\[-12pt]
\end{split}}
so that $\mathbf{\tilde{C}}_{{\color[rgb]{0,0,0}\mathbf{s}}}$ are labelled by the irreducible representations $\mathbf{s}\!\in\!\{\mathbf{10},\mathbf{12}\}$.

The requisite Clebsch tensors appearing in the definitions are given in closed-form by \funL{a1Clebsch}\brace{$\var{w_1},\var{w_2},\var{w_3}$}---with negative numbers indicating conjugation---so that, for example, 
\mathematicaBox{
\mathematicaSequence{clebschVertices=\{\funL[1]{a1Clebsch}\brace{4,14,-10},\funL[1]{a1Clebsch}\brace{10,7,5}\}\\[-10pt]}{\fig[-2pt]{0.91}{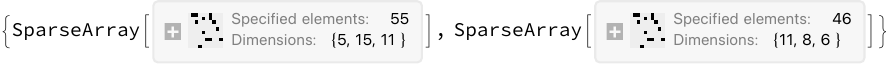}\vspace{-5pt}}
\mathematicaSequence{\funL[0]{nice}@\%\\[-10pt]}{\fig[-2pt]{1}{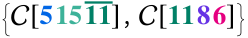}\vspace{-2pt}}
\mathematicaSequence[1]{\funL[0]{dot}@@clebschVertices;\\
\{\%,\funL[0]{drawTensors}@\%\}\\[-12pt]}{\{\fig[-2pt]{1}{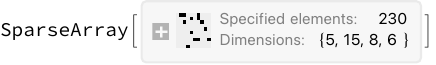},\fig[-2pt]{1}{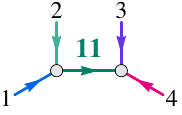}\}\vspace{-2pt}}
%\mathematicaSequence{\funL[1]{drawTensors}@\%\\[-10pt]}{\fig[-2pt]{1}{a1Clebsch_pair_eg_1_dotted_drawn}\vspace{-2pt}}
}

More directly, these bases can be enumerated via \funL{a1ClebschBasis} and constructed using \funL{buildTensors}:
\mathematicaBox{
\mathematicaSequence{basis1=\funL[1]{a1ClebschBasis}\brace{\{4,14,7,5\}}\\[-12pt]}{\{\tensor{clebschTensor}\brace{10,\{4,14,7,5\}},\tensor{clebschTensor}\brace{12,\{4,14,7,5\}}\}}
\mathematicaSequence{\funL[1]{drawTensors}@\%\\[-10pt]}{\{\fig[-2pt]{1}{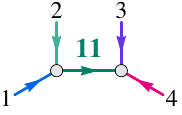},\fig[-2pt]{1}{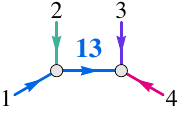}\}\vspace{-2pt}}
\mathematicaSequence{\funL[0]{buildTensors}\brace{\algL{a}\brace{1}}\brace{basis1}\\[-10pt]}{\fig[-2pt]{0.89}{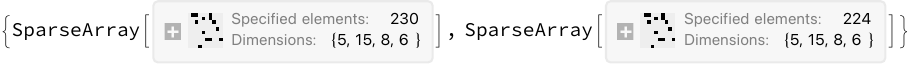}\vspace{-2pt}}
\mathematicaSequence{basis2=\funL[1]{a1ClebschBasis}\brace{\{4,14,7,5\},\{1,3,2,4\}}\\[-12pt]}{\{\tensor{clebschTensor}\brace{9,\{4,7,14,5\},\{1,3,2,4\}},\\\phantom{\{}\mbox{\tensor{clebschTensor}\brace{11,\{4,7,14,5\},\{1,3,2,4\}}}\}}
\mathematicaSequence{\funL[1]{drawTensors}@\%\\[-10pt]}{\{\fig[-2pt]{1}{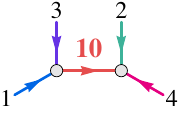},\fig[-2pt]{1}{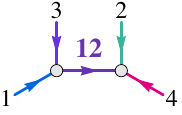}\}\vspace{-2pt}}
\mathematicaSequence{\funL[1]{buildTensors}\brace{\algL{a}\brace{1}}\brace{basis2}\\[-10pt]}{\fig[-2pt]{0.89}{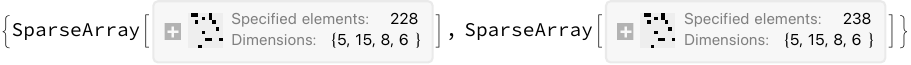}\vspace{-2pt}}
} 

We can verify that these two bases of tensors can be expanded into each other using the function \funL{tensorRelations}: 
\mathematicaBox{
\mathematicaSequence[3]{basis1=\funL[1]{a1ClebschBasis}\brace{\{4,14,7,5\}};\\
basis2=\funL[1]{a1ClebschBasis}\brace{\{4,14,7,5\},\{1,3,2,4\}};\\
\funL[0]{tensorRelations}\brace{\algL{a}\brace{1}}\brace{basis2,basis1};\\
\built{Column}\brace{\funL[1]{drawTensors}@\%}\\[-10pt]}{\fig[-2pt]{1}{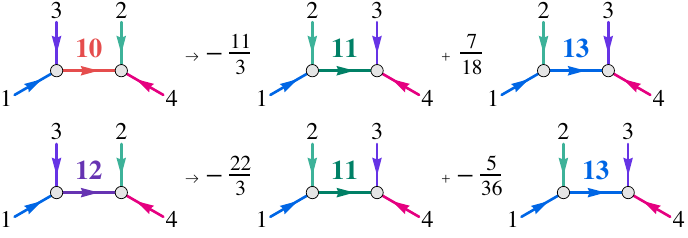}\vspace{-4pt}}
}

And we can recognize the coefficients of these expansions by computing the \built{NullSpace} (or \funL{nullSpace}) of their contractions computed via \funL{tensorOverlap}\brace{$\mathfrak{a}_{\r{1}}$}:
\mathematicaBox{
\mathematicaSequence[3]{basis1=\funL[1]{a1ClebschBasis}\brace{\{4,14,7,5\}};\\
basis2=\funL[1]{a1ClebschBasis}\brace{\{4,14,7,5\},\{1,3,2,4\}};\\
\funL[1]{tensorOverlap}\brace{\algL{a}\brace{1}}\brace{basis2,basis1};\\
\funL[1]{nice}@\%\\[-10pt]}{\fig[-2pt]{1}{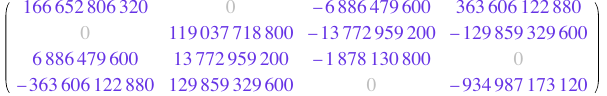}\vspace{-4pt}}
\mathematicaSequence{\funL[1]{nice}@\built{RowReduce}\brace{\built{NullSpace}@\%\%}\\[-10pt]}{\fig[-2pt]{1}{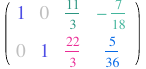}\vspace{-2pt}}
}~\\[-4pt]
This can also be computed using pairwise contractions:
\mathematicaBox{
\mathematicaSequence[3]{basis1=\funL[1]{a1ClebschBasis}\brace{\{4,14,7,5\}}\\
basis2=\funL[1]{a1ClebschBasis}\brace{\{4,14,7,5\},\{1,3,2,4\}};\\
pairwiseContractions=\funL[1]{contractTensors}\brace{\algL{a}\brace{1}}@@\{basis2,basis1\};\\
\funL[1]{nice}@\%\\[-10pt]}{\fig[-2pt]{1}{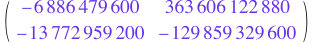}\vspace{-2pt}}
\mathematicaSequence[1]{selfContractions=\funL[1]{contractTensors}\brace{\algL{a}\brace{1}}@@\{basis1,basis1\};\\
\funL[1]{nice}@\%\\[-10pt]}{\fig[-2pt]{1}{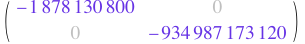}\vspace{-2pt}}
\mathematicaSequence[1]{\funL[0]{dot}\brace{pairwiseContractions\funL[0]{inverse}\brace{selfContractions}}\\
\funL[1]{nice}@\%\\[-10pt]}{\fig[-2pt]{1}{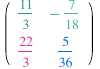}\vspace{-1pt}}
}~\\[-4pt]
%

%\subsection{Linear Relations among Colour Tensors}\label{subsec:linear_relations_and_overlaps}\vspace{-4pt}

\newpage
\section{\texorpdfstring{Starting the \rpackage~Package}{Starting the Colour_Tensors Package}}\label{sec:code}\vspace{-4pt}

The main contribution of this work is the included \rpackage~ package. In our submission to the \texttt{arXiv}, we have attached the package as an ancillary file obtainable from the `Ancillary files' section of the upper-right panel on this work's abstract page. We have also included an (admittedly verbose and thorough) walkthrough notebook \mbox{\textbf{\texttt{demonstrating}\rule[-1.05pt]{7.5pt}{.75pt}\texttt{colour}\rule[-1.05pt]{7.5pt}{.75pt}\texttt{tensor}\rule[-1.05pt]{7.5pt}{.75pt}\texttt{tools}.\texttt{nb}}.} Complete documentation for the functions defined by \rpackage~can be found in the \hyperlink{context_organization_of_appendix}{appendix}, and we made efforts to include illustrative examples throughout the earlier sections.\\[-10pt]

Although there are a number of symbolic representations of tensors, the package \rpackage~makes sure to generate most output for concrete tensors (including representations) directly as \built{SparseArray} objects. These are displayed directly in \textsc{Mathematica} in a way that indicates the dimension spanned by entries in each \built{Slot} and also the number of non-vanishing entries defining the array. For an additional layer of human-readability, we have endowed \emph{most} concrete (and abstract) tensors with \emph{stylized} formatting provided by \funL{nice}, and \emph{diagrammatic} representations provided by \funL{drawTensors}. For both stylized and graphical displays, additional information regarding these objects can be obtained via \emph{mouse-over} in \textsc{Mathematica}. We hope the users find these layers of interpretation helpful.

\subsection{Initializing and Installing the Package}\label{subsec:obtaining_and_installing}\vspace{-4pt}

To use the package, one need only download the \rpackage\built{.m} file and place it in the same directory as any saved notebook (so that the \built{\$Path} is obvious). Provided this is the case, the package can be loaded by running the following: 
\mathematicaBox{
\mathematicaSequence[1]{\smash{\built{SetDirectory}\brace{\built{NotebookDirectory}\brace{}};}\\
\texttt{<\hspace{-1pt}<\,\rpackage\built{.m}}\\[-10pt]}{\fig[-2pt]{0.9}{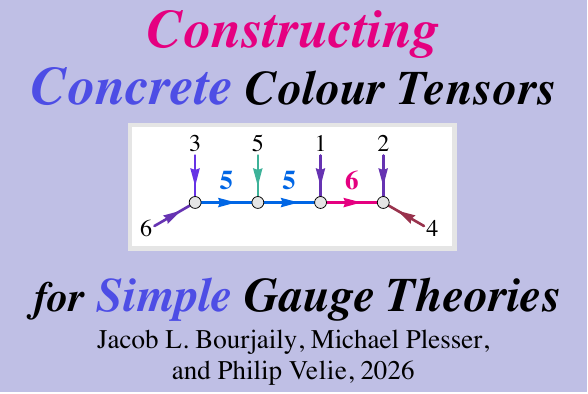}}
}\\
The package can be made more globally-accessible by adding the \rpackage\built{.m} file to the user's \built{\$Path} using the function \funL{installColourTensors}.

Many of the functions and much of the functionality provided are similar to those of other \textsc{Mathematica} packages, most notably \cite{Bourjaily:2023uln} (see also \cite{Bourjaily:2010wh,Bourjaily:2012gy,Bourjaily:2013mma,Bourjaily:2015jna,Bourjaily:2026adf}).

\newpage
\section{Conclusions and Commentary}\label{sec:conclusion}\vspace{-4pt}

In this work, we have described the functionality of the \rpackage~package for \textsc{Mathematica}. This package provides concrete, defining representations of all the simple Lie algebras, and also the tools required to construct arbitrary other representations as well as concrete colour tensors from them. 

The bases for representations we have chosen do not reflect the ordinary preferences for Hermitian generators (equivalently, unitary group elements); rather, we have chosen bases for our representations which most directly simplify the algebra required for tensor contractions. This choice has no effect on a number of important representation-theoretic, basis-independent questions. These include contracting colour tensors to determine colour-interference between partial amplitudes, determining the linear independence of a set of colour tensors, or finding their linear relations. Although the literal colour tensors needed by amplitudes \emph{do} require unitary representations, these can always be constructed \emph{in post} from those provided by \rpackage.

Among the colour tensor bases that have appeared in the physics literature, the tensor bases described in \cite{Bourjaily:2025hvq} have notably \emph{not} been included in \rpackage, beyond the special case of $\mathfrak{a}_{\r{1}}$ gauge theory. This is because such bases require constructing \emph{concrete} Clebsch-Gordan coefficients involved in tensor products of arbitrary representations of simple Lie algebras. Although these can be constructed with relative ease using the tools provided by \rpackage~in most cases of greatest interest (\emph{e.g.} for all the examples discussed in \cite{Bourjaily:2025hvq}), the completely \emph{general} strategy for their determination is less straightforward. For classical algebras, reasonably efficient general methods based on Gelfand-Tsetlin patterns and Young tableaux exist, see \cite{Alex:2011}. For \emph{any} simple Lie algebra, the Clebsch-Gordan coefficients can be computed by solving a linear system related to the action of the representation generators, but this becomes increasingly computationally costly as the size of the representations grow. The method via projecting onto Casimir eigenspaces is computationally superior to either of these, but is insufficient when the subspaces contain irreps with multiplicity. It would be very valuable to have an efficient resolution to this issue, and we hope this work encourages some algorithmically-minded representation theorists to build such a tool in the future. 
% Among the colour tensor bases that have appeared in the physics literature, the general tensor bases described in \cite{Bourjaily:2025hvq} have notably \emph{not} been included in \rpackage, beyond the special case of $\mathfrak{a}_{\r{1}}$ gauge theory. This is because there does not seem to exist any general algorithm for constructing the requisite \emph{concrete} Clebsch-Gordan coefficients involved in tensor products of arbitrary representations of simple Lie algebras. Although these can be constructed with relative ease using the tools provided by \rpackage~in most cases of greatest interest (\emph{e.g.} for all the examples discussed in \cite{Bourjaily:2025hvq}), we see no \emph{general} strategy for their determination. We hope this work encourages some algorithmically-minded representation theorists to build such a tool in the future. 

%================================================================================================================
%    Acknowledgments 
%================================================================================================================
\vspace{\fill}\vspace{-4pt}
\section*{Acknowledgments}%
\vspace{-4pt}
\noindent The authors gratefully acknowledge fruitful discussion and early collaboration with Cristian Vergu; MP thanks Matthew Mitchell for helpful conversations during and after the TASI 2025 summer school. This work was supported in part by a grant from the US Department of Energy (No.\ DE-SC00019066), and was performed in part at the Aspen Center for Physics, supported by National Science Foundation grant PHY-2210452.

\newpage
\addtocontents{toc}{\protect~\\[-22pt]\mbox{\vspace{-30pt}}\protect\hrulefill\par~\\[-36pt]}
\appendix
%
%================================================================================================================
\setcounter{section}{-1}
\renewcommand{\@seccntformat}[1]{\csname the #1\endcsname\;}
\hypertarget{context_organization_of_appendix}{}
\vspace{-0pt}\section[\mbox{\hspace{-18pt}Context-Organized List of Functions Provided by the Package}]{\hspace{-5.2pt}Context-Organized List of Functions Provided}%\label{context_organization_of_appendix}
\vspace{-4pt}
\renewcommand{\@seccntformat}[1]{\csname the#1\endcsname.\;}

%================================================================================================================
\vspace{-4pt}\sectionAppendix{Protected Symbols and Symbolic Expressions}{appendix:protected_symbols}\vspace{-4pt}
%================================================================================================================ 

\vspace{-0pt}\subsectionAppendix{Naming of Simple Lie Algebras}{naming_of_lie_algebras}\vspace{-4pt}

\vspace{-0pt}\subsubsectionAppendix{\emph{Standard} (Cartan's) Naming of Simple Lie Algebras}{cartan_naming_of_lie_algebras}\vspace{-1pt}

%Algebras
\defnBox[5]{a}{\var{$k$}\pattern}{denotes the $\mathfrak{a}$-series classical, simple Lie algebra of rank \var{$k$}$\,\geq\!1$.\\Equivalent to \algL{su}\brace{\var{$k$}+1}.}

\defnBox[5]{b}{\var{$k$}\pattern}{denotes the $\mathfrak{b}$-series classical, simple Lie algebra of rank \var{$k$}$\,\geq\!2$.\\Equivalent to \mbox{\algL{so}\brace{2\var{$k$}+1}}. (\textbf{Note}: $\mathfrak{so}_{3}\!\simeq\mathfrak{su}_2$, so \algL{b}\brace{1}$\!\equivR\!$\algL{a}\brace{1}.)}

\defnBox[5]{c}{\var{$k$}\pattern}{denotes the $\mathfrak{c}$-series classical, simple Lie algebra of rank \var{$k$}$\,\geq\!3$.\\Equivalent to \algL{sp}\brace{2\var{$k$}}. (\textbf{Note}: $\mathfrak{sp}_{4}\!\simeq\mathfrak{so}_{5}$, so \algL{c}\brace{2}$\!\equivR\!$\algL{b}\brace{2}.)}

\defnBox[5]{d}{\var{$k$}\pattern}{denotes the $\mathfrak{d}$-series classical, simple Lie algebra of rank \var{$k$}$\,\geq\!4$.\\Equivalent to \algL{so}\brace{2\var{$k$}}. (\textbf{Note}: $\mathfrak{so}_6\!\simeq\mathfrak{su}_{4}$, so \algL{d}\brace{3}$\!\equivR\!$\algL{a}\brace{3}.)}

\defnBox[5]{e}{\var{$k$}\pattern}{denotes the exceptional, simple Lie algebra $\mathfrak{e}_{\var{k}}$ with \var{$k$}$\,\in\!\{6,7,8\}$.\\
(Some authors refer to the simple algebras \algL{e}\brace{{$4$}}$\!\equivR\!$\algL{a}\brace{{$4$}} and \algL{e}\brace{{$5$}}$\!\equivR\!$\algL{d}\brace{{$5$}}.)}

\defnBox[5]{f}{\var{$4$}}{denotes the exceptional, simple Lie algebra $\mathfrak{f}_{\var{4}}$.}

\defnBox[5]{g}{\var{$2$}}{denotes the exceptional, simple Lie algebra $\mathfrak{g}_{\var{2}}$.}

\subsubsectionAppendix{Other Supported Names for Identifying Simple Lie Algebras}{alternative_naming_of_algebras}

\defnBox[5]{su}{\var{$n$}\pattern}{denotes the simple Lie algebra $\mathfrak{su}_{\var{n}}$ for \var{$n$}$\,\geq\!2$. Equivalent to \mbox{\algL{a}\brace{\var{$n$}-1}}.}

\defnBox[5]{so}{\var{$n$}\pattern}{denotes the simple Lie algebra $\mathfrak{so}_{\var{n}}$ for \var{$n$}$\,\geq\!5$.\\
For odd \var{$n$},\hspace{4pt} \algL{so}\brace{\var{$n$}} is defined to be \mbox{\algL{b}\brace{(\var{$n$}-1)/2}}.\\
For even \var{$n$}, \algL{so}\brace{\var{$n$}} is defined to be \mbox{\algL{d}\brace{2\var{$n$}}}.\\
\mbox{}\hspace{-10pt}\textbf{Note}: $\mathfrak{so}_{4}\!\simeq\mathfrak{a}_1\!\times\!\mathfrak{a}_1$ is \emph{not} \emph{\textbf{simple}}; and $\mathfrak{so}_3\!\simeq\mathfrak{su}_2\!\simeq\mathfrak{a}_1$, so \algL{so}\brace{\var{$3$}}$\!\equivR\!$\algL{a}\brace{\var{$1$}}.}

\defnBox[5]{sp}{2\var{$n$}\pattern}{denotes the algebra $\mathfrak{sp}_{2\var{n}}$ for \var{$n$}$\,\geq\!3$. Equivalent to \mbox{\algL{c}\brace{\var{$n$}}}.\\
\mbox{}\hspace{-10pt}\textbf{Note}: $\mathfrak{sp}_4\!\simeq\mathfrak{so}_5$, so \algL{sp}\brace{\var{$4$}}$\!\equivR\!$\algL{b}\brace{\var{$2$}}.}

\subsectionAppendix{\textbf{Dynkin Labels} and Syntax for \emph{Irreducible} Representations}{dynkin_labels_for_irreps}

\defnBoxTwo[2]{w}{\var{cartanType}\pattern}{\var{highestWeightLabel}\patternTwo}{denotes the \emph{Dynkin label} for the irreducible representation of the Lie algebra of type `\var{cartanType}' whose highest-weight vector is given by the \built{Integer} \built{Sequence} \var{highestWeightLabel}.\\[-12pt]

\mbox{}\hspace{-10pt}\textbf{Note}: the \mbox{\var{cartanType}}$\,\in\!\{\algL{a},\algL{b},\algL{c},\algL{d},\algL{e},\algL{f},\algL{g}\}$, and the \var{highestWeightLabel}$\,\in\!\mathbb{Z}_{\geq0}^\var{k}$ for a rank-\var{$k$} Lie algebra.
}

\defnBox[6]{ad}{\var{algebra}\pattern}{abstractly denotes the \textbf{adjoint representation} of the simple Lie algebra \var{algebra}.}

\defnBox[6]{F}{\var{algebra}\pattern}{abstractly denotes the `\textbf{fundamental}' (or `\emph{defining}') representation of the simple Lie algebra \var{algebra}.}

\newpage
\subsectionAppendix{Naming of Arbitrary Representations of Simple Lie Algebras}{naming_arbitrary_reps}

Arbitrary representations $\mathbf{\b{R}}$ of any Lie algebra can be decomposed into an outer sum over \emph{irreducible representations} `$\t{\lambda}$' with \emph{multiplicities} `$m\indices{\mathbf{\b{R}}}{\t{\lambda}}\!$'$\in\!\mathbb{Z}_{\geq0}$ which are uniquely identified by their \hyperlink{label:w}{\emph{Dynkin labels}}:
\eq{\mathbf{\b{R}}\simeq\bigoplus_{\hspace{-5pt}\t{\lambda}\in\text{irreps}\hspace{-0pt}}\underbrace{(\t{\lambda}\oplus\cdots\oplus\t{\lambda})}_{m\indices{\mathbf{\b{R}}}{\t{\lambda}}\text{ times}}\equivL\bigoplus_{\hspace{-5pt}\t{\lambda}\in\text{irreps}\hspace{-5pt}}\hspace{-4pt}\t{\lambda}^{\oplus m\indices{\mathbf{\b{R}}}{\t{\lambda}}}\equivL\bigoplus_{\hspace{-5pt}\t{\lambda}\in\text{irreps}\hspace{-5pt}}\hspace{-1pt}m\indices{\mathbf{\b{R}}}{\t{\lambda}}\t{\lambda}\,.\vspace{-4pt}}
To encode these decompositions, we use the following syntax.

%Functions and wrappers
\defnBox[3]{plus}{\var{args}\patternTwo}{denotes the representation constructed out of representations specified by their Dynkin labels in \var{args} according to the outer sum
\eq{\bigoplus_{\lambda\in\text{\var{ags}}}\lambda\,.\vspace{-16pt}}
}

\defnBox[3]{times}{\var{$n$}\pattern\built{Integer}, \var{dynkinLabel}\pattern}{denotes the representation constructed from the \emph{irreducible} representation labelled by \var{dynkinLabel} by taking an \var{$n$}-fold sum:\\[-10pt]
\eq{\symbolL{times}\text{\brace{\var{$n$},\var{dynkinLabel}}}\equivR\symbolL{plus}\text{\brace{$\underbrace{\text{\var{dynkinLabel}},\ldots,\text{\var{dynkinLabel}}}_{\var{n}\text{ times}}$}}\,.\vspace{-10pt}}
}

\defnBox[3]{power}{\var{dynkinLabel}\pattern, \var{$n$}\pattern}{denotes the representation constructed from the representation labelled by \var{dynkinLabel} by taking an \var{$n$}-fold tensor product:\\[-10pt]
\eq{\symbolL{power}\text{\brace{\var{dynkinLabel},\var{$n$}}}\equivR\underbrace{(\text{\var{dynkinLabel}})\otimes\cdots\otimes(\text{\var{dynkinLabel}})}_{\var{n}\text{ times}}\,.\vspace{-10pt}}
}

\defnBox[3]{powers}{\var{dynkinLabel}\pattern, \var{$n$}\pattern}{denotes the representation constructed from the representation labelled by \var{dynkinLabel} by taking an \var{$n$}-fold \emph{symmetric} tensor product:\\[-10pt]
\eq{\symbolL{powers}\text{\brace{\var{dynkinLabel},\var{$n$}}}\equivR\underbrace{(\text{\var{dynkinLabel}})\odot\cdots\odot(\text{\var{dynkinLabel}})}_{\var{n}\text{ times}}\,.\vspace{-10pt}}
}

\defnBox[3]{powera}{\var{dynkinLabel}\pattern, \var{$n$}\pattern}{denotes the representation constructed from the representation labelled by \var{dynkinLabel} by taking an \var{$n$}-fold \emph{anti-symmetric} tensor product:\\[-10pt]
\eq{\symbolL{powers}\text{\brace{\var{dynkinLabel},\var{$n$}}}\equivR\underbrace{(\text{\var{dynkinLabel}})\wedge\cdots\wedge(\text{\var{dynkinLabel}})}_{\var{n}\text{ times}}\,.\vspace{-10pt}}
}

\subsectionAppendix{Abstract Colour Tensors and Other Objects Related to Amplitudes}{naming_colour_tensors}
\defnBox[7]{amp}{\var{legOrdering}\patternTwo\built{Integer}}{\emph{abstractly} denotes the \emph{ordered}, partial amplitude (`primitive') involving external particles with momenta labelled by the \built{List} appearing in \var{legOrdering}.%
}

\defnBox[4]{trR}{\var{indexList}\patternTwo\built{List}}{\emph{abstractly} denotes the tensor constructed as an outer-product of traces over generators `$\mathbf{T}^{\,\,\r{a}}_{\mathbf{\b{R}}}$'$\equivR
\mathbf{\b{R}}\indices{\b{[r]}\,\r{a}}{\b{[r]}}$ of an \emph{arbitrary} representation `$\mathbf{\b{R}}$'.\\[-10pt]
\eq{\mathbf{tr}_{\b{\textbf{R}}}\textbf{[}\r{\vec{\sigma}_1}|\cdots|\r{\vec{\sigma}_{\text{-}1}}\textbf{]}\equivR\bigotimes_i\mathbf{tr}_{\b{\textbf{R}}}\textbf{[}\r{\vec{\sigma}_i}\textbf{]}\vspace{-2pt}}
where
\eq{\mathbf{tr}_{\b{\mathbf{R}}}\textbf{[}\r{\sigma_i^1},\r{\ldots},\r{\sigma_i^{\text{-}1}}\textbf{]}\equivR\mathbf{tr}\big(\mathbf{T}_{\mathbf{\b{R}}}^{\r{\sigma_i^1}}\cdots\mathbf{T}_{\mathbf{\b{R}}}^{\r{\sigma_i^{\text{-}1}}}\big)\,.\vspace{-5pt}}
\mathematicaBox{
\mathematicaSequence{\{\funL[1]{nice}\brace{\var{\texttt{\#\hspace{-1pt}\,}}},\funL[1]{drawTensors}\brace{\var{\texttt{\#\hspace{-1pt}\,}}}\}\&@\tensor[4]{trR}\brace{\built{\{}1,2,3,4,5,6\built{\}}}}{
\fig[-4pt]{1}{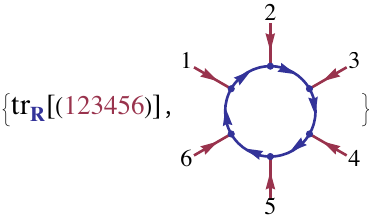}\vspace{-6pt}}
\mathematicaSequence{\{\funL[1]{nice}\brace{\var{\texttt{\#\hspace{-1pt}\,}}},\funL[1]{drawTensors}\brace{\var{\texttt{\#\hspace{-1pt}\,}}}\}\&@\tensor[4]{trR}\brace{\built{\{}1,4,5\built{\},\{}2,3,6\built{\}}}}{
\fig[-4pt]{1}{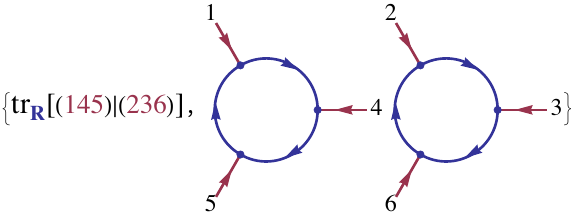}\vspace{-6pt}}
}
}

\defnBox[4]{trF}{\var{indexList}\patternTwo\built{List}}{same as \tensor[4]{trR}\brace{} above, but where the representation `$\mathbf{\b{R}}$' is taken as `$\mathbf{\b{F}}$': the \emph{fundamental} (or `defining') representation.}

\defnBox[4]{trAd}{\var{indexList}\patternTwo\built{List}}{same as \tensor[4]{trR}\brace{} above, but where the representation `$\mathbf{\b{R}}$' is taken as `$\mathbf{\r{ad}}$': the \emph{adjoint} representation.}

\defnBox[4]{ddmTensor}{\var{indexSequence}\patternTwo\built{Integer}}{\emph{abstractly} represents the tensor defined by\\[-8pt]
\eq{\tensor[4]{ddmTensor}\text{\brace{$\r{\sigma_1},\r{\ldots},\r{\sigma_n}$}}\equivR \sum_{\t{c_i}\in\t{[\mathfrak{g}]}}\big(\mathbf{\r{ad}}\indices{\r{\sigma_1}\,\r{\sigma_2}}{\t{c_1}}\mathbf{\r{ad}}\indices{\t{c_1}\,\r{\sigma_3}}{\t{c_2}}\cdots\mathbf{\r{ad}}\indices{\t{c_{\text{-2}}}\,\r{\sigma_{\text{-}2}}}{\t{c_{\text{-}1}}}\big)\mathcal{\r{K}}^{\smash{\t{c_{\text{-}1}}\,\r{\sigma_{\text{-}1}}}}\vspace{-1pt}}
where `$\r{\mathbf{ad}}$' denotes the adjoint and $\mathcal{\r{K}}$ the Cartan-Killing form (or `Killing metric') \funAL{killingMetric}\brace{\var{algebra}}. If some of the \var{indexSequence} are negative, then these index slots are lowered using the inverse of the Killing metric, \funAL{killingMetricInverse}\brace{\var{algebra}}.
\mathematicaBox{
\mathematicaSequence{\{\funL[1]{nice}\brace{\var{\texttt{\#\hspace{-1pt}\,}}},\funL[1]{drawTensors}\brace{\var{\texttt{\#\hspace{-1pt}\,}}}\}\&@\tensor[4]{ddmTensor}\brace{5,1,3,2,4}}{
\fig[-5pt]{1}{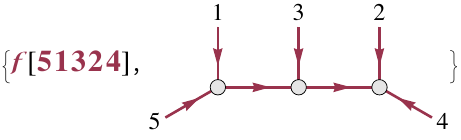}\vspace{-8pt}}
\mathematicaSequence{\{\funL[1]{nice}\brace{\var{\texttt{\#\hspace{-1pt}\,}}},\funL[1]{drawTensors}\brace{\var{\texttt{\#\hspace{-1pt}\,}}}\}\&@\tensor[4]{ddmTensor}\brace{5,-1,3,-2,4}}{
\fig[-5pt]{1}{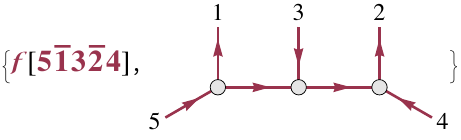}\vspace{-8pt}}
}

These tensors appear in the colour decomposition of tree amplitudes described by Del Duca, Dixon and Maltoni in \cite{DelDuca:1999rs}, and were denoted abstractly by `\fun{colourTensor}\brace{}' in the package \cite{Bourjaily:2026adf} and `\fun{colorFactor}\brace{}' in \cite{Bourjaily:2023uln}.
}

\defnBox[4]{joTensor}{\var{arguments}\patternTwo}{\emph{abstractly} represents the colour tensor described by Johansson and Ochirov in \cite{Johansson:2015oia} (see also \cite{Kosower:1988kh,Melia:2015ika,Ochirov:2019mtf}) appearing in the expansion of tree amplitudes involving some number of (identically) charged Fermions and gauge bosons in terms of the partial amplitudes described by Melia \cite{Melia:2013bta,Melia:2013epa,Melia:2013xok}. These tensors were denoted (abstractly) with the \built{Head} `\fun{colourTensor}' in the work of \cite{Bourjaily:2026adf}; other than this change in \built{Head}, our package uses identical syntax for \var{arguments}.%
\mathematicaBox{
\mathematicaSequence{egTensor=\built{RandomChoice}\brace{\funL[0]{fermionicTreeAmpTensors}\brace{3,2}}}{
\tensor[4]{joTensor}\brace{\{\fun{f}\brace{\hspace{-0pt}1\hspace{-0pt}}\hspace{-2pt},\hspace{-1pt}\fun{g}\brace{4}\hspace{-2pt},\hspace{-1pt}\fun{f}\brace{3}\hspace{-2pt},\hspace{-1pt}\fun{fb}\brace{2}\hspace{-2pt},\hspace{-1pt}\fun{f}\brace{2}\hspace{-2pt},\hspace{-1pt}\fun{g}\brace{5}\hspace{-2pt},\hspace{-1pt}\fun{fb}\brace{3}\hspace{-2pt},\hspace{-1pt}\fun{fb}\brace{1}\}\hspace{-2pt},\hspace{-2pt}\{\!\{1\hspace{-2pt},\hspace{-2pt}8\}\hspace{-2pt},\hspace{-2pt}\{5\hspace{-2pt},\hspace{-2pt}7\}\hspace{-2pt},\hspace{-2pt}\{3\hspace{-2pt},\hspace{-2pt}4\}\!\}}
}
\mathematicaSequence{\funL[0]{nice}\brace{egTensor}}{\fig[-2pt]{1}{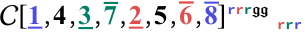}\vspace{-2pt}}
\mathematicaSequence{\funL[0]{drawTensors}\brace{egTensor}}{\fig[4pt]{1}{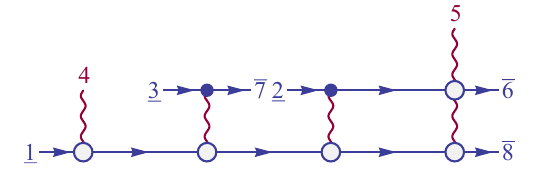}\vspace{-6pt}}
}\vspace{-20pt}
}

\defnBox[4]{clebschTensor}{\var{arguments}\patternTwo}{\emph{abstractly} encodes a Clebsch colour tensor (as defined in \mbox{ref.~\cite{Bourjaily:2026adf}}) relevant to \mbox{\algL{a}\brace{1}} gauge theory. The first (and only \emph{necessary}) part of \var{arguments} must consist of an \built{Integer} \built{Sequence} encoding the Dynkin weights of irreducible representations appearing along internal edges in the diagram; if no further arguments are given, then all external representations are taken to be the adjoint, \dynkinLabel{a}{2}:
\mathematicaBox{
\mathematicaSequence{egTensor=\built{RandomChoice}\brace{\funL[0]{a1ClebschBasis}\brace{8}}}{\tensor[4]{clebschTensor}\brace{4,2,0,2,0}}
\mathematicaSequence{\funL[0]{nice}\brace{egTensor}}{\fig[-2pt]{1}{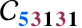}\vspace{-2pt}}
\mathematicaSequence{\funL[0]{drawTensors}\brace{egTensor}}{\fig[-4pt]{1}{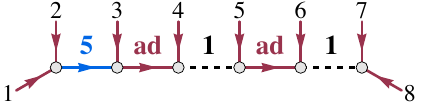}\vspace{-6pt}}
}

If, after the \built{Integer} \built{Sequence} encoding internal representations, a \built{List} is given as input, this is understood to encode either the weights of \emph{external particles} as ordered around the diagram:
\mathematicaBox{
\mathematicaSequence{egTensor=\built{RandomChoice}\brace{\funL[0]{a1ClebschBasis}\brace{\{3,5,-9,7,2,-4,6\}}}}{\tensor[4]{clebschTensor}\brace{4,9,4,6,\{3,5,-9,7,2,-4,6\}}}
\mathematicaSequence{\funL[0]{nice}\brace{egTensor}}{\fig[-2pt]{1}{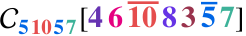}\vspace{-2pt}}
\mathematicaSequence{\funL[0]{drawTensors}\brace{egTensor}}{\fig[-4pt]{1}{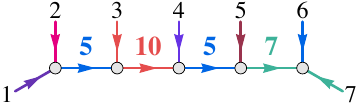}\vspace{-6pt}}
}

\noindent ---or, if a \built{List} encoding a permutation, the ordering of external particle labels\\[-12pt]
\mathematicaBox{
\mathematicaSequence{egTensor=\built{RandomChoice}\brace{\funL[0]{a1ClebschBasis}\brace{\{8,7,2,3,1,5,6,4\}}}}{\tensor[4]{clebschTensor}\brace{4,2,0,2,4,\{8,7,2,3,1,5,6,4\}}}
\mathematicaSequence{\funL[0]{drawTensors}\brace{egTensor}}{\fig[-4pt]{1}{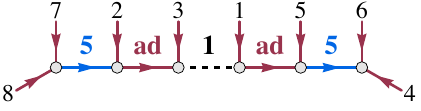}\vspace{-6pt}}
}

These cases may be disambiguated by including both optional arguments:
\mathematicaBox{
\mathematicaSequence{egTensor=\built{RandomChoice}\brace{\funL[0]{a1ClebschBasis}\brace{\{3\hspace{-2pt},\hspace{-2pt}5\hspace{-2pt},\hspace{-2pt}-9\hspace{-2pt},\hspace{-2pt}7\hspace{-2pt},\hspace{-2pt}2\hspace{-2pt},\hspace{-2pt}-4\hspace{-2pt},\hspace{-2pt}6\}\hspace{-2pt},\hspace{-2pt}\{3\hspace{-2pt},\hspace{-2pt}1\hspace{-2pt},\hspace{-2pt}6\hspace{-2pt},\hspace{-2pt}2\hspace{-2pt},\hspace{-2pt}7\hspace{-2pt},\hspace{-2pt}4\hspace{-2pt},\hspace{-2pt}5\}}}}{\tensor[4]{clebschTensor}\brace{12,10,7,5,\{-9,3,-4,5,6,7,2\},\{3,1,6,2,7,4,5\}}\hspace{-1pt}}
\mathematicaSequence{\funL[0]{nice}\brace{egTensor}}{\fig[-2pt]{1}{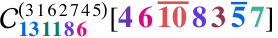}\vspace{-2pt}}
\mathematicaSequence{\funL[0]{drawTensors}\brace{egTensor}}{\fig[-4pt]{1}{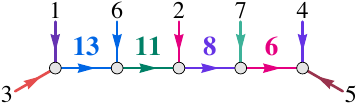}\vspace{-6pt}}
}
}

%================================================================================================================
%\newpage
\sectionAppendix{Formatted and Graphical Representations of Expressions}{appendix:formatted_output_and_graphics}\vspace{-4pt}
%================================================================================================================ 

\subsectionAppendix{Formatted Representations of Symbolic Expressions}{formatted_output}

\defnBox{nice}{\var{expression}\pattern}{converts many symbols and objects (including concrete \built{SparseArray} objects) into more human-readable, stylized form. Examples appear throughout this documentation and in this work's associated demonstration \built{Notebook}\brace{} \textbf{\texttt{demonstrating}\rule[-1.05pt]{7.5pt}{.75pt}\texttt{colour}\rule[-1.05pt]{7.5pt}{.75pt}\texttt{tensor}\rule[-1.05pt]{7.5pt}{.75pt}\texttt{tools}.\texttt{nb}}.}

\defnBox{niceTime}{\var{timeInSeconds}\pattern}{returns a more human-readable, approximated form of \var{timeInSeconds}.%
\mathematicaBox{
\mathematicaSequence{\built{RandomReal}\brace{1000}}{116.683}
\mathematicaSequence{\funL[1]{niceTime}@\%}{\fig[-2pt]{1}{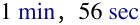}}
}\vspace{-6pt}
}

\defnBox{byteCount}{\var{expression}\pattern}{returns a stylized version of the output of the built-in \textsc{Mathematica} function \built{ByteCount}.
\mathematicaBox{
\mathematicaSequence{\funL[1]{fundamentalRep}\brace{\algL{e}\brace{8}}\\[-10pt]}{\fig[-2.5pt]{1}{fundamentalRep_e8}\vspace{-2pt}}
\mathematicaSequence{\funL[1]{byteCount}@\%}{\fig[-2pt]{1}{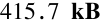}}
}\vspace{-6pt}
}

\defnBox{matrixForm}{\var{matrix}\pattern}{similar to \textsc{Mathematica}'s \built{MatrixForm}, but with some additional choices made for stylization.%
\mathematicaBox{
\mathematicaSequence{\funL[1]{matrixForm}@\built{RandomInteger}\brace{\{0,5\},\{10,10\}}\\[-8pt]}{\fig[-4pt]{1}{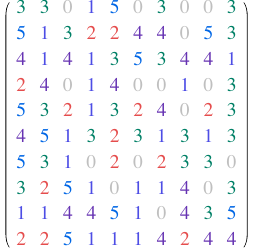}}
}\vspace{-6pt}
}

\defnBox{heatMap}{\var{matrix}\pattern}{similar to \funL[4]{matrixForm}, but where the elements of \var{matrix} are coloured according to their absolute magnitudes.\\[-10pt]

One application of interest to visually observe that mis-aligned (multi-)trace tensors become increasingly orthogonal for $\mathfrak{su}_{N_c}$ gauge theory in the limit of large rank:
\mathematicaBox{
\mathematicaSequence{\mbox{overlaps=\funL[0]{tensorOverlap}\brace{\repL{F}\brace{\algL{a}\brace{\var{\texttt{\#\hspace{-1pt}\,}}}\hspace{-1pt}}\hspace{-1pt}}\brace{\funL[0]{realMultiTraceTensors}\brace{4}\hspace{-1pt}}\&/@\{4\hspace{-1pt},\hspace{-2pt}15\}}\\[-12pt]}{\textnormal{\!\!\{}\fig[-2pt]{0.88}{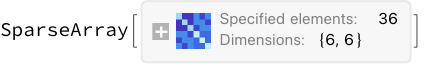},\fig[-2pt]{0.88}{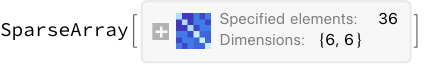}\textnormal{\}}\vspace{-4pt}}
\mathematicaSequence{\funL[1]{heatMap}/@overlaps\\[-8pt]}{\textnormal{\!\!\{}\fig[-5pt]{0.92}{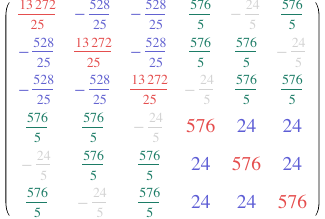},\fig[-5pt]{0.92}{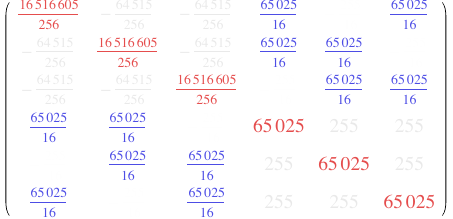}\textnormal{\}}\vspace{-0pt}}
}\vspace{-8pt}
}\vspace{-10pt}

\newpage
\subsectionAppendix{Graphical Representation of Concrete \& Symbolic Colour Tensors}{graphical_tensors}\vspace{-4pt}

\defnBox{drawTensors}{\var{expression}\pattern}{converts many symbols and concrete objects (including \built{SparseArray} objects) into graphical representations of tensor diagrams. While conventions used in these representations vary slightly depending on context (for example, the graphical representation of \mbox{\tensor[4]{joTensor}\brace{}} objects follows the standards described in \mbox{ref.\ \cite{Bourjaily:2026adf}}) there are several important features to note.\\[-10pt]

In these diagrams, the edges represent representations of some Lie algebra, with external edges representing the various arguments of the associated tensor. The labels attached to each external edge denote the \built{Slot} \built{Position}  of that representation's index range in the corresponding \built{SparseArray}. 
\mathematicaBox{
\mathematicaSequence{egTensor=\funL[0]{casimirC2}\brace{\funL[0]{fundamentalRep}\brace{\algL{f}\brace{4}}}}{\fig[-2pt]{1}{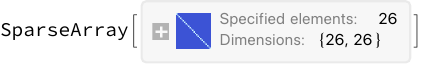}}
\mathematicaSequence{\funL[1]{drawTensors}\brace{egTesnor}}{\fig[-2pt]{1}{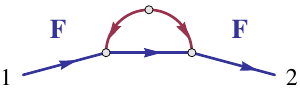}}
\mathematicaSequence{\funL[1]{nice}\brace{egTesnor}}{\fig[-2pt]{1}{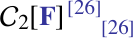}}
}~\\[-2pt]
The \emph{orientation} of edges distinguish representations from their conjugates; orientations of external edges encode whether the corresponding tensor's indices are raised or lowered.\\[-10pt]

The colours chosen for edges is largely arbitrary, but edges involving the adjoint representation are mostly distinguished from other representations by being coloured \r{red}; more precise information about the particular representations involved is shown upon mouse-over in \textsc{Mathematica} (which, unfortunately, is not easily demonstrated the examples shown here). 
\vspace{\fill}
}

\newpage
%================================================================================================================
%\vspace{8pt}
\sectionAppendix{Weight Systems and Dynkin Labels for Representations}{appendix:abstract_rep_operations}\vspace{-4pt}
%================================================================================================================ 

Functions in this section act at the level of weights and lattices, taking in algebras and weight vectors (as opposed to concrete representations). For these functions, any variable `\var{algebra}' should be given in the form described in \mbox{appendix~\ref{cartan_naming_of_lie_algebras}}, and any `\var{dynkinLabel}' for an \emph{irreducible} representation must be given in the form \dynkinLabel{\var{type}\pattern}{\var{highestWeightLabelSequence}\patternTwo} (or, if supported, in terms of representations constructed from such labels using \symbolL{plus} and \symbolL{times}).

\subsectionAppendix{Formatted Display of Lie Algebra Root System Information}{formatted_lattice_data}

\subsubsectionAppendix{Summarizing Conventions for Lie Algebra Root Systems}{summary_of_adjoint_roots}

\defnBox[21]{showLatticeData}{\var{algebra}\pattern}{%
displays a variety of useful information about the specified \var{algebra}, including: the Dynkin diagram; the Dynkin labels of the fundamental (\emph{defining}) `$\mathbf{\b{F}}$', adjoint `$\mathbf{\r{ad}}$', and other representations of note; the Cartan matrix; and the weight metric (also called the `quadratic form'). These objects all depend on an ordering of simple roots for the algebra; as these conventions vary considerably among authors, we summarize this collection of objects here. 
\mathematicaBox{\mathematicaSequence{\funAL[1]{showLatticeData}\brace{\algL{f}\brace{\var{$4$}}}\\[-8pt]}{$\begin{array}{@{}c@{}}\includegraphics[scale=1]{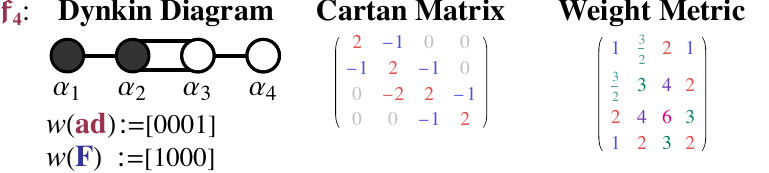}\end{array}$}
}\vspace{-6pt}%
}

\defnBox[21]{showFundamentalIrrepLabels}{\var{algebra}\pattern}{%
displays a formatted summary of Dynkin labels of irreducible representations of `fundamental weight'. As these assignments depend on how the simple roots are ordered (which is variable to an unfortunate extent across the literature), this summary should allow the user to rapidly translate between various conventions. 
\mathematicaBox{\mathematicaSequence{\funL[1]{showFundamentalIrrepLabels}\brace{\algL{d}\brace{\var{$5$}}}
}{$\begin{array}{@{}c@{}}\includegraphics[scale=1]{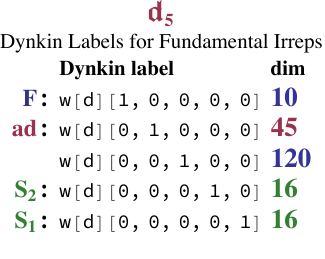}\end{array}$\\[-12pt]}
}\vspace{-6pt}%
}

\defnBox[21]{dynkinDiagram}{\var{algebra}\pattern}{shows the Dynkin diagram associated to the specified algebra, with nodes labelled according to our conventions for ordering the simple roots. Nodes associated to long roots are coloured white, and the nodes of short roots are coloured black.
\mathematicaBox{\mathematicaSequence{\funL[1]{dynkinDiagram}\brace{\algL{e}\brace{{$7$}}}
}{\fig[6pt]{1}{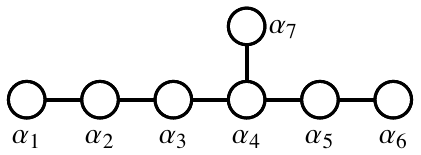}\\[-4pt]}
}\vspace{-6pt}%
}

\subsectionAppendix{Weight Systems for Representations of Simple Lie Algebras}{irrep_weight_systems}

\subsubsectionAppendix{Primary Objects Defined for Root Systems of Lie Algebras}{weights_and_dynkin_labels_primary}

\defnBox[21]{adjointDynkinLabel}{\var{algebra}\pattern}{returns the \hyperlink{label:w}{Dynkin label} for the adjoint representation of the specified \var{algebra}.
\mathematicaBox{
\mathematicaSequence{\funL[1]{adjointDynkinLabel}\brace{\algL{e}\brace{{$7$}}}}{
\dynkinLabel{e}{0,0,0,0,0,1,0}}
}\vspace{-6pt}
}

\defnBox[21]{cartan}{\var{algebra}\pattern}{returns the Cartan matrix for the specified \var{algebra}.
\mathematicaBox{
\mathematicaSequence{\funL[1]{cartan}\brace{\algL{e}\brace{$7$}}}{
\fig[-2pt]{1}{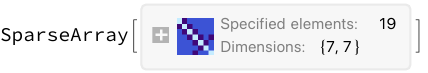}\vspace{-10pt}
}
\mathematicaSequence{\funL[0]{nice}@\%}{
\fig[-2.5pt]{1}{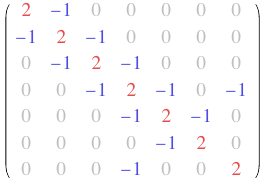}
}
}\vspace{-6pt}
}

\defnBox[21]{weightMetric}{\var{algebra}\pattern}{%
returns the weight metric for the specified \var{algebra}.
\mathematicaBox{
\mathematicaSequence{\funL[1]{weightMetric}\brace{\algL{f}\brace{$4$}}}{
\fig[-2pt]{1}{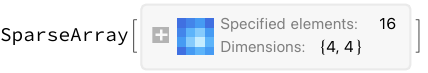}\vspace{-10pt}
}
\mathematicaSequence{\funL[1]{nice}@\%}{%
\fig[-2.5pt]{1}{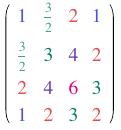}\vspace{-2pt}
}
}\vspace{-4pt}
}

\defnBox[21]{positiveRoots}{\var{algebra}\pattern}{returns a \built{SparseArray} corresponding to the (coefficients of the) positive roots of the adjoint lattice for the Lie algebra \var{algebra}. These are ordered from highest to lowest, with the \emph{simple} positive roots at the end. As all positive roots consist of sums over the \emph{simple roots} with non-negative, integer coefficients, this array is positive semi-definite.%
\mathematicaBox{
\mathematicaSequence{\funL[1]{positiveRoots}\brace{\algL{g}\brace{2}}}{\fig[-2pt]{1}{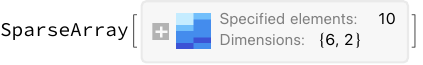}\vspace{-6pt}}
\mathematicaSequence{\funL[1]{nice}@\%\\[-8pt]}{\fig[-2pt]{1}{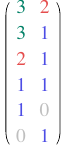}}
}\vspace{-4pt}
}

\defnBox[21]{positiveWeights}{\var{algebra}\pattern}{returns a \built{SparseArray} corresponding to the weights of the `positive' part of adjoint lattice for the Lie algebra \var{algebra}.\\[-12pt]

\mbox{}\hspace{-10pt}\textbf{Note}: these are defined via
\vspace{-8pt}\eq{\fwboxL{200pt}{\hspace{-100pt}\text{\funL[4]{positiveWeights}\brace{\var{algebra}}$\equivR$\funL[4]{positiveRoots}\brace{\var{algebra}}.\funL[4]{cartan}\brace{\var{algebra}}.}}\vspace{-20pt}}
\mathematicaBox{
\mathematicaSequence{\funL[1]{positiveWeights}\brace{\algL{g}\brace{2}}}{\fig[-2pt]{1}{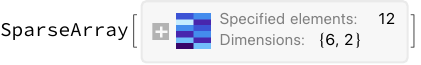}\vspace{-6pt}}
\mathematicaSequence{\funL[1]{nice}@\%\\[-8pt]}{\fig[-2pt]{1}{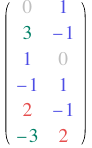}}
}\vspace{-4pt}
}

\defnBox[21]{coxeter}{\var{algebra}\pattern}{returns the Coxeter number for the specified \var{algebra}.%
\mathematicaBox{
\mathematicaSequence{\funL[1]{coxeter}/@\{\algL{a}\brace{8},\algL{b}\brace{8},\algL{c}\brace{8},\algL{d}\brace{8},\algL{e}\brace{6},\algL{e}\brace{7},\algL{e}\brace{8},\algL{f}\brace{4},\algL{g}\brace{2}\}}{
\{9,16,16,14,12,18,30,12,6\}}
}\vspace{-4pt}
}

\defnBox[21]{dualCoxeter}{\var{algebra}\pattern}{returns the dual-Coxeter number for \var{algebra}.%
\mathematicaBox{
\mathematicaSequence{\funL[1]{dualCoxeter}/@\{\algL{a}\brace{8},\algL{b}\brace{8},\algL{c}\brace{8},\algL{d}\brace{8},\algL{e}\brace{6},\algL{e}\brace{7},\algL{e}\brace{8},\algL{f}\brace{4},\algL{g}\brace{2}\}}{
\{9,15,9,14,12,18,30,9,4\}}
}\vspace{-4pt}
}

\newpage
\subsubsectionAppendix{Weight Systems for Irreducible Representations}{irrep_weight_systems}

\defnBoxTwo{irrepDimensionToDynkinLabels}{\var{algebra}\pattern}{\var{$d$}\pattern}{gives a \built{List} of \hyperlink{label:w}{\emph{Dynkin labels}} for  irreducible representations of \var{algebra} of the specified dimension \var{$d$}.
\mathematicaBox{
\mathematicaSequence{\funL[1]{irrepDimensionToDynkinLabels}\brace{\algL{e}\brace{6}}\brace{351}}{
\{\dynkinLabel{e}{2\hspace{-1.01pt},\hspace{-1.01pt}0\hspace{-1.01pt},\hspace{-1.01pt}0\hspace{-1.01pt},\hspace{-1.01pt}0\hspace{-1.01pt},\hspace{-1.01pt}0\hspace{-1.01pt},\hspace{-1.01pt}0}\hspace{-1.01pt},\hspace{-1.01pt}\dynkinLabel{e}{0\hspace{-1.01pt},\hspace{-1.01pt}1\hspace{-1.01pt},\hspace{-1.01pt}0\hspace{-1.01pt},\hspace{-1.01pt}0\hspace{-1.01pt},\hspace{-1.01pt}0\hspace{-1.01pt},\hspace{-1.01pt}0}\hspace{-1.01pt},\hspace{-1.01pt}\dynkinLabel{e}{0\hspace{-1.01pt},\hspace{-1.01pt}0\hspace{-1.01pt},\hspace{-1.01pt}0\hspace{-1.01pt},\hspace{-1.01pt}1\hspace{-1.01pt},\hspace{-1.01pt}0\hspace{-1.01pt},\hspace{-1.01pt}0}\hspace{-1.01pt},\hspace{-1.01pt}\dynkinLabel{e}{0\hspace{-1.01pt},\hspace{-1.01pt}0\hspace{-1.01pt},\hspace{-1.01pt}0\hspace{-1.01pt},\hspace{-1.01pt}0\hspace{-1.01pt},\hspace{-1.01pt}2\hspace{-1.01pt},\hspace{-1.01pt}0}\}}
}\vspace{-4pt}
}

\defnBox[8]{representationDimension}{\var{dynkinLabel}\pattern}{
returns the dimension of the representation indicated by \var{dynkinLabel}---which can be reducible or irreducible.\\[-12pt]

\mbox{}\hspace{-10pt}\textbf{Note}: one can also use \mbox{\repL{ad}\brace{\var{algebra}\pattern}} or \mbox{\repL{F}\brace{\var{algebra}\pattern}}.%
\mathematicaBox{
\mathematicaSequence{egIrrep=\hyperlink{label:w}{\fun{w}{\color{black}\brace{\algL{e}}}}@@\built{RandomInteger}\brace{\{1,5\},\{6\}}}{
\dynkinLabel{e}{1,2,3,1,5,5}}
\mathematicaSequence{\funL[1]{representationDimension}@\%}{129\hspace{1pt}122\hspace{1pt}300\hspace{1pt}461\hspace{1pt}794\hspace{1pt}003\hspace{1pt}000}
}\vspace{-4pt}
}

\defnBox[8]{conjugateRep}{\var{dynkinLabel}\pattern}{returns the Dynkin label of the \emph{conjugate} representation to that labelled by \var{dynkinLabel}.
\mathematicaBox{
\mathematicaSequence[1]{\{\var{\texttt{\#\hspace{-1pt}\,}},``$\leftrightarrow$'',\funDL[1]{conjugateRep}\brace{\var{\texttt{\#\hspace{-1pt}\,}}}\}\&/@\hyperlink{label:w}{\fun{w}{\color{black}\brace{\fun{e}}}}@@@\built{IdentityMatrix}\brace{6};\\
\{\built{Grid}\brace{\funL[0]{nice}/@\%},\built{Grid}\brace{\%}\}}{
$\Big\{\begin{array}{@{}c@{\leftrightarrow}c@{}}
\rule[-1pt]{0pt}{13pt}\textnormal{\textbf{\b{F}}}&\b{\bar{\textnormal{\textbf{\b{F}}}}}\\[-4pt]
\rule[-1pt]{0pt}{13pt}\textnormal{\textbf{\b{351}}}&\b{\textnormal{\textbf{\b{351}}}}\\[-4pt]
\rule[-1pt]{0pt}{13pt}\textnormal{\textbf{\b{2925}}}&\b{\textnormal{\textbf{\b{2925}}}}\\[-4pt]
\rule[-1pt]{0pt}{13pt}\textnormal{\textbf{\b{351}}}&\b{\textnormal{\textbf{\b{351}}}}\\[-4pt]
\rule[-1pt]{0pt}{13pt}\b{\bar{\textnormal{\textbf{\b{F}}}}}&\b{\textnormal{\textbf{\b{F}}}}\\[-4pt]
\rule[-1pt]{0pt}{13pt}\textnormal{\textbf{\r{ad}}}&\r{\textnormal{\textbf{\r{ad}}}}
\end{array},\begin{array}{@{}c@{\leftrightarrow}c@{}}
\rule[-1pt]{0pt}{13pt}\dynkinLabel{e}{1,{\color{dim}0},{\color{dim}0},{\color{dim}0},{\color{dim}0},{\color{dim}0}}&\dynkinLabel{e}{{\color{dim}0},{\color{dim}0},{\color{dim}0},{\color{dim}0},1,{\color{dim}0}}\\[-4pt]
\rule[-1pt]{0pt}{13pt}\dynkinLabel{e}{{\color{dim}0},1,{\color{dim}0},{\color{dim}0},{\color{dim}0},{\color{dim}0}}&\dynkinLabel{e}{{\color{dim}0},{\color{dim}0},{\color{dim}0},1,{\color{dim}0},{\color{dim}0}}\\[-4pt]
\rule[-1pt]{0pt}{13pt}\dynkinLabel{e}{{\color{dim}0},{\color{dim}0},1,{\color{dim}0},{\color{dim}0},{\color{dim}0}}&\dynkinLabel{e}{{\color{dim}0},{\color{dim}0},1,{\color{dim}0},{\color{dim}0},{\color{dim}0}}\\[-4pt]
\rule[-1pt]{0pt}{13pt}\dynkinLabel{e}{{\color{dim}0},{\color{dim}0},{\color{dim}0},1,{\color{dim}0},{\color{dim}0}}&\dynkinLabel{e}{{\color{dim}0},1,{\color{dim}0},{\color{dim}0},{\color{dim}0},{\color{dim}0}}\\[-4pt]
\rule[-1pt]{0pt}{13pt}\dynkinLabel{e}{{\color{dim}0},{\color{dim}0},{\color{dim}0},{\color{dim}0},1,{\color{dim}0}}&\dynkinLabel{e}{1,{\color{dim}0},{\color{dim}0},{\color{dim}0},{\color{dim}0},{\color{dim}0}}\\[-4pt]
\rule[-1pt]{0pt}{13pt}\dynkinLabel{e}{{\color{dim}0},{\color{dim}0},{\color{dim}0},{\color{dim}0},{\color{dim}0},1}&\dynkinLabel{e}{{\color{dim}0},{\color{dim}0},{\color{dim}0},{\color{dim}0},{\color{dim}0},1}
\end{array}\Big\}$
}
}\vspace{-4pt}
}

\defnBox[8]{realRepresentationQ}{\var{dynkinLabel}\pattern}{returns \built{True} or \built{False} depending on whether \var{dynkinLabel}==\funDL[4]{conjugateRep}\brace{\var{dynkinLabel}}.%
\mathematicaBox{
\mathematicaSequence{\funL[1]{realRepresentationQ}/\@(\hyperlink{label:w}{\fun{w}{\color{black}\brace{\algL{e}}}}@@@\built{IdentityMatrix}\brace{6})}{\{\built{False},\built{False},\built{True},\built{False},\built{False},\built{True}\}}
}\vspace{-4pt}
}

\defnBox[8]{pseudoRealRepresentationQ}{\var{dynkinLabel}\pattern}{returns \built{True} if the representation is \textbf{\emph{both}} \textbf{\emph{real}} \emph{and} the decomposition of the \emph{antisymmetric} tensor product of this representation with itself includes the trivial representation. The former criterion is tested by \funDL[4]{realRepresentationQ}\brace{\var{dynkinLabel}}, and the later can be tested using \funL[2]{antiSymPowerDecomposition}\brace{\var{dynkinLabel},2}.
\mathematicaBox{
\mathematicaSequence{\built{Select}\brace{\hyperlink{label:w}{\fun{w}{\color{black}\brace{\algL{c}}}}@@@\built{IdentityMatrix}\brace{6}),\funRL[1]{pseudoRealRepresentationQ}}}{
\{\dynkinLabel{c}{1,0,0,0,0,0},\dynkinLabel{c}{0,0,1,0,0,0},\dynkinLabel{c}{0,0,0,0,1,0}\}
}
\mathematicaSequence{\built{Column}\brace{\funL[1]{nice}\brace{\funL[1]{antiSymPowerDecomposition}\brace{\var{\texttt{\#\hspace{-1pt}\,}},2}}\&/@\%}\\[-8pt]}{\fig[-4pt]{1}{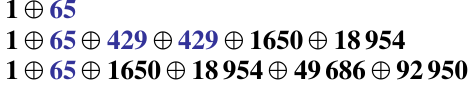}\vspace{-4pt}}
\mathematicaSequence{\built{Column}\brace{\funL[1]{nice}\brace{\funL[1]{symPowerDecomposition}\brace{\var{\texttt{\#\hspace{-1pt}\,}},2}}\&/@\%\%}\\[-8pt]}{\fig[-4pt]{1}{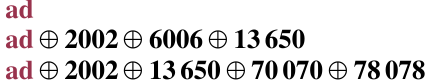}\vspace{-4pt}}
}\vspace{-6pt}
}

\defnBox[8]{showWeightLattice}{\var{dynkinLabel}\pattern}{%
returns a \emph{formatted} (`spindle-shaped') display of the collection of weights for any \emph{irreducible} representation identified by \var{dynkinLabel}.\\[-14pt]

\mbox{}\hspace{-10pt}\textbf{Note}: negative integers are shown with over-bars: $\bar{q}\equivR\text{-}q$.
\mathematicaBox{\mathematicaSequence{\funDL[1]{showWeightLattice}\brace{\dynkinLabel{g}{1,2}}
}{\fig[-3.5pt]{0.9}{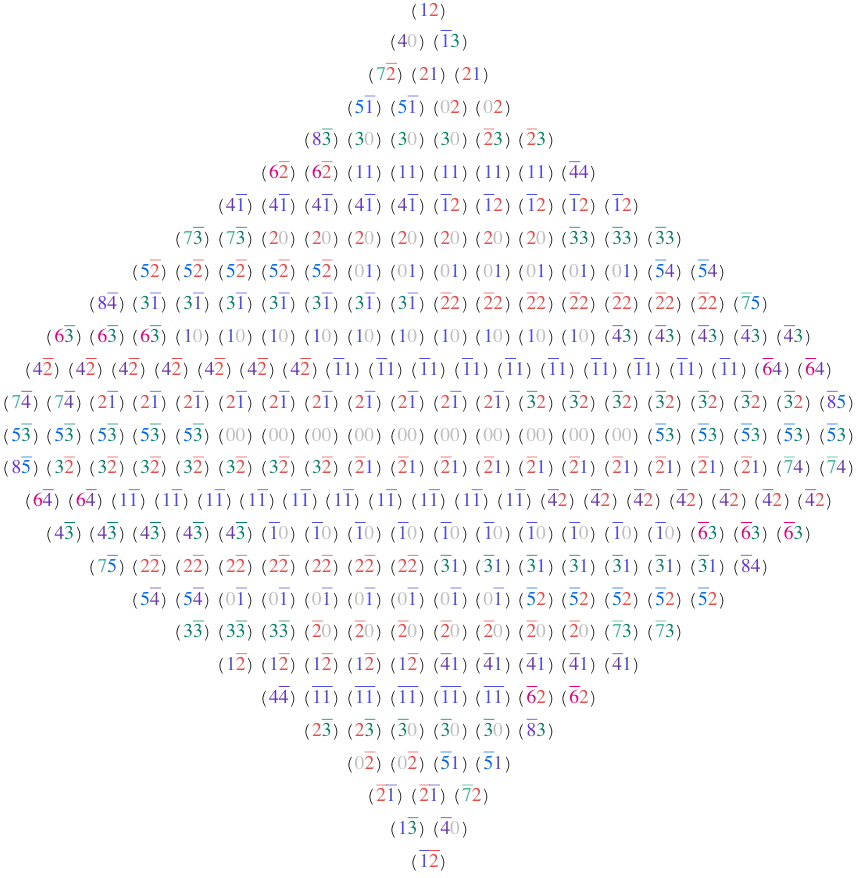}\\[-2pt]}
}\vspace{-4pt}%
}

\defnBox[8]{representationWeights}{\var{dynkinLabel}\pattern}{
returns the \emph{unformatted} collection of weights (with duplication for multiplicity) for the lattice corresponding to the irreducible representation labelled by \var{dynkinLabel}.
\mathematicaBox{%
\mathematicaSequence{\funDL[1]{representationWeights}\brace{\dynkinLabel{g}{1,2}}}{
\fig[-2.5pt]{1}{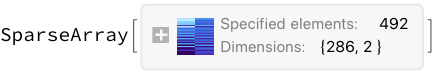}\vspace{-2pt}
}
}\vspace{-0pt}
}

\defnBox[8]{showRootLattice}{\var{dynkinLabel}\pattern}{%
returns a formatted (`spindle-shaped') display of the collection of roots (or, more precisely, \emph{marks}: the \emph{coefficients} of the weights as expanded in terms of simple roots) for any \emph{irreducible} representation identified by \var{dynkinLabel}. 
\\[-14pt]

\mbox{}\hspace{-10pt}\textbf{Note}: negative numbers are shown with over-bars: $\bar{q}\equivR\text{-}q$.
\mathematicaBox{\mathematicaSequence{\funDL[1]{showRootLattice}\brace{\dynkinLabel{g}{1,2}}
}{\fig[-18pt]{0.9}{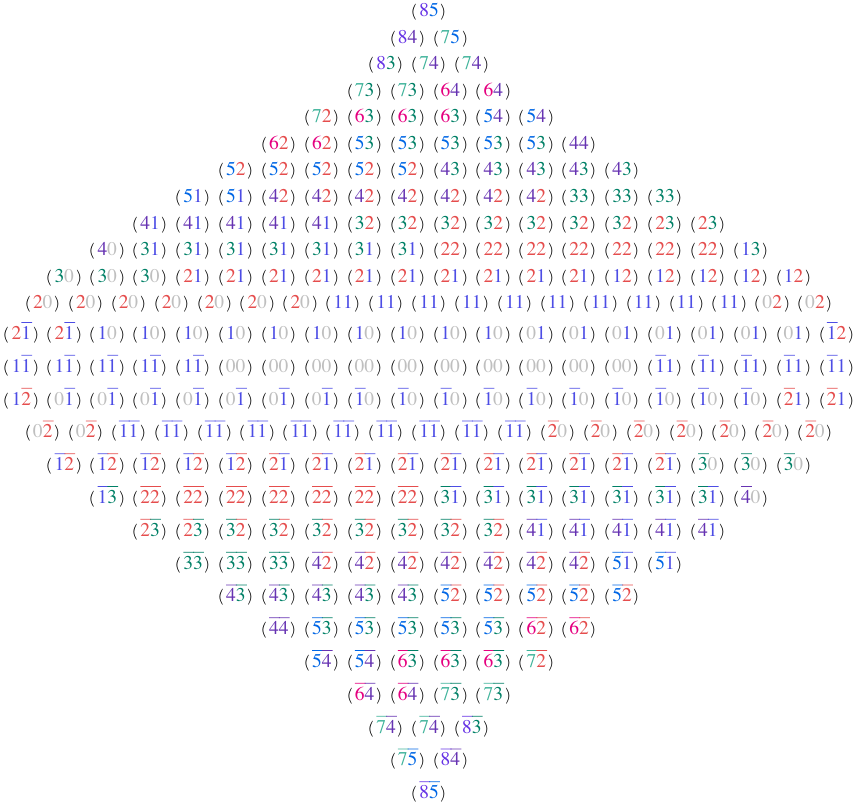}\\[-2pt]}
}\vspace{-4pt}
}

\defnBox[8]{representationRoots}{\var{dynkinLabel}\pattern}{
returns the \emph{unformatted} collection of roots (or, more precisely, \emph{marks}: the \emph{coefficients} of weights as expanded in terms of simple roots) (with duplication for multiplicity) for the lattice corresponding to the irreducible representation labelled by \var{dynkinLabel}.
\mathematicaBox{%
\mathematicaSequence{\funDL[1]{representationRoots}\brace{\dynkinLabel{g}{1,2}}}{
\rule{0pt}{16pt}\fig[-2.5pt]{1}{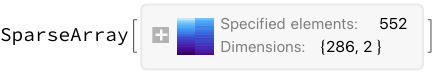}\vspace{-2pt}
}
}\vspace{-4pt}
}

\defnBox[8]{dynkinIndex}{\var{dynkinLabel}\pattern}{returns the Dynkin index `$T(\mathbf{\b{R}})$' for the irreducible representation $\mathbf{\b{R}}$ labelled by \var{dynkinLabel}. It is defined relative to the (conventional) scaling of the Killing form:
\eq{T(\mathbf{\b{R}})\r{\mathcal{K}}^{\r{a\,b}}\equivR\mathbf{tr}_{\mathbf{\b{R}}}(\r{a\,b})\,.}

\mbox{}\hspace{-10pt}\textbf{Note}: this eigenvalue is subject to conventional ambiguity related to that of \mbox{\funL[4]{casimirEigenvalue}\brace{}}:
\eq{T(\mathbf{\b{R}})\,\mathrm{dim}(\mathfrak{\r{g}})=C_2(\mathbf{\b{R}})\,\mathrm{dim}(\mathbf{\b{R}})\,.\vspace{-16pt}\label{casimir_and_index}}
}

\defnBox[81]{casimirEigenvalue}{\var{dynkinLabel}\pattern}{returns the \emph{eigenvalue} of the quadratic Casimir `$C_2(\mathbf{\b{R}})$' for the irreducible representation $\mathbf{\b{R}}$ specified by \var{dynkinLabel}.\\[-14pt]

\mbox{}\hspace{-10pt}\textbf{Note}: this eigenvalue is subject to conventional ambiguity related to that of \mbox{\funDL[4]{dynkinIndex}\brace{}}; see equation (\ref{casimir_and_index}).
\mathematicaBox{
\mathematicaSequence{egIrrep=\dynkinLabel{e}{\var{\texttt{\#\hspace{-1pt}\#\hspace{-1pt}}}}\&@@\built{RandomInteger}\brace{\{0,4\},\{6\}}}{
\dynkinLabel{e}{4,1,0,3,1,3}}
\mathematicaSequence{\mbox{\funL[1]{casimirEigenvalue}\brace{egIrrep}\,\funDL[1]{representationDimension}\brace{egIrrep}}}{
2049465600}
\mathematicaSequence{\mbox{\funDL[1]{dynkinIndex}\brace{egIrrep}\,\funDL[1]{representationDimension}\brace{\repL{ad}\brace{egIrrep}}}}{
2049465600}
}\vspace{-4pt}
}

\subsubsectionAppendix{Weight Systems for Representations \emph{Built from} Irreducible Ones}{building_lattices}

\defnBox{tensorProductDecomposition}{\var{dynkinLabelA}\pattern,\var{dynkinLabelB}\pattern}{returns the decomposition of the tensor-product-representation of the (not necessarily irreducible) representations labelled by \var{dynkinLabelA} and \var{dynkinLabelB} into irreducible representations with multiplicity.
\mathematicaBox{
\mathematicaSequence{\funL[1]{tensorProductDecomposition}\brace{\dynkinLabel{\algL{f}}{1,0,0,0},\dynkinLabel{\algL{f}}{0,1,0,0}}}{
\symbolL{plus}\!\texttt{[}%
\mbox{\symbolL{times}\brace{\hspace{-1pt}1\hspace{-1.02pt},\hspace{-1.02pt}\dynkinLabel{\algL{f}}{1\hspace{-1.17pt},\hspace{-1.17pt}0\hspace{-1.17pt},\hspace{-1.17pt}0\hspace{-1.17pt},\hspace{-1.17pt}0\hspace{-0.91pt}}\hspace{-0.9pt}}}\hspace{-1.11pt},%
\hspace{-1.11pt}\mbox{\symbolL{times}\brace{\hspace{-1pt}1\hspace{-1.02pt},\hspace{-1.02pt}\dynkinLabel{\algL{f}}{0\hspace{-1.17pt},\hspace{-1.17pt}0\hspace{-1.17pt},\hspace{-1.17pt}0\hspace{-1.17pt},\hspace{-1.17pt}1\hspace{-0.91pt}}\hspace{-0.9pt}}}\hspace{-1.11pt},%
\hspace{-1.11pt}\mbox{\symbolL{times}\brace{\hspace{-1pt}1\hspace{-1.02pt},\hspace{-1.02pt}\dynkinLabel{\algL{f}}{0\hspace{-1.17pt},\hspace{-1.17pt}1\hspace{-1.17pt},\hspace{-1.17pt}0\hspace{-1.17pt},\hspace{-1.17pt}0\hspace{-0.91pt}}\hspace{-0.9pt}}}\hspace{-1.11pt},\\%
\hspace{-1.11pt}\mbox{\symbolL{times}\brace{\hspace{-1pt}1\hspace{-1.02pt},\hspace{-1.02pt}\dynkinLabel{\algL{f}}{2\hspace{-1.17pt},\hspace{-1.17pt}0\hspace{-1.17pt},\hspace{-1.17pt}0\hspace{-1.17pt},\hspace{-1.17pt}0\hspace{-0.91pt}}\hspace{-0.9pt}}}\hspace{-1.11pt},%
\hspace{-1.11pt}\mbox{\symbolL{times}\brace{\hspace{-1pt}1\hspace{-1.02pt},\hspace{-1.02pt}\dynkinLabel{\algL{f}}{1\hspace{-1.17pt},\hspace{-1.17pt}0\hspace{-1.17pt},\hspace{-1.17pt}0\hspace{-1.17pt},\hspace{-1.17pt}1\hspace{-0.91pt}}\hspace{-0.9pt}}}\hspace{-1.11pt},%
\hspace{-1.11pt}\mbox{\symbolL{times}\brace{\hspace{-1pt}1\hspace{-1.02pt},\hspace{-1.02pt}\dynkinLabel{\algL{f}}{0\hspace{-1.17pt},\hspace{-1.17pt}0\hspace{-1.17pt},\hspace{-1.17pt}1\hspace{-1.17pt},\hspace{-1.17pt}0\hspace{-0.91pt}}\hspace{-0.9pt}}}\hspace{-1.11pt},%
\hspace{-1.11pt}\mbox{\symbolL{times}\brace{\hspace{-1pt}1\hspace{-1.02pt},\hspace{-1.02pt}\dynkinLabel{\algL{f}}{1\hspace{-1.17pt},\hspace{-1.17pt}1\hspace{-1.17pt},\hspace{-1.17pt}0\hspace{-1.17pt},\hspace{-1.17pt}0\hspace{-0.91pt}}\hspace{-0.9pt}}}\texttt{]}
}
\mathematicaSequence{\funL[1]{nice}@\%}{\fig[-2pt]{1}{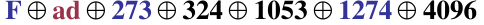}
}
}\vspace{-4pt}
}

\defnBox{tensorPowerDecomposition}{\var{dynkinLabel}\pattern,\var{$n$}\pattern\built{Integer}}{returns the decomposition of the tensor-power-representation of the (not necessarily irreducible) representation labelled by \var{dynkinLabel} into irreducible representations with multiplicity.
\mathematicaBox{
\mathematicaSequence{\funL[1]{tensorPowerDecomposition}\brace{\repL{F}\brace{\algL{g}\brace{2}},3}}{
\symbolL{plus}\!\texttt{[}%
\mbox{\symbolL{times}\brace{\hspace{-1pt}1\hspace{-1.02pt},\hspace{-1.02pt}\dynkinLabel{\algL{g}}{0\hspace{-1.17pt},\hspace{-1.17pt}0\hspace{-0.91pt}}\hspace{-0.9pt}}}\hspace{-1.11pt},%
\hspace{-1.11pt}\mbox{\symbolL{times}\brace{\hspace{-1pt}4\hspace{-1.02pt},\hspace{-1.02pt}\dynkinLabel{\algL{g}}{1\hspace{-1.17pt},\hspace{-1.17pt}0\hspace{-0.91pt}}\hspace{-0.9pt}}}\hspace{-1.11pt},%
\hspace{-1.11pt}\mbox{\symbolL{times}\brace{\hspace{-1pt}2\hspace{-1.02pt},\hspace{-1.02pt}\dynkinLabel{\algL{g}}{0\hspace{-1.17pt},\hspace{-1.17pt}1\hspace{-0.91pt}}\hspace{-0.9pt}}}\hspace{-1.11pt},%
\hspace{-1.11pt}\mbox{\symbolL{times}\brace{\hspace{-1pt}3\hspace{-1.02pt},\hspace{-1.02pt}\dynkinLabel{\algL{g}}{2\hspace{-1.17pt},\hspace{-1.17pt}0\hspace{-0.91pt}}\hspace{-0.9pt}}}\hspace{-1.11pt},%
\hspace{-1.11pt}\mbox{\symbolL{times}\brace{\hspace{-1pt}2\hspace{-1.02pt},\hspace{-1.02pt}\dynkinLabel{\algL{g}}{1\hspace{-1.17pt},\hspace{-1.17pt}1\hspace{-0.91pt}}\hspace{-0.9pt}}}\hspace{-1.11pt},%
\hspace{-1.11pt}\mbox{\symbolL{times}\brace{\hspace{-1pt}1\hspace{-1.02pt},\hspace{-1.02pt}\dynkinLabel{\algL{g}}{3\hspace{-1.17pt},\hspace{-1.17pt}0\hspace{-0.91pt}}\hspace{-0.9pt}}}\texttt{]}
}
\mathematicaSequence{\funL[1]{nice}@\%}{\fig[-0pt]{1}{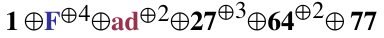}\vspace{-4pt}
}
}\vspace{-4pt}
}

\defnBox{symPowerDecomposition}{\var{dynkinLabel}\pattern,\var{$n$}\pattern\built{Integer}}{returns the decomposition of the symmetric tensor-power-representation of the (not necessarily irreducible) representation labelled by \var{dynkinLabel} into irreducible representations with multiplicity.
\mathematicaBox{
\mathematicaSequence{\funL[1]{symPowerDecomposition}\brace{\repL{ad}\brace{\algL{g}\brace{2}},2}}{
\symbolL{plus}\!\texttt{[}%
\mbox{\symbolL{times}\brace{\hspace{-1pt}1\hspace{-1.02pt},\hspace{-1.02pt}\dynkinLabel{\algL{g}}{0\hspace{-1.17pt},\hspace{-1.17pt}0\hspace{-0.91pt}}\hspace{-0.9pt}}}\hspace{-1.11pt},%
\hspace{-1.11pt}\mbox{\symbolL{times}\brace{\hspace{-1pt}1\hspace{-1.02pt},\hspace{-1.02pt}\dynkinLabel{\algL{g}}{2\hspace{-1.17pt},\hspace{-1.17pt}0\hspace{-0.91pt}}\hspace{-0.9pt}}}\hspace{-1.11pt},%
\hspace{-1.11pt}\mbox{\symbolL{times}\brace{\hspace{-1pt}1\hspace{-1.02pt},\hspace{-1.02pt}\dynkinLabel{\algL{g}}{0\hspace{-1.17pt},\hspace{-1.17pt}2\hspace{-0.91pt}}\hspace{-0.9pt}}}\texttt{]}
}
\mathematicaSequence{\funL[1]{nice}@\%}{\fig[-2pt]{1}{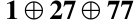}
}
}\vspace{-4pt}
}

\defnBox{antiSymPowerDecomposition}{\var{dynkinLabel}\pattern,\var{$n$}\pattern\built{Integer}}{returns the decomposition of the antisymmetric tensor-power-representation of the (not necessarily irreducible) representation labelled by \var{dynkinLabel} into irreducible representations with multiplicity.
\mathematicaBox{
\mathematicaSequence{\funL[1]{antiSymPowerDecomposition}\brace{\repL{ad}\brace{\algL{g}\brace{2}},2}}{
\symbolL{plus}\!\texttt{[}%
\mbox{\symbolL{times}\brace{\hspace{-1pt}1\hspace{-1.02pt},\hspace{-1.02pt}\dynkinLabel{\algL{g}}{0\hspace{-1.17pt},\hspace{-1.17pt}1\hspace{-0.91pt}}\hspace{-0.9pt}}}\hspace{-1.11pt},%
\hspace{-1.11pt}\mbox{\symbolL{times}\brace{\hspace{-1pt}1\hspace{-1.02pt},\hspace{-1.02pt}\dynkinLabel{\algL{g}}{3\hspace{-1.17pt},\hspace{-1.17pt}0\hspace{-0.91pt}}\hspace{-0.9pt}}}\texttt{]}
}
\mathematicaSequence{\funL[1]{nice}@\%}{\fig[-2pt]{1}{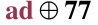}
}
}\vspace{-4pt}
}

\subsectionAppendix{Counting Independent Colour Tensors for Amplitudes}{counting_colour_tensor_bases}

\defnBox{numberOfIndependentColourTensors}{\var{stateReps}\patternTwo}{returns the number of linearly independent colour tensors involving irreducible representations encoded by \var{stateReps}. See \mbox{ref.~\cite{Bourjaily:2024jbt}}. 
\mathematicaBox{
\mathematicaSequence{\funL[1]{numberOfIndependentColourTensors}\brace{\var{\texttt{\#\hspace{-1pt}\,}}\repL{ad}\brace{\algL{a}\brace{1}}}\&/@\built{Range}\brace{3,10}}{
\{1,3,6,15,36,91,232,603\}
}
\mathematicaSequence{\funL[1]{numberOfIndependentColourTensors}\brace{\var{\texttt{\#\hspace{-1pt}\,}}\repL{ad}\brace{\algL{e}\brace{8}}}\&/@\built{Range}\brace{3,10}}{
\{1,5,16,79,421,2674,19244,156612\}
}
}

\mbox{}\hspace{-10pt}\textbf{Note}: in the interpretation of \var{stateReps}, any representations \emph{other than the adjoint} will be considered to label that of a Fermion-$\bar{\text{Fermion}}$ \emph{pair}. Thus, for example,
\mathematicaBox{
\mathematicaSequence{\funL[1]{numberOfIndependentColourTensors}\brace{3\repL{ad}\brace{\algL{a}\brace{3}},3\repL{F}\brace{\algL{a}\brace{3}}}}{
403}
\mathematicaSequence[1]{\funL[1]{tensorProductDecomposition}\brace{\symbolL{power}\brace{\repL{ad}\brace{\algL{a}\brace{3}},3},\symbolL{power}\brace{\repL{F}\brace{\algL{a}\brace{3}},3}}\!;\\
\built{Cases}\brace{\%,\symbolL{times}\brace{\var{mult}\pattern,\fun{w}\brace{\pattern}\brace{0..}}$\mapsto$\var{mult}}}{\{\}}
\mathematicaSequence[1]{\funL[1]{tensorProductDecomposition}\brace{\symbolL{power}\brace{\repL{ad}\brace{\algL{a}\brace{3}},3},\symbolL{power}\brace{\repL{F}\brace{\algL{a}\brace{3}},3}, \symbolL{power}\brace{\funL[1]{conjugateRep}@\repL{F}\brace{\algL{a}\brace{3}},3}};\\
\built{Cases}\brace{\%,\symbolL{times}\brace{\var{mult}\pattern,\fun{w}\brace{\pattern}\brace{0..}}$\mapsto$\var{mult}}}{\{403\}}
}
}

\newpage
%================================================================================================================
\vspace{8pt}\sectionAppendix{\emph{Concrete} Representations of Simple Lie Algebras}{appendix:concrete_reps_of_simple_lie_algebras}\vspace{-0pt}
%================================================================================================================ 

\subsectionAppendix{Primary, Concrete Representations of Simple Lie Algebras}{main_irreps}

\subsubsectionAppendix{\emph{Defining} and Related Representations of Simple Lie Algebras}{fundamental_reps}

\defnBox{fundamentalRep}{\var{algebra}\pattern}{returns the \emph{defining} representation of the simple Lie algebra \var{algebra} as a \built{SparseArray} object encoding the rank-(2,1) tensor $\mathbf{\b{R}}\indices{\b{[r]}\,\r{[\mathfrak{g}]}}{\b{[r]}}.$
\mathematicaBox{
\mathematicaSequence{\funL[1]{fundamentalRep}\brace{\algL{f}\brace{4}}}{\fig[-2pt]{1}{fundamentalRep_f4}\vspace{-4pt}}
\mathematicaSequence{\{\funL[0]{nice}@\%,\funL[0]{drawTensors}@\%\}}{\fig[-2pt]{1}{fundamentalRep_f4_nice}\vspace{-4pt}}
}\vspace{-4pt}
}

\defnBox{adjointRep}{\var{algebra}\pattern}{returns the \emph{adjoint representation} for the corresponding simple Lie algebra. Equivalent to \mbox{\funL[4]{inducedAdjoint}\brace{\funL[4]{fundamentalRep}\brace{\var{algebra}\hspace{-1pt}}\hspace{-1pt}}\hspace{-1pt}.\hspace{-10pt}}}

\defnBox{trivialRep}{\var{algebra}\pattern}{returns an empty \built{SparseArray} with \built{Dimensions} equal to $\{1,\mathrm{dim}(\text{\var{algebra}}),1\}$.
\mathematicaBox{
\mathematicaSequence{\funL[1]{trivialRep}\brace{\algL{f}\brace{4}}}{\fig[-2pt]{1}{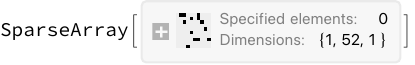}\vspace{-4pt}}
\mathematicaSequence{\funL[1]{nice}@\%}{\fig[-2pt]{1}{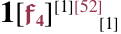}\vspace{-4pt}}
}\vspace{-4pt}
}

\subsubsectionAppendix{Other Irreducible Representations of `\emph{Fundamental Weight}'}{other_fundamental_irreps}

\defnBox[31]{fundamentalRep}{\var{algebra}\pattern,\var{$j$}\pattern}{returns the representation of \emph{fundamental \emph{(}Dynkin-\emph{)}weight} with a `1' in the $\var{j}^{\text{th}}$ position---that is, the irreducible representation with \hyperlink{label:w}{\emph{Dynkin index}} specified by \hyperlink{label:w}{\fun{w}{\color{black}\brace{\var{type}}\brace{$\underset{1}{0},\underset{\cdots}{\ldots},\underset{\var{j}}{1},\underset{\cdots}{\ldots},\underset{k}{0}$}}}.\\[-12pt]
\mathematicaBox{
\mathematicaSequence{egRep=\funL[1]{fundamentalRep}\brace{\algL{e}\brace{6},4}}{\fig[-2pt]{1}{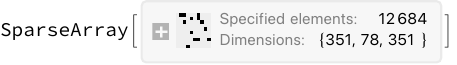}\vspace{-4pt}}
\mathematicaSequence{\funL[1]{nice}\brace{egRep}}{\fig[-2pt]{1}{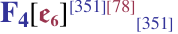}\vspace{-4pt}}
\mathematicaSequence{\funL[1]{dynkinLabel}\brace{egRep}}{\dynkinLabel{e}{0,0,0,1,0,0}}
}\vspace{-4pt}
}

\defnBox{spinorRep}{\var{algebra}\pattern}{for \algL{so}-type algebras (of rank$\leq\!8$), returns the \emph{spinor} representation as a \built{SparseArray}. These representations are conventionally defined to correspond to the irreducible representation labelled by \dynkinLabel{b}{0,\ldots,0,1} or \dynkinLabel{d}{0,\ldots,0,1}.
\mathematicaBox{
\mathematicaSequence{egRep=\funL[1]{spinorRep}\brace{\algL{so}\brace{9}}}{\fig[-2pt]{1}{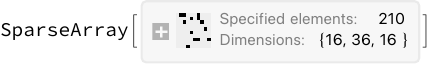}\vspace{-6pt}}
\mathematicaSequence{\funL[1]{nice}\brace{egRep}}{\fig[-2pt]{1}{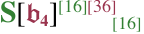}\vspace{-2pt}}
\mathematicaSequence{\funL[1]{dynkinLabel}\brace{egRep}}{\dynkinLabel{b}{0,0,0,1}}
\mathematicaSequence{egRep=\funL[1]{spinorRep}\brace{\algL{so}\brace{10}}}{\fig[-2pt]{1}{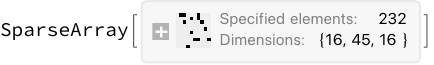}\vspace{-6pt}}
\mathematicaSequence{\funL[1]{nice}\brace{egRep}}{\fig[-2pt]{1}{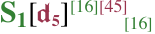}\vspace{-2pt}}
\mathematicaSequence{\funL[1]{dynkinLabel}\brace{egRep}}{\dynkinLabel{d}{0,0,0,0,1}}
}\vspace{-4pt}
}

\defnBox{spinor2Rep}{\var{algebra}\pattern}{for \algL{d}-type algebras (of rank$\leq\!8$), returns the second spinor representation as a \built{SparseArray}. These representations are conventionally defined to correspond to the irreducible representation labelled by \dynkinLabel{d}{0,\ldots,0,1,0}.
\mathematicaBox{
\mathematicaSequence{egRep=\funL[1]{spinor2Rep}\brace{\algL{so}\brace{10}}}{\fig[-2pt]{1}{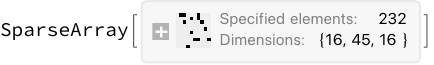}\vspace{-6pt}}
\mathematicaSequence{\funL[1]{nice}\brace{egRep}}{\fig[-2pt]{1}{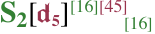}\vspace{-2pt}}
\mathematicaSequence{\funL[1]{dynkinLabel}\brace{egRep}}{\dynkinLabel{d}{0,0,0,1,0}}
}\vspace{-4pt}
}

\newpage
\subsubsectionAppendix{Arbitrary, Irreducible Representations of $\mathfrak{\algL{a}}$\brace{{{1}}}}{a1_irreps}

\defnBox{a1Rep}{\var{$w$}\pattern\built{Integer}}{returns a concrete \built{SparseArray} manifestation of the irreducible representation of highest weight `$\var{w}$' in the Chevalley basis. This representation is ($\var{w}${+}1)-dimensional.%
\mathematicaBox{
\mathematicaSequence{egRep=\funL[1]{a1Rep}\brace{8}}{\fig[-2pt]{1}{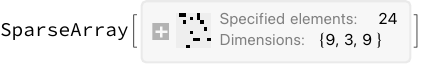}}
\mathematicaSequence{\funL[0]{showGenerators}\brace{egRep}\\[-8pt]}{\fig[-2pt]{0.875}{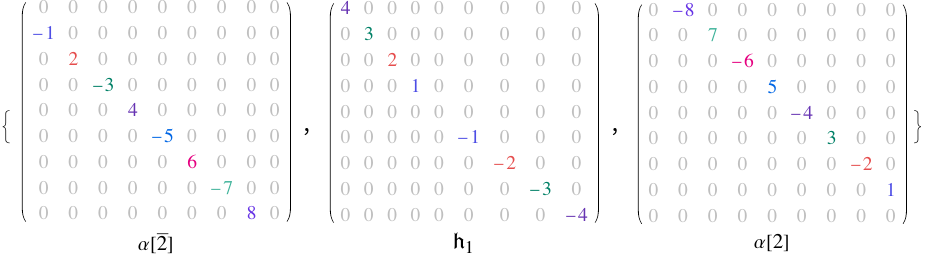}}
\mathematicaSequence{\{\funL[0]{nice}\brace{egRep},\funL[0]{drawTensors}\brace{egRep}\}}{\fig[-2pt]{0.875}{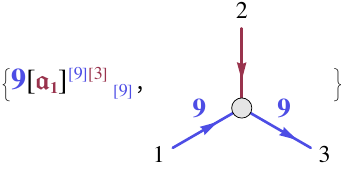}}
}
}

\newpage

\subsectionAppendix{General Aspects of Concrete Representations}{general_info_on_reps}\vspace{-4pt}

\subsubsectionAppendix{\emph{Generators} of Concrete Representations}{generators_of_concrete_reps}

\defnBox{showGenerators}{\var{subset}\pattern\optArg,\,\var{rep}\pattern\built{SparseArray}}{%
returns a \emph{formatted} display of the generators of the concrete representation \var{rep}, encoded as a \built{SparseArray} object. See (\ref{organization_of_generators}) for our conventions in ordering these. The \built{List} of \emph{generators} of \var{rep} would be given by \built{List@@}\built{Transpose@}\var{rep}.\\[-14pt]
\mathematicaBox{%
\mathematicaSequence{\funL[1]{showGenerators}\brace{\funL[1]{fundamentalRep}\brace{\algL{a}\brace{{$1$}}}}\\[-10pt]}{%
\fig[-2pt]{1}{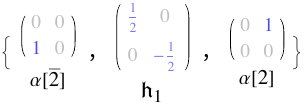}\vspace{-2pt}
}
}

Notice that each generator is labelled by its \emph{root} (corresponding to an element of the \emph{adjoint representation}'s root lattice, with negative integers indicated by `$\bar{q}$'$\equivR\text{-}q$), from \emph{lowest}-weight to \emph{highest}-weight; elements of the \emph{Cartan subalgebra} are indicated by $\mathfrak{h}_{a}$ for $a\!\in\![k]$, where $k$ is the \emph{rank} of the algebra.\\[-8pt]

In the Chevalley basis, the generators of a representation are organized into those of the `lowering' (`\var{$f$}'), `raising' (`\var{$e$}'), and Cartan subalgebra (`\var{$h$}') subsets. If the optional argument \var{subset} is set to \var{$f$}, \var{$e$}, or \var{$h$}, the corresponding subset of generators will be shown. 
\mathematicaBox{%
\mathematicaSequence{\funL[1]{showGenerators}\brace{\var{$e$},\funL[1]{fundamentalRep}\brace{\algL{g}\brace{{$2$}}}}\\[-8pt]}{%
\{\fig[-10pt]{1}{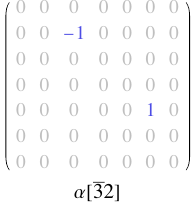},\fig[-10pt]{1}{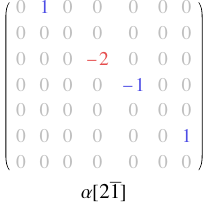},\fig[-10pt]{1}{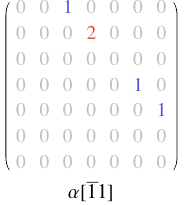},\\
\phantom{\{}\fig[-10pt]{1}{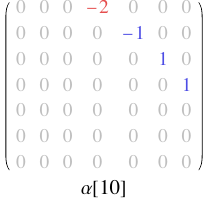},\fig[-10pt]{1}{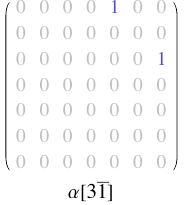},\fig[-10pt]{1}{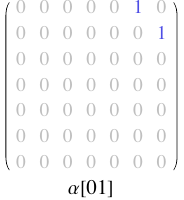}\}
}
\mathematicaSequence{\funL[1]{showGenerators}\brace{\var{$h$},\funL[1]{fundamentalRep}\brace{\algL{b}\brace{{$2$}}}}\\[-10pt]}{%
\fig[-2pt]{1}{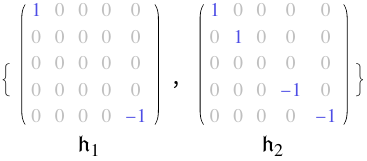}\vspace{-2pt}
}
}
}

\defnBox{cartanGenerators}{\var{rep}\pattern\built{SparseArray}}{returns the \built{Part}s of \var{rep} corresponding to the generators of the Cartan sub-algebra.}

\subsectionAppendix{Tensors Related to Concrete Representations}{other_tensors}

\defnBox{inducedAdjoint}{\var{rep}\pattern\built{SparseArray}}{returns the \emph{induced} \textbf{adjoint representation} for a concrete \var{rep} as a \built{SparseArray} object. That is, given any concrete representation $\mathbf{\b{R}}\indices{\b{[r]}\,\r{[\mathfrak{g}]}}{\b{[r]}}$, it constructs the tensor $\mathbf{\r{ad}}\indices{\r{[\mathfrak{g}]\,[\mathfrak{g}]}}{\r{[\mathfrak{g}]}}$ such that 
\eq{[\mathbf{\b{R}}\indices{\r{a}}{},\mathbf{\b{R}}\indices{\r{b}}{}]=\sum_{\t{c}\in\t{[\mathfrak{g}]}}\mathbf{\r{ad}}\indices{\r{a\,b}}{\t{c}}\,\mathbf{\b{R}}^{\t{c}}\,\vspace{-10pt}}
for all values of $\r{a,b}\!\in\!\r{[\mathfrak{g}]}$.

Importantly, \mbox{\funL[4]{inducedAdjoint}\brace{\var{rep}}=\funL[4]{inducedAdjoint}\brace{\funL[4]{inducedAdjoint}\brace{\var{rep}}}.}\\[-20pt]
}

\defnBox[9]{conjugateRep}{\var{rep}\pattern\built{SparseArray}}{returns -\built{Transpose}\brace{\var{rep},\built{\{}3,2,1\built{\}}}.}

\defnBox[21]{killingMetric}{\var{algebra}\pattern}{returns the rank-(2,0) Cartan-Killing form or `Killing metric' for \var{algebra}, denoted `$\mathcal{\r{K}}\indices{\r{[\mathfrak{g}]}\,\r{[\mathfrak{g}]}}{}$'.\\[-8pt]

\mbox{}\hspace{-10pt}\textbf{Note}: the overall scaling of the Killing metric is subject to convention; we choose to \emph{define} the Killing metric as
\eq{\mathcal{\r{K}}\indices{\r{[\mathfrak{g}]}\,\r{[\mathfrak{g}]}}{}\equivR\mathbf{tr}_{\mathbf{\b{F}}}(\r{1\,2})\,\vspace{-6pt}}
where the fundamental representation `$\mathbf{\b{F}}$' is that defined by \mbox{\funL{fundamentalRep}.}\\[-12pt]
\mathematicaBox{
\mathematicaSequence{\funL[1]{killingMetric}\brace{\algL{g}\brace{2}}}{\fig[-2pt]{1}{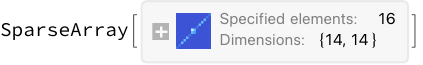}\vspace{-4pt}}
\mathematicaSequence{\{\funL[0]{matrixForm}@\%,\funL[0]{nice}@\%,\funL[0]{drawTensors}@\%\}\\[-10pt]}{\{\fig[-2pt]{1}{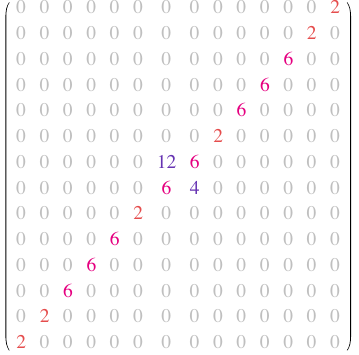},\fig[-2pt]{1}{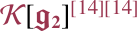},\fig[-2pt]{1}{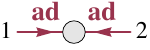}\}}
}\vspace{-4pt}
}

\defnBox[21]{killingMetricInverse}{\var{algebra}\pattern}{returns the rank-(0,2) tensor defined as \mbox{\funL{inverse}\brace{\funL[4]{killingMetric}\brace{\var{algebra}}\hspace{-1.5pt}}.}
\mathematicaBox{
\mathematicaSequence{\funL[1]{killingMetricInverse}\brace{\algL{g}\brace{2}}}{\fig[-2pt]{1}{killingMetric_g2}\vspace{-4pt}}
\mathematicaSequence{\{\funL[0]{nice}@\%,\funL[0]{drawTensors}@\%\}\\[-10pt]}{\{\fig[-2pt]{1}{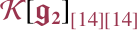},\fig[-2pt]{1}{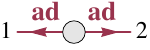}\}}
}\vspace{-4pt}
}

\defnBox{structureConstants}{\var{algebra}\pattern}{returns the totally-antisymmetric, rank-(3,0) tensor defined by
\eq{\begin{split}\mathbf{\r{ad}}\indices{\r{[\mathfrak{g}]\,[\mathfrak{g}]\,[\mathfrak{g}]}}{}&\equivR\mathbf{tr}_{\mathbf{\b{F}}}(\r{1\,2\,3}){-}\mathbf{tr}_{\mathbf{\b{F}}}(\r{2\,1\,3})=\frac{1}{T(\mathbf{\b{R}})}\big(\mathbf{tr}_{\mathbf{\b{R}}}(\r{1\,2\,3}){-}\mathbf{tr}_{\mathbf{\b{R}}}(\r{2\,1\,3})\big)\\
&=\sum_{\t{a}\in\t{[\mathfrak{g}]}}\mathbf{\b{\r{ad}}}\indices{\r{[\mathfrak{g}]\,[\mathfrak{g}]}}{\t{a}}\mathcal{\r{K}}\indices{\t{a}\,\r{[\mathfrak{g}]}}{}\\[-8pt]\end{split}}
where $\mathcal{\r{K}}\indices{\r{[\mathfrak{g}]\,[\mathfrak{g}]}}{}$ is the Cartan-Killing form. 
}

\defnBox{selfDualityMetric}{\var{rep}\pattern\built{SparseArray}}{for a \emph{real} representation \var{rep}, returns the rank-two tensor $\mathbf{M}\indices{\b{[r]}}{\t{[\bar{r}]}}\equivL\mathbf{M}\indices{\b{[r]}\,\b{[r]}}{}$ (given as a \built{SparseArray} object) encoding the \emph{similarity transformation} between $\mathbf{\b{R}}$ and its \funRL{conjugateRep}\brace{\var{rep}} $\t{\bar{\mathbf{R}}}$ such that 
\eq{\sum_{\b{r_i}\in\b{[r]}}\big(\mathbf{M}^{\text{-}1}\big)\indices{\t{[\bar{r}]}}{\b{r_1}}\mathbf{\b{R}}\indices{\b{r_1}\r{[\mathfrak{g}]}}{\b{r_2}}\mathbf{M}\indices{\b{r_2}}{\t{[\bar{r}]}}=\mathbf{\t{\bar{R}}}\indices{\t{[\bar{r}]}\,\r{[\mathfrak{g}]}}{\t{[\bar{r}]}}\,.}
\mathematicaBox{
\mathematicaSequence[1]{egRep=\funL[1]{fundamentalRep}\brace{\algL{g}\brace{2}};\\
egDuality=\funL[1]{selfDualityMetric}\brace{egRep}}{\fig[-2pt]{1}{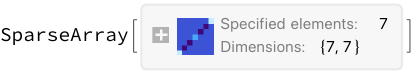}}
\mathematicaSequence{\{\funL[0]{nice}@\%,\funL[0]{drawTensors}@\%,\funL[0]{matrixForm}@\%\}}{\fig[-4pt]{1}{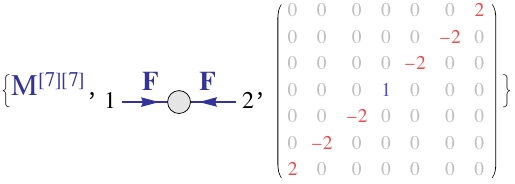}}
\mathematicaSequence{\funL[1]{dot}\brace{\funL[1]{inverse}\brace{egDuality},egRep,egDuality}==\funRL[1]{conjugateRep}\brace{egRep}}{\texttt{\textbf{True}}}
}
}

\defnBox{casimirC2}{\var{rep}\pattern\built{SparseArray}}{returns the $\mathrm{dim}(\mathbf{\b{R}})\!\times\!\mathrm{dim}(\mathbf{\b{R}})$, rank-(1,1) tensor constructed from \var{rep} according to 
\eq{C_2(\mathbf{\b{R}})\indices{\b{[r]}}{\b{[r]}}\equivR\sum_{\substack{\t{r}\in\t{[r]}\\
\r{a_i}\in\r{[\mathfrak{g}]}}}\mathbf{\b{R}}\indices{\b{[r]}\,\r{a_1}}{\t{r}}\,\mathbf{\b{R}}\indices{\t{r}\,\r{a_2}}{\b{[r]}}\,\r{\mathcal{K}_{\smash{\r{a_1}\,\r{a_2}}}}\,.\vspace{-10pt}}
\mathematicaBox{
\mathematicaSequence{\funL[1]{casimirC2}\brace{\funL[0]{symPower}\brace{\funL[1]{fundamentalRep}\brace{\algL{g}\brace{2}},2}}}{
\fig[-2pt]{1}{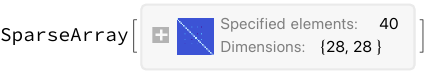}}
\mathematicaSequence{\{\funL[0]{nice}@\%,\funL[0]{drawTensors}@\%\}}{
\fig[-5pt]{1}{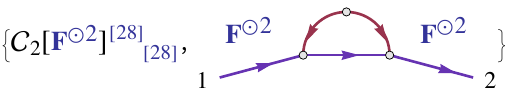}}
}
}

\subsubsectionAppendix{Clebsch Tensors Involving Representations of $\mathfrak{\algL{a}}$\brace{{{1}}}}{a1_irreps}

\defnBox{a1Clebsch}{\var{$w_1$}\pattern,\var{$w_2$}\pattern,\var{$w_3$}\pattern}{returns the Clebsch-Gordan tensor involving the irreducible representations of \algL{a}\brace{1} labelled by highest-weight integers $\var{w_i}$. A negative weight $\var{w_i}$ is understood as the conjugate representation (an outgoing arrow in the tensor diagram).
\mathematicaBox{
\mathematicaSequence{\funL[1]{a1Clebsch}\brace{8,6,4}}{\fig[-2pt]{1}{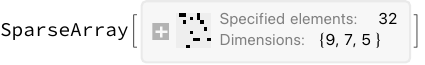}}
\mathematicaSequence{\{\funL[0]{nice}@\%,\funL[0]{drawTensors}@\%\}}{\fig[-2pt]{1}{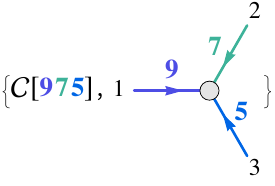}}
\mathematicaSequence{\funL[1]{a1Clebsch}\brace{2,-5,-7}}{\fig[-2pt]{1}{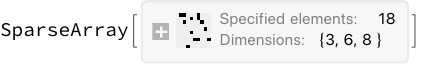}}
\mathematicaSequence{\{\funL[0]{nice}@\%,\funL[0]{drawTensors}@\%\}}{\fig[-2pt]{1}{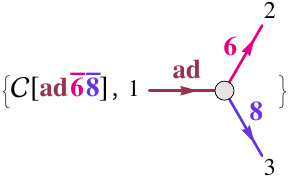}}
}
}

\newpage

\subsectionAppendix{Analyzing Aspects of Concrete Representations}{analysis_of_concrete_reps}

\defnBox[9]{representationDimension}{\var{rep}\pattern\built{SparseArray}}{returns the dimension of the concrete representation. Equivalent to \built{First}\brace{\built{Dimensions}\brace{\var{rep}}}.}

\defnBox[9]{dynkinIndex}{\var{rep}\pattern\built{SparseArray}}{returns the \emph{Dynkin index} `$T(\mathbf{\b{R}})$' of the concrete representation $\mathbf{\b{R}}$ encoded by the \built{SparseArray} \var{rep}.\\
\mbox{}\hspace{-10pt}\textbf{Note}: the Dynkin index may be arbitrarily chosen for any \emph{single} representation, after which it is determined for all others. Our convention is to take $T(\mathbf{\b{F}})\equivR1$, where $\mathbf{\b{F}}$ is the \built{Output} of \funL{fundamentalRep}.
}

\defnBox{irreducibleRepresentationQ}{\var{rep}\pattern\built{SparseArray}}{returns \built{True} if the concrete representation \var{rep} is \emph{irreducible} and returns \built{False} otherwise. Specifically, this is tested by determining if \funRL[4]{representationDimension}\brace{\var{rep}} is equal to that of the irreducible representation labelled by the highest weight of \funRL{representationWeights}\brace{\var{rep}}.}

\defnBox[9]{realRepresentationQ}{\var{rep}\pattern\built{SparseArray}}{returns \built{True} if the concrete representation \var{rep} is \emph{real} (\emph{self-conjugate}) and returns \built{False} otherwise. 
}

\defnBox[9]{pseudoRealRepresentationQ}{\var{rep}\pattern\built{SparseArray}}{returns \built{True} if the concrete representation \var{rep} is \textbf{both} \emph{real} and \emph{pseudoreal}, and returns \built{False} otherwise.
}

\subsectionAppendix{Weight Systems for Concrete Representations}{weight_systems}

\defnBox[9]{dynkinLabel}{\var{rep}\pattern\built{SparseArray}}{returns the \hyperlink{label:w}{\emph{Dynkin label}} of the concrete representation \var{rep}; if \var{rep} is not irreducible, it determines its decomposition into irreducible representations with multiplicity.
\mathematicaBox{
\mathematicaSequence{egRep=\funL[0]{repPower}\brace{\funL[0]{fundamentalRep}\brace{\algL{f}\brace{4}},3}\\[-8pt]}{\fig[-2pt]{1}{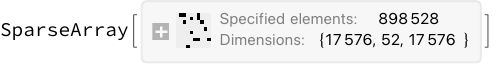}\vspace{-4pt}}
\mathematicaSequence{\funL[0]{nice}\brace{egRep}\\[-10pt]}{\fig[-2pt]{1}{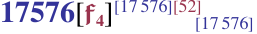}\vspace{-2pt}}
\mathematicaSequence{\funL[1]{dynkinLabel}\brace{egRep}}{
\symbolL{plus}\texttt{[}\mbox{\symbolL{times}\brace{\hspace{-1pt}1\hspace{-1.02pt},\hspace{-1.02pt}\dynkinLabel{\algL{f}}{0\hspace{-1.17pt},\hspace{-1.17pt}0\hspace{-1.17pt},\hspace{-1.17pt}0\hspace{-1.17pt},\hspace{-1.17pt}0\hspace{-0.91pt}}\hspace{-0.9pt}}}\hspace{-1.11pt},%
\hspace{-1.11pt}\mbox{\symbolL{times}\brace{\hspace{-1pt}5\hspace{-1.02pt},\hspace{-1.02pt}\dynkinLabel{\algL{f}}{1\hspace{-1.17pt},\hspace{-1.17pt}0\hspace{-1.17pt},\hspace{-1.17pt}0\hspace{-1.17pt},\hspace{-1.17pt}0\hspace{-0.91pt}}\hspace{-0.9pt}}}\hspace{-1.11pt},%
\hspace{-1.11pt}\mbox{\symbolL{times}\brace{\hspace{-1pt}2\hspace{-1.02pt},\hspace{-1.02pt}\dynkinLabel{\algL{f}}{0\hspace{-1.17pt},\hspace{-1.17pt}0\hspace{-1.17pt},\hspace{-1.17pt}0\hspace{-1.17pt},\hspace{-1.17pt}1\hspace{-0.91pt}}\hspace{-0.9pt}}}\hspace{-1.11pt},\\%
\hspace{-1.11pt}\mbox{\symbolL{times}\brace{\hspace{-1pt}4\hspace{-1.02pt},\hspace{-1.02pt}\dynkinLabel{\algL{f}}{0\hspace{-1.17pt},\hspace{-1.17pt}1\hspace{-1.17pt},\hspace{-1.17pt}0\hspace{-1.17pt},\hspace{-1.17pt}0\hspace{-0.91pt}}\hspace{-0.9pt}}}\hspace{-1.11pt},%
\hspace{-1.11pt}\mbox{\symbolL{times}\brace{\hspace{-1pt}3\hspace{-1.02pt},\hspace{-1.02pt}\dynkinLabel{\algL{f}}{2\hspace{-1.17pt},\hspace{-1.17pt}0\hspace{-1.17pt},\hspace{-1.17pt}0\hspace{-1.17pt},\hspace{-1.17pt}0\hspace{-0.91pt}}\hspace{-0.9pt}}}\hspace{-1.11pt},%
\hspace{-1.11pt}\mbox{\symbolL{times}\brace{\hspace{-1pt}3\hspace{-1.02pt},\hspace{-1.02pt}\dynkinLabel{\algL{f}}{1\hspace{-1.17pt},\hspace{-1.17pt}0\hspace{-1.17pt},\hspace{-1.17pt}0\hspace{-1.17pt},\hspace{-1.17pt}1\hspace{-0.91pt}}\hspace{-0.9pt}}}\hspace{-1.11pt},%
\hspace{-1.11pt}\mbox{\symbolL{times}\brace{\hspace{-1pt}1\hspace{-1.02pt},\hspace{-1.02pt}\dynkinLabel{\algL{f}}{0\hspace{-1.17pt},\hspace{-1.17pt}0\hspace{-1.17pt},\hspace{-1.17pt}1\hspace{-1.17pt},\hspace{-1.17pt}0\hspace{-0.91pt}}\hspace{-0.9pt}}}\hspace{-1.11pt},%
\hspace{-1.11pt}\mbox{\symbolL{times}\brace{\hspace{-1pt}1\hspace{-1.02pt},\hspace{-1.02pt}\dynkinLabel{\algL{f}}{3\hspace{-1.17pt},\hspace{-1.17pt}0\hspace{-1.17pt},\hspace{-1.17pt}0\hspace{-1.17pt},\hspace{-1.17pt}0\hspace{-0.91pt}}\hspace{-0.9pt}}}\hspace{-1.11pt},%
\hspace{-1.11pt}\mbox{\symbolL{times}\brace{\hspace{-1pt}2\hspace{-1.02pt},\hspace{-1.02pt}\dynkinLabel{\algL{f}}{1\hspace{-1.17pt},\hspace{-1.17pt}1\hspace{-1.17pt},\hspace{-1.17pt}0\hspace{-1.17pt},\hspace{-1.17pt}0\hspace{-0.91pt}}\hspace{-0.9pt}}}\texttt{]}}
\mathematicaSequence{\funL[0]{nice}@\%}{\fig[-2pt]{1}{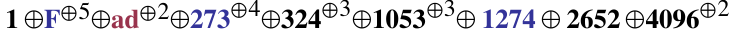}}
}\vspace{-4pt}
}

\defnBox[9]{representationAlgebra}{\var{rep}\pattern\built{SparseArray}}{returns the Lie algebra (according to the Cartan naming conventions) of the concrete representation \var{rep}.}

\defnBox[9]{showWeightLattice}{\var{rep}\pattern\built{SparseArray},\var{intoIrrepsQ}\pattern\optArg{\built{False}}}{%
returns a formatted (`spindle-shaped') display of the collection of weights for any \emph{not necessarily irreducible} concrete representation \var{rep}. Notice that negative integers are displayed with over-lines: $\bar{q}\equivR\text{-}q$.\\[-6pt]

If the second, \emph{optional} argument \var{intoIrrepsQ} is set to \built{True}, then the set of weights is decomposed into subsets according to the lattices of irreducible representations. 
\mathematicaBox{%
\mathematicaSequence[1]{egRep=\funL[0]{symPower}\brace{\funL[0]{antiSymPower}\brace{\funL[0]{fundamentalRep}\brace{\algL{b}\brace{{$2$}}},2},2};\\
\funRL[1]{showWeightLattice}\brace{egRep}
}{
\fig[-4pt]{1}{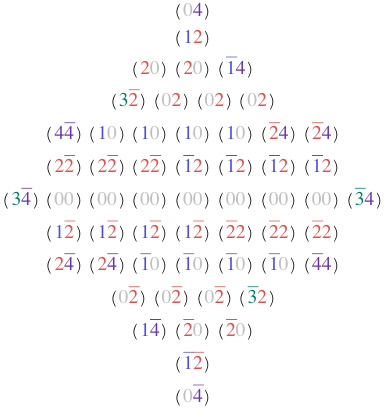}\vspace{-2pt}}
\mathematicaSequence{\funRL[1]{showWeightLattice}\brace{egRep,\built{True}}
}{
\fig[-2pt]{1}{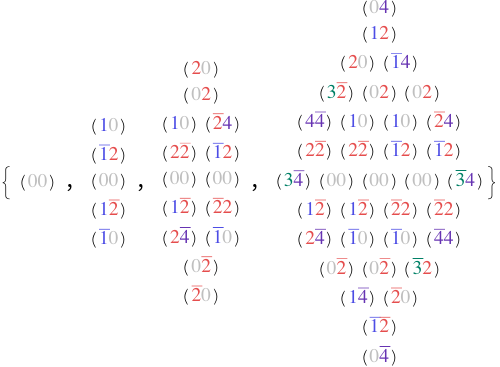}\vspace{-2pt}}
\mathematicaSequence{\funRL[0]{dynkinLabel}\brace{egRep}
}{\scalebox{0.9725}{\mbox{\symbolL{plus}\brace{\hspace{-1pt}\symbolL{times}\brace{\!1\!,\!\dynkinLabel{b}{\hspace{-1pt}0\!,\!0\hspace{-1pt}}\hspace{-1pt}}\!,\!\symbolL{times}\brace{\!1\!,\!\dynkinLabel{b}{\hspace{-1pt}1\!,\!0\hspace{-1pt}}\hspace{-1pt}}\!,\!\symbolL{times}\brace{\!1\!,\!\dynkinLabel{b}{\hspace{-1pt}2\!,\!0\hspace{-1pt}}\hspace{-1pt}}\!,\symbolL{times}\brace{\!1\!,\!\dynkinLabel{b}{\hspace{-1pt}0\!,\!4\hspace{-1pt}}\hspace{-1pt}}\hspace{-1pt}}}}}
\vspace{-4pt}
\mathematicaSequence{\funL[0]{nice}@\%
}{\fig[-4pt]{1}{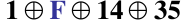}\vspace{-8pt}}
}
}

\defnBox[9]{representationWeights}{\var{rep}\pattern\built{SparseArray}}{%
returns the \emph{unformatted} collection of weights with duplications for multiplicity for the representation \var{rep}.
\mathematicaBox{%
\mathematicaSequence[1]{egRep=\funL[0]{symPower}\brace{\funL[0]{antiSymPower}\brace{\funL[0]{fundamentalRep}\brace{\algL{b}\brace{{$2$}}},2},2};\\
\funRL[1]{representationWeights}\brace{egRep}
}{
\fig[-2.5pt]{1}{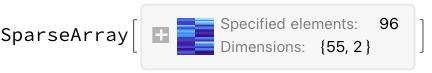}\vspace{-2pt}
}
}
}

\defnBox[9]{showRootLattice}{\var{rep}\pattern\built{SparseArray},\var{intoIrrepsQ}\pattern\optArg{\built{False}}}{%
returns a formatted (`spindle-shaped') display of the collection of roots (or, more formally `marks': the coefficients of simple roots) for any \emph{not necessarily irreducible}, concrete representation \var{rep}. Negative numbers are displayed with over-lines: $\bar{q}\equivR\text{-}q$.\\[-10pt]

If the second, \emph{optional} argument \var{intoIrrepsQ} is set to \built{True}, then the set of roots is decomposed into subsets according to irreducible representations. 
\mathematicaBox{%
\mathematicaSequence[1]{egRep=\funL[0]{symPower}\brace{\funL[0]{antiSymPower}\brace{\funL[0]{fundamentalRep}\brace{\algL{b}\brace{{$2$}}},2},2};\\
\funRL[1]{showRootLattice}\brace{egRep}
}{
\fig[-10pt]{1}{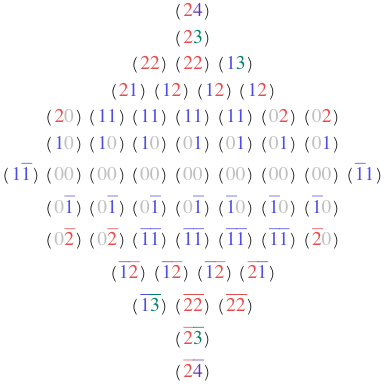}\vspace{-6pt}}
\mathematicaSequence{\funRL[1]{showRootLattice}\brace{egRep,\built{True}}
}{
\fig[-2pt]{1}{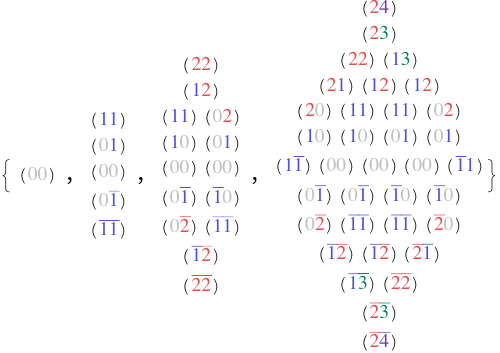}\vspace{-2pt}}
}\vspace{-4pt}
}

\defnBox[9]{representationRoots}{\var{rep}\pattern\built{SparseArray},\var{sortedQ}\pattern\optArg\built{False}}{
returns the \emph{unformatted} collection of roots with duplications for multiplicity for the representation \var{rep}.
\mathematicaBox{%
\mathematicaSequence[1]{egRep=\funL[0]{symPower}\brace{\funL[0]{antiSymPower}\brace{\funL[0]{fundamentalRep}\brace{\algL{b}\brace{{$2$}}},2},2};\\
\funRL[1]{representationRoots}\brace{egRep}
}{
\fig[-2.5pt]{1}{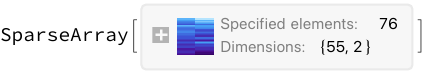}\vspace{-2pt}
}
}\vspace{-4pt}
}

\newpage
%================================================================================================================
\vspace{8pt}\sectionAppendix{Building Concrete Representations from Representations}{appendix:building_concrete_reps}\vspace{-4pt}
%================================================================================================================ 
\subsectionAppendix{Concrete Representations Defined via Concrete Representations}{higher_rep_constructors}

\defnBox{repSum}{\var{repA}\pattern\built{SparseArray},\var{repB}\pattern\built{SparseArray}}{returns the concrete representation generated by the \emph{outer sum} of the representations \var{repA} and \var{repB}.
\mathematicaBox{
\mathematicaSequence{egRep=\funL[1]{repSum}\brace{\funL[0]{spinorRep}\brace{\algL{d}\brace{4}},\funL[0]{fundamentalRep}\brace{\algL{d}\brace{4}}}\\[-6pt]}{\fig[-2pt]{1}{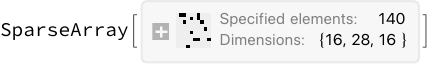}}
\mathematicaSequence{\{\funL[0]{nice}\brace{egRep},\funL[0]{drawTensors}\brace{egRep}\}}{\fig[-2pt]{1}{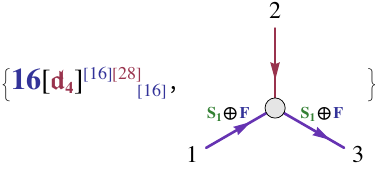}}
\mathematicaSequence{\funRL[0]{dynkinLabel}\brace{egRep}}{\symbolL{plus}\brace{\mbox{\symbolL{times}\brace{\hspace{-1pt}1\hspace{-1.02pt},\hspace{-1.02pt}\dynkinLabel{\algL{d}}{0\hspace{-1.17pt},\hspace{-1.17pt}0\hspace{-1.17pt},\hspace{-1.17pt}0\hspace{-1.17pt},\hspace{-1.17pt}1\hspace{-0.91pt}}\hspace{-0.9pt}}}\hspace{-1.11pt},%
\hspace{-1.11pt}\mbox{\symbolL{times}\brace{\hspace{-1pt}1\hspace{-1.02pt},\hspace{-1.02pt}\dynkinLabel{\algL{d}}{1\hspace{-1.17pt},\hspace{-1.17pt}0\hspace{-1.17pt},\hspace{-1.17pt}0\hspace{-1.17pt},\hspace{-1.17pt}0\hspace{-0.91pt}}\hspace{-0.9pt}}}}}
\mathematicaSequence{\funL[0]{nice}@\%}{\fig[-2pt]{1}{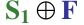}}
}\vspace{-4pt}
}

\defnBox[31]{repSum}{\var{$n$}\pattern\built{Integer},\var{rep}\pattern\built{SparseArray}}{returns the concrete representation generated by the \emph{outer sum} of \var{$n$} copies of the representation \var{rep}.
\mathematicaBox{
\mathematicaSequence{egRep=\funL[1]{repSum}\brace{4,\funL[0]{fundamentalRep}\brace{\algL{e}\brace{6}}}\\[-6pt]}{\fig[-2pt]{1}{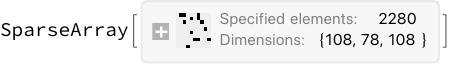}}
\mathematicaSequence{\{\funL[0]{nice}\brace{egRep},\funL[0]{drawTensors}\brace{egRep}\}}{\fig[-2pt]{1}{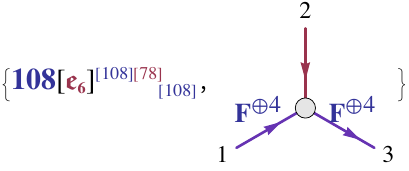}}
\mathematicaSequence{\funRL[0]{dynkinLabel}\brace{egRep}}{\symbolL{plus}\brace{\mbox{\symbolL{times}\brace{\hspace{-1pt}4\hspace{-1.02pt},\hspace{-1.02pt}\dynkinLabel{\algL{e}}{1\hspace{-1.17pt},\hspace{-1.17pt}0\hspace{-1.17pt},\hspace{-1.17pt}0\hspace{-1.17pt},\hspace{-1.17pt}0\hspace{-1.17pt},\hspace{-1.17pt}0\hspace{-1.17pt},\hspace{-1.17pt}0\hspace{-0.91pt}}\hspace{-0.9pt}}}}}
\mathematicaSequence{\funL[0]{nice}@\%}{\fig[-2pt]{1}{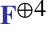}}
}\vspace{-4pt}
}

\defnBox{productRep}{\var{repA}\pattern\built{SparseArray},\var{repB}\pattern\built{SparseArray}}{returns the tensor-product-representation denoted `\var{repA}$\otimes$\var{repB}'.
\mathematicaBox{
\mathematicaSequence{egRep=\funL[1]{productRep}\brace{\funL[0]{spinorRep}\brace{\algL{b}\brace{4}},\funL[0]{adjointRep}\brace{\algL{b}\brace{4}}}\\[-6pt]}{\fig[-2pt]{1}{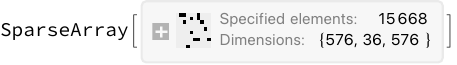}}
\mathematicaSequence{\{\funL[0]{nice}\brace{egRep},\funL[0]{drawTensors}\brace{egRep}\}}{\fig[-2pt]{1}{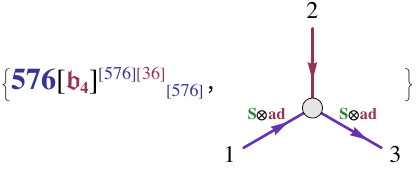}}
\mathematicaSequence{\funRL[0]{dynkinLabel}\brace{egRep}}{\symbolL{plus}\brace{%
\mbox{\symbolL{times}\brace{\hspace{-1pt}1\hspace{-1.02pt},\hspace{-1.02pt}\dynkinLabel{\algL{b}}{0\hspace{-1.17pt},\hspace{-1.17pt}0\hspace{-1.17pt},\hspace{-1.17pt}0\hspace{-1.17pt},\hspace{-1.17pt}1\hspace{-0.91pt}}\hspace{-0.9pt}}}\hspace{-1.11pt},%
\hspace{-1.11pt}\mbox{\symbolL{times}\brace{\hspace{-1pt}1\hspace{-1.02pt},\hspace{-1.02pt}\dynkinLabel{\algL{b}}{1\hspace{-1.17pt},\hspace{-1.17pt}0\hspace{-1.17pt},\hspace{-1.17pt}0\hspace{-1.17pt},\hspace{-1.17pt}1\hspace{-0.91pt}}\hspace{-0.9pt}}}\hspace{-1.11pt},%
\hspace{-1.11pt}\mbox{\symbolL{times}\brace{\hspace{-1pt}1\hspace{-1.02pt},\hspace{-1.02pt}\dynkinLabel{\algL{b}}{0\hspace{-1.17pt},\hspace{-1.17pt}1\hspace{-1.17pt},\hspace{-1.17pt}0\hspace{-1.17pt},\hspace{-1.17pt}1\hspace{-0.91pt}}\hspace{-0.9pt}}}%
}}
\mathematicaSequence{\funL[0]{nice}@\%}{\fig[-2pt]{1}{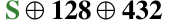}}
}\vspace{-4pt}
}

\defnBox{repPower}{\var{rep}\pattern\built{SparseArray},\var{$n$}\pattern\built{Integer}}{returns the \var{$n$}-fold tensor-product-representation of \var{rep} with itself.
\mathematicaBox{
\mathematicaSequence{egRep=\funL[1]{repPower}\brace{\funL[0]{fundamentalRep}\brace{\algL{g}\brace{2}},3}\\[-6pt]}{\fig[-2pt]{1}{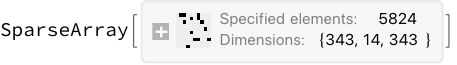}}
\mathematicaSequence{\{\funL[0]{nice}\brace{egRep},\funL[0]{drawTensors}\brace{egRep}\}}{\fig[-2pt]{1}{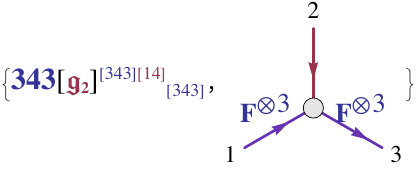}}
\mathematicaSequence{\funRL[0]{dynkinLabel}\brace{egRep}}{\symbolL{plus}\texttt{[}%
\mbox{\symbolL{times}\brace{\hspace{-1pt}1\hspace{-1.02pt},\hspace{-1.02pt}\dynkinLabel{\algL{g}}{0\hspace{-1.17pt},\hspace{-1.17pt}0\hspace{-0.91pt}}\hspace{-0.9pt}}}\hspace{-1.11pt},%
\hspace{-1.11pt}\mbox{\symbolL{times}\brace{\hspace{-1pt}4\hspace{-1.02pt},\hspace{-1.02pt}\dynkinLabel{\algL{g}}{1\hspace{-1.17pt},\hspace{-1.17pt}0\hspace{-0.91pt}}\hspace{-0.9pt}}}\hspace{-1.11pt},%
\hspace{-1.11pt}\mbox{\symbolL{times}\brace{\hspace{-1pt}2\hspace{-1.02pt},\hspace{-1.02pt}\dynkinLabel{\algL{g}}{0\hspace{-1.17pt},\hspace{-1.17pt}1\hspace{-0.91pt}}\hspace{-0.9pt}}}\hspace{-1.11pt},%
\hspace{-1.11pt}\mbox{\symbolL{times}\brace{\hspace{-1pt}3\hspace{-1.02pt},\hspace{-1.02pt}\dynkinLabel{\algL{g}}{2\hspace{-1.17pt},\hspace{-1.17pt}0\hspace{-0.91pt}}\hspace{-0.9pt}}}\hspace{-1.11pt},%
\hspace{-1.11pt}\mbox{\symbolL{times}\brace{\hspace{-1pt}2\hspace{-1.02pt},\hspace{-1.02pt}\dynkinLabel{\algL{g}}{1\hspace{-1.17pt},\hspace{-1.17pt}1\hspace{-0.91pt}}\hspace{-0.9pt}}}\hspace{-1.11pt},%
\hspace{-1.11pt}\mbox{\symbolL{times}\brace{\hspace{-1pt}1\hspace{-1.02pt},\hspace{-1.02pt}\dynkinLabel{\algL{g}}{3\hspace{-1.17pt},\hspace{-1.17pt}0\hspace{-0.91pt}}\hspace{-0.9pt}}}%
\texttt{]}
}
\mathematicaSequence{\funL[0]{nice}@\%}{\fig[-2pt]{1}{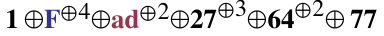}}
}\vspace{-4pt}
}

\defnBox{symPower}{\var{rep}\pattern\built{SparseArray},\var{$n$}\pattern\built{Integer}}{returns the \var{$n$}-fold \emph{symmetric} tensor-product-representation of \var{rep} with itself.
\mathematicaBox{
\mathematicaSequence{egRep=\funL[1]{symPower}\brace{\funL[0]{fundamentalRep}\brace{\algL{f}\brace{4}},3}\\[-6pt]}{\fig[-2pt]{1}{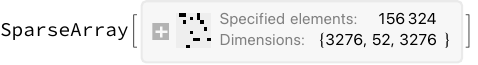}}
\mathematicaSequence{\{\funL[0]{nice}\brace{egRep},\funL[0]{drawTensors}\brace{egRep}\}}{\fig[-2pt]{1}{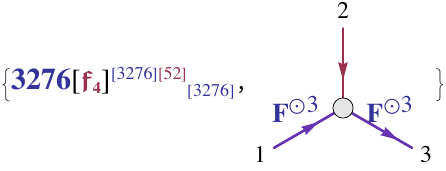}}
\mathematicaSequence{\funRL[0]{dynkinLabel}\brace{egRep}}{\symbolL{plus}\texttt{[}%
\mbox{\symbolL{times}\brace{\hspace{-1pt}1\hspace{-1.02pt},\hspace{-1.02pt}\dynkinLabel{\algL{f}}{0\hspace{-1.17pt},\hspace{-1.17pt}0\hspace{-1.17pt},\hspace{-1.17pt}0\hspace{-1.17pt},\hspace{-1.17pt}0\hspace{-0.91pt}}\hspace{-0.9pt}}}\hspace{-1.11pt},%
\hspace{-1.11pt}\mbox{\symbolL{times}\brace{\hspace{-1pt}1\hspace{-1.02pt},\hspace{-1.02pt}\dynkinLabel{\algL{f}}{1\hspace{-1.17pt},\hspace{-1.17pt}0\hspace{-1.17pt},\hspace{-1.17pt}0\hspace{-1.17pt},\hspace{-1.17pt}0\hspace{-0.91pt}}\hspace{-0.9pt}}}\hspace{-1.11pt},%
\hspace{-1.11pt}\mbox{\symbolL{times}\brace{\hspace{-1pt}1\hspace{-1.02pt},\hspace{-1.02pt}\dynkinLabel{\algL{f}}{0\hspace{-1.17pt},\hspace{-1.17pt}1\hspace{-1.17pt},\hspace{-1.17pt}0\hspace{-1.17pt},\hspace{-1.17pt}0\hspace{-0.91pt}}\hspace{-0.9pt}}}\hspace{-1.11pt},%
\hspace{-1.11pt}\mbox{\symbolL{times}\brace{\hspace{-1pt}1\hspace{-1.02pt},\hspace{-1.02pt}\dynkinLabel{\algL{f}}{2\hspace{-1.17pt},\hspace{-1.17pt}0\hspace{-1.17pt},\hspace{-1.17pt}0\hspace{-1.17pt},\hspace{-1.17pt}0\hspace{-0.91pt}}\hspace{-0.9pt}}}\hspace{-1.11pt},%
\hspace{-1.11pt}\mbox{\symbolL{times}\brace{\hspace{-1pt}1\hspace{-1.02pt},\hspace{-1.02pt}\dynkinLabel{\algL{f}}{3\hspace{-1.17pt},\hspace{-1.17pt}0\hspace{-1.17pt},\hspace{-1.17pt}0\hspace{-1.17pt},\hspace{-1.17pt}0\hspace{-0.91pt}}\hspace{-0.9pt}}}%
\texttt{]}
}
\mathematicaSequence{\funL[0]{nice}@\%}{\fig[-2pt]{1}{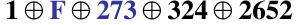}}
}\vspace{-4pt}
}

\defnBox{antiSymPower}{\var{rep}\pattern\built{SparseArray},\var{$n$}\pattern\built{Integer}}{returns the \var{$n$}-fold \emph{antisymmetric} tensor-product-representation of \var{rep} with itself.
\mathematicaBox{
\mathematicaSequence{egRep=\funL[1]{antiSymPower}\brace{\funL[0]{spinorRep}\brace{\algL{d}\brace{5}},3}\\[-6pt]}{\fig[-2pt]{1}{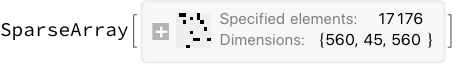}}
\mathematicaSequence{\{\funL[1]{nice}\brace{egRep},\funL[1]{drawTensors}\brace{egRep}\}}{\fig[-2pt]{1}{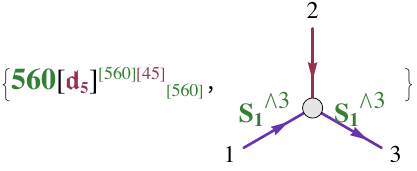}}
\mathematicaSequence{\funRL[1]{dynkinLabel}\brace{egRep}}{\symbolL{plus}\texttt{[}%
\mbox{\symbolL{times}\brace{\hspace{-1pt}1\hspace{-1.02pt},\hspace{-1.02pt}\dynkinLabel{\algL{d}}{0\hspace{-1.17pt},\hspace{-1.17pt}1\hspace{-1.17pt},\hspace{-1.17pt}0\hspace{-1.17pt},\hspace{-1.17pt}1\hspace{-1.17pt},\hspace{-1.17pt}0\hspace{-0.91pt}}\hspace{-0.9pt}}}%
\texttt{]}
}
\mathematicaSequence{\funL[1]{nice}@\%}{\fig[-2pt]{1}{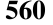}}
}\vspace{-4pt}
}

\newpage

\subsectionAppendix{Projections of Product Representations into (Eigen-)Subspaces}{projectors}

\defnBox{productRepProjectors}{\var{repA}\pattern\built{SparseArray},\var{repB}\pattern\built{SparseArray}}{given the pair of concrete representations \var{repA} and \var{repB}, returns data encoding the \emph{projectors} of the tensor-product-representation \var{repA}$\otimes$\var{repB} into eigenspaces of the quadratic Casimir \funL[4]{casimirC2}\brace{\funL[4]{productRep}\brace{\var{repA},\var{repB}}}. Specifically, \funL[4]{productRepProjectors} returns a \built{List} of data for each distinct \built{Eigenvalue} of the quadratic Casimir; for each, this data consists of a \built{List} whose \built{First} entry gives a \built{List} of Dynkin labels for irreducible representations appearing appearing in the decomposition of the product representation and share a given \built{Eigenvalue}; this is followed by a smaller-in-dimension concrete representation obtained by projecting the tensor product into this subspace; and this is followed by the \emph{Clebsch} for this projector.\\[-10pt]

This admittedly idiosyncratic information is best illustrated by example.
\mathematicaBox{
\mathematicaSequence[1]{egReps=\{\funL[0]{adjointRep}\brace{\algL{g}\brace{2}},\funL[0]{fundamentalRep}\brace{\algL{g}\brace{2}}\};\\
\funL[1]{nice}@\%}{\fig[-2pt]{1}{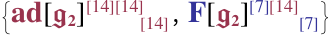}}
%\mathematicaSequence{\funL[0]{nice}@\%}{\fig[-2pt]{1}{productRepProjectors_egReps_nice}}
\mathematicaSequence{egProduct=\funL[0]{productRep}@@egReps}{\fig[-2pt]{1}{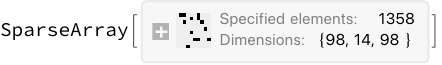}}
\mathematicaSequence{\funL[0]{drawTensors}@\%}{\fig[-2pt]{1}{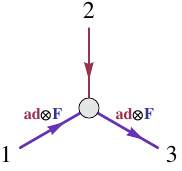}}
\mathematicaSequence[1]{projectionData=\funL[1]{productRepProjectors}@@egReps;\\
\{\var{\texttt{\#1\hspace{-1pt}\,}},\funL[1]{nice}\brace{\var{\texttt{\#2\hspace{-1pt}\,}}},\var{\texttt{\#3\hspace{-1pt}\,}}\}\&@@@\%}{\big\{\{\{\dynkinLabel{g}{1,0}\},\fig[-2pt]{1}{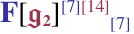},\fig[-2pt]{1}{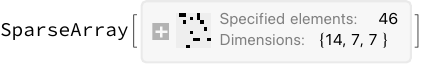}\},
\phantom{\big\{}\{\{\dynkinLabel{g}{2,0}\},\fig[-2pt]{1}{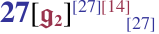},\fig[-2pt]{1}{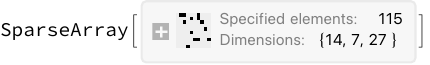}\},
\phantom{\big\{}\{\{\dynkinLabel{g}{1,1}\},\fig[-2pt]{1}{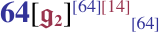},\fig[-2pt]{1}{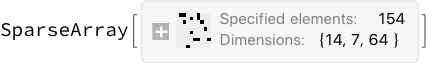}\}%
\big\}}
}\vspace{-4pt}
}

\defnBox{fromClebschToProjector}{\var{clebschTensor}\pattern\built{SparseArray}}{given a \var{clebschTensor} as generated by \funL[4]{productRepProjectors}, returns the $\mathrm{dim}(\mathbf{\b{R}}\!\otimes\!\mathbf{\t{S}})\!\times\!\mathrm{dim}(\mathbf{\b{R}}\!\otimes\!\mathbf{\t{S}})$ \built{SparseArray} denoted `$\mathbf{P}$' with the property that $\mathbf{P}.\mathbf{P}\!=\!\mathbf{P}$ and $\mathrm{tr}(\mathbf{P})\!=\!\mathrm{rank}(\mathbf{P})$.\\[-10pt]

Output equals $\funL{dot}@@\built{Reverse}\brace{\funL[4]{fromClebschToLRprojectors}\brace{\text{\var{clebschTensor}}}}$.\\[-12pt]
\mathematicaBox{
\mathematicaSequence[2]{egReps=\{\funL[0]{adjointRep}\brace{\algL{g}\brace{2}},\funL[0]{fundamentalRep}\brace{\algL{g}\brace{2}}\};\\
egProduct=\funL[0]{productRep}@@egReps;\\
projectionData=\funL[1]{productRepProjectors}@@egRep;}{}
\mathematicaSequence{largeProjectors=\funL[1]{fromClebschToProjector}/@projectionData\brace{\hspace{-1pt}\brace{\built{All}\hspace{-1pt},\hspace{-1pt}\texttt{-1}}\hspace{-1.75pt}}}{\{\fig[-2pt]{1}{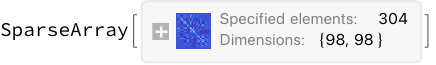},\\
\phantom{\{}\fig[-2pt]{1}{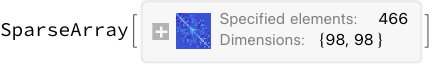},\\
\phantom{\{}\fig[-2pt]{1}{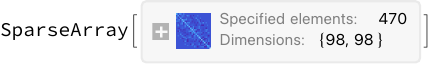}\}}
\mathematicaSequence{\{\built{Tr}\brace{\var{\texttt{\#\hspace{-1pt}\,}}},\built{MatrixRank}\brace{\var{\texttt{\#\hspace{-1pt}\,}}}\}\&/@largeProjectors}{\{\{7,7\},\{27,27\},\{64,64\}\}}
\mathematicaSequence{largeProjectedReps=\funL[1]{dot}\brace{\var{\texttt{\#\hspace{-1pt}\,}},egProduct,\var{\texttt{\#\hspace{-1pt}\,}}}\&/@largeProjectors}{\{\fig[-2pt]{1}{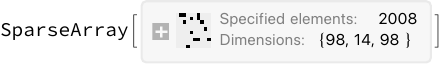},\\
\phantom{\{}\fig[-2pt]{1}{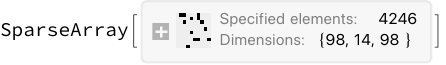},\\
\phantom{\{}\fig[-2pt]{1}{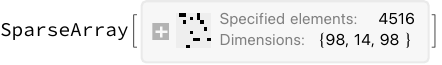}\}}
\mathematicaSequence{\funL[1]{inducedAdjoint}\brace{\var{\texttt{\#\hspace{-1pt}\,}}}===\funL[1]{adjointRep}\brace{\algL{g}\brace{2}}\&/@largeProjectedReps}{\{\built{True},\built{True},\built{True}\}}
}%\vspace{-4pt}

Notice that the `large' projected representations all involve generators of size $(98\!\times\!98)$, even if they are isomorphic to representations of smaller dimension. 
}

\defnBox{fromClebschToLRprojectors}{\var{clebschTensor}\pattern\built{SparseArray}}{for a \var{clebschTensor} of \built{Dimensions} equal to $\{\b{r},\t{s},\r{t}\}$, returns a pair of \built{SparseArray} objects denoted $\{\mathbf{L},\mathbf{R}\}$ having \built{Dimensions} $\{\r{t},\b{r}\!\times\!\t{s}\}$ and $\{\b{r}\!\times\!\t{s},\r{t}\}$, respectively such that the rank-(2,1) tensor given by $\mathbf{L}.(\mathbf{\b{R}}\!\otimes\!\mathbf{\t{T}}).\mathbf{R}$ is a concrete representation (satisfying commutation relations with \emph{identical} coefficients as the representations $\mathbf{\b{R}}$ and $\mathbf{\t{S}}$) but now having dimension equal to $\r{t}$.
\mathematicaBox{
\mathematicaSequence[2]{egReps=\{\funL[0]{adjointRep}\brace{\algL{g}\brace{2}},\funL[0]{fundamentalRep}\brace{\algL{g}\brace{2}}\};\\
egProduct=\funL[0]{productRep}@@egReps;\\
projectionData=\funL[1]{productRepProjectors}@@egRep;}{}
\mathematicaSequence{lrProjectors=\funL[1]{fromClebschToLRprojectors}/@projectionData\brace{\hspace{-1pt}\brace{\built{All}\hspace{-1pt},\hspace{-1pt}\texttt{-1}}\hspace{-1.75pt}}}{\{\fig[-2pt]{0.85}{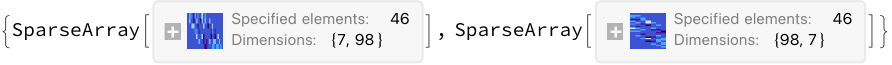},\\
\phantom{\{}\fig[-2pt]{0.85}{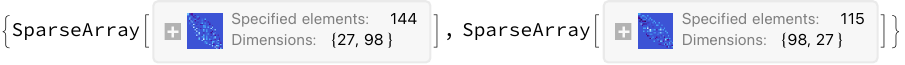},\\
\phantom{\{}\fig[-2pt]{0.85}{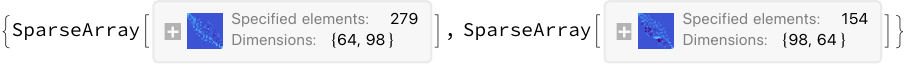}\}}
\mathematicaSequence{\funL[1]{dot}\brace{\var{\texttt{\#\hspace{-1pt}1\,}},egProduct,\var{\texttt{\#\hspace{-1pt}2\,}}}\&@@@lrProjectors}{\{\fig[-2pt]{1}{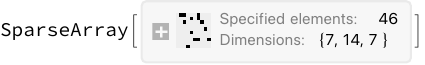},\\
\phantom{\{}\fig[-2pt]{1}{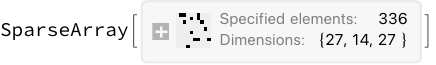},\\
\phantom{\{}\fig[-2pt]{1}{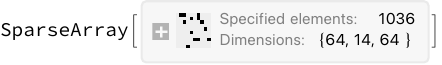}\}}
\mathematicaSequence{\funL[1]{nice}@\%}{\{\fig[-2pt]{1}{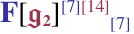},\fig[-2pt]{1}{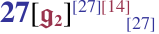},\fig[-2pt]{1}{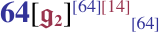}\}}
}\vspace{-4pt}
}

\newpage
%================================================================================================================
\vspace{8pt}\sectionAppendix{Building Colour Tensors from Concrete Representations}{appendix:building_colour_tensors}\vspace{-4pt}
%================================================================================================================ 
\subsectionAppendix{Tensors Defined as (`Dot'-)Powers of Representations}{dot_powers}

\defnBox{dotPower}{\var{rep}\pattern\built{SparseArray},\var{$n$}\pattern\built{Integer}}{returns the \built{SparseArray} object encoding the rank-($1{+}\var{n}$,1) tensor obtained by taking the dot-product of \var{$n$} copies of the concrete representation $\mathbf{\b{R}}$ encoded by the \built{SparseArray} \var{rep}.
\mathematicaBox{
\mathematicaSequence{\funL[1]{dotPower}\brace{\funL[1]{fundamentalRep}\brace{\algL{g}\brace{2}},4}\\[-8pt]}{\fig[-2pt]{1}{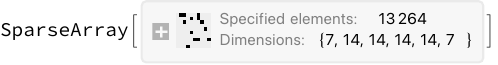}}
\mathematicaSequence{\{\funL[1]{nice}@\%,\funL[1]{drawTensors}@\%\}}{\fig[-2pt]{1}{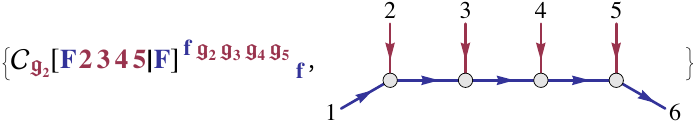}}
}\vspace{-4pt}
}

\defnBox[10]{dotPower}{\var{rep}\pattern\built{SparseArray},\built{\{}\var{ordering}\patternTwo\built{Integer}\built{\}}}{returns the \built{SparseArray} obtained by transposing the \emph{middle} \built{Length}\brace{\built{\{}\var{ordering}\built{\}}} slots. Note: \var{ordering} must represent a permutation of $\geq\!2$ indices.
\mathematicaBox{
\mathematicaSequence{\funL[1]{dotPower}\brace{\funL[1]{fundamentalRep}\brace{\algL{g}\brace{2}},\{2,4,1,3\}}\\[-8pt]}{\fig[-2pt]{1}{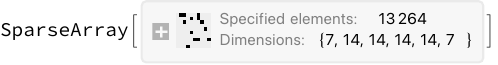}}
\mathematicaSequence{\{\funL[1]{nice}@\%,\funL[1]{drawTensors}@\%\}}{\fig[-2pt]{1}{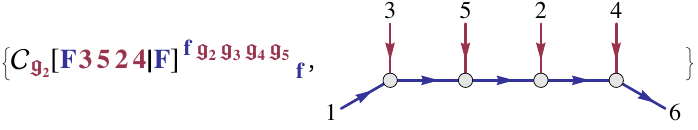}}
}\vspace{-4pt}
}

\newpage
\subsectionAppendix{Tracing Over the Generators of Concrete Representations}{traces}

\defnBox{repTrace}{\var{rep}\pattern\built{SparseArray},\var{$n$}\pattern\built{Integer}}{returns the \built{SparseArray} object encoding the rank-(\var{$n$},1) tensor obtained by tracing over the first and last indices of \funL[4]{dotPower}\brace{\var{rep},\var{$n$}}.
\mathematicaBox{
\mathematicaSequence{\funL[1]{repTrace}\brace{\funL[1]{spinorRep}\brace{\algL{d}\brace{5}},4}\\[-8pt]}{\fig[-2pt]{1}{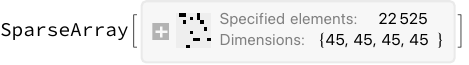}}
\mathematicaSequence{\{\funL[1]{nice}@\%,\funL[1]{drawTensors}@\%\}}{\fig[-2pt]{1}{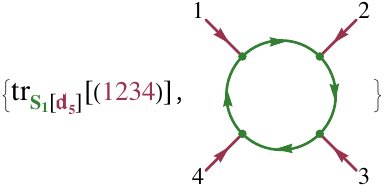}}
}\vspace{-4pt}
}

\defnBox[31]{repTrace}{\var{rep}\pattern\built{SparseArray},\built{\{}\var{indexSets}\patternTwo\built{List}\built{\}}}{returns the tensor product of \funL[4]{repTrace} tensors with \built{Slot} sequences specified by \var{indexSets}.
\mathematicaBox{
\mathematicaSequence{\funL[1]{repTrace}\brace{\funL[1]{spinorRep}\brace{\algL{d}\brace{5}},\{\{1,5,3\},\{2,4,6\}\}}\\[-8pt]}{\fig[-2pt]{1}{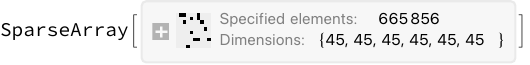}}
\mathematicaSequence{\{\funL[1]{nice}@\%,\funL[1]{drawTensors}@\%\}}{\fig[-2pt]{1}{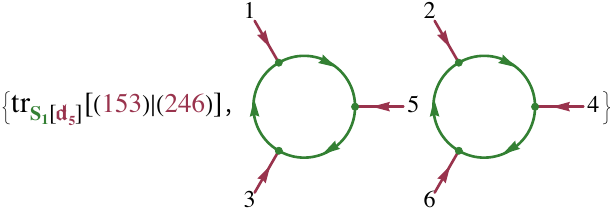}}
}\vspace{-4pt}
}

\newpage
%================================================================================================================
\vspace{8pt}\sectionAppendix{Symbolic Colour Tensors Related to Amplitudes}{appendix:abstract_colour_tensors}\vspace{-0pt}
%================================================================================================================ 
\subsectionAppendix{Colour Tensors for Scattering of \var{$n$} Adjoint-Charged Particles}{adjoint_scattering_amplitudes}

\subsubsectionAppendix{(Multi-)Trace Tensors}{multitraceTensors}

\defnBox{treeAmpTraceExpansion}{\var{$n$}\pattern,\var{fundamentalRepQ}\pattern\optArg{\built{True}}\hspace{1pt},\var{realRepQ}\pattern\optArg{\built{False}}}{~\\returns the expansion of the tree-level amplitude involving $\var{n}$ adjoint-charged particles in terms of kinematic partial amplitudes (`primitives') \mbox{\hyperlink{amp:amp}{\fun{amp}}\brace{}} and \mbox{(single-)}traces over the generators of the defining representation $\mathbf{\b{F}}$.\\[-10pt]

The role of the \emph{fundamental} or defining representation $\mathbf{\b{F}}$ in this expression is merely that we have conventionally defined the Dynkin index of this representation to be $T(\mathbf{\b{F}})\equivR1$. Any other representation may be used at the cost of introducing the compensatory factor of $T(\mathbf{\b{R}})$ into the expression. If the representation chosen were real, then the dihedral symmetry of traces would allow for a more compact expression. Such variations can be obtained from \funL[4]{treeAmpTraceExpansion} by setting the optional second argument \var{fundamentalRepQ} to \built{False} (which replaces all instances of \tensor[4]{trF} with \tensor[4]{trR}) or by setting the optional third argument, \var{realRepQ}, to \built{True}. 
}

\defnBox{singleTraceTensors}{\var{$n$}\pattern}{returns a \built{List} of abstract \tensor[4]{trR} objects signifying \emph{cyclically}-distinct single-trace tensors of rank-($\var{n}$,0) for a generic representation $\mathbf{\b{R}}$. Assuming an arbitrary (in particular, not assumed real) representation $\mathbf{\b{R}}$ results in a \built{List} of these objects of \built{Length} $(\var{n}{-}1)!$. These are the tensors appearing in \funL[4]{treeAmpTraceExpansion}\brace{\var{$n$}}.}

\defnBox{realSingleTraceTensors}{\var{$n$}\pattern}{returns a \built{List} of abstract \tensor[4]{trR} objects signifying \emph{dihedrally}-distinct single-trace tensors of rank-($\var{n}$,0) for a generic \emph{real} representation $\mathbf{\b{R}}$. In particular, this returns a \built{List} of these objects of \built{Length} $(\var{n}{-}1)!/2$.}

\defnBox{multiTraceTensors}{\var{$n$}\pattern}{returns a \built{List} of abstract \emph{multi-}trace tensors involving \emph{cyclically}-distinct sets of arguments. The number of such tensors is counted by the number of \emph{derangements} of $\var{n}$ letters, denoted $!\var{n}\equivR\mathrm{Round}\left[\var{n}!/e\right]$.}

\defnBox{realMultiTraceTensors}{\var{$n$}\pattern}{returns a \built{List} of abstract \emph{multi-}trace tensors involving \emph{dihedrally}-distinct sets of arguments. The number of such tensors is counted by the number of \emph{dihedral derangements} of $\var{n}$ letters, denoted $!!\var{n}$, \cite{Bourjaily:2024jbt}.}

\defnBox{adjointMultiTraceTensors}{\var{$n$}\pattern}{equivalent to \funL[4]{realMultiTraceTensors}\brace{\var{$n$}}, but where instances of \tensor[4]{trR} have been replaced with \tensor[4]{trAd}.}

\newpage
\subsubsectionAppendix{Del Duca, Dixon, and Maltoni's Tensors for Tree Amplitudes}{ddmTensors}

\defnBox{treeAmpDDMExpansion}{\var{$n$}\pattern,\var{anchors}\pattern\optArg{}}{returns the expansion of the tree-level amplitude involving $\var{n}$ adjoint-charged particles in terms of kinematic partial amplitudes (`primitives') \mbox{\hyperlink{amp:amp}{\fun{amp}}\brace{}} and \mbox{\tensor{ddmTensor}\brace{}} objects (considered abstractly) as given by Del Duca, Dixon and Maltoni in \cite{DelDuca:1999rs}.\\[-10pt]

By default, the (`KK'-)basis chosen for this expansion uses the `anchors' $\{1,\var{n}\}$---that is, a sum over products of the form \mbox{\hyperlink{amp:amp}{\fun{amp}}\brace{1,\patternTwo,\var{$n$}}}\mbox{\tensor{ddmTensor}\brace{1,\patternTwo,\var{$n$}}}; this choice can be changed by specifying any pair of labels $\{a,b\}\!\in\![\var{n}]$ as the optional argument \var{anchors}.
}

\defnBox{ddmTreeAmpTensors}{\var{$n$}\pattern,\var{anchors}\pattern\optArg{}}{returns a \built{List} of \tensor{ddmTensor}\brace{} objects appearing in \funL[4]{treeAmpDDMExpansion}\brace{\var{$n$},\var{anchors}}.}

\subsectionAppendix{Relations Among Tensors and Amplitudes for Adjoint Scattering}{adjoint_scattering_relations}

\defnBox{ddmToTrace}{\var{fundamentalRepQ}\pattern\optArg{\built{True}}\hspace{1pt},\var{expression}\pattern}{converts any \mbox{\tensor[4]{ddmTensor}\brace{}} objects appearing in \var{expression} into a sum over single-trace tensors \mbox{\tensor[4]{trF}\brace{}} or \mbox{\tensor[4]{trR}\brace{}} objects depending on whether the optional leading argument \var{fundamentalRepQ} is set to \built{True} (default) or \built{False}, respectively.%
}

\defnBox{jacobiReduce}{\var{anchors}\pattern\optArg{},\var{expression}\pattern}{converts any \tensor[4]{ddmTensor}\brace{} objects appearing in \var{expression} into a linear combination of those of \mbox{\funL[4]{ddmTreeAmpTensors}\brace{\var{$n$},\var{anchors}}}. Importantly, \emph{only} the (generally valid) Jacobi relation is used in this replacement---and therefore the reduction will be valid for all Lie algebras. See also the package of \cite{Bourjaily:2025hvq}.%
}

\defnBox{kkReduce}{\var{anchors}\pattern\optArg{},\var{expression}\pattern}{converts any partial amplitudes represented by \hyperlink{amp:amp}{\fun{amp}}\brace{} objects appearing in \var{expression} into those appearing in \mbox{\funL[4]{treeAmpDDMExpansion}\brace{\var{$n$},\var{anchors}}}. This replacement makes use of the so-called `KK' relations \cite{KK} (see also \cite{Bourjaily:2023uln,Bourjaily:2026adf}). 
}

\subsectionAppendix{Tree-Amplitudes involving Identically-Charged Fermions \& Adjoints}{fermionic_amp_tensors}

\defnBox{fermionicTreeAmpTensors}{\var{$n_{F}$}\pattern,\var{$n_{{G}}$}\pattern\optArg{\built{0}}}{returns a \built{List} of \mbox{\tensor{joTensor}\brace{}} objects relevant to the scattering of $\var{n_F}$ \emph{pairs} of \emph{indistinguishable} charged Fermions and \var{$n_G$} gauge bosons. See reference \cite{Bourjaily:2026adf} for more details (see also refs.\ \cite{Johansson:2015oia,Kosower:1988kh,Melia:2015ika,Ochirov:2019mtf}).%
}

\newpage
\subsectionAppendix{Colour Tensor Bases for Scattering in $\mathfrak{\algL{a}}$\brace{{{$1$}}} Gauge Theory}{a1_tensors}

\subsubsectionAppendix{Clebsch Colour Tensor Bases for Arbitrary Representations}{abstract_bases}

\defnBox{a1ClebschBasis}{\var{$n$}\pattern,\var{legOrdering}\pattern\built{List}\!\optArg{}}{returns a \built{List} of \mbox{\tensor{clebschTensor}\brace{}} objects which form a linearly-independent, colour-orthogonal \emph{basis} for colour tensors relevant to the scattering of \var{$n$} adjoint-charged particles in \algL{a}\brace{$1$} gauge theory. Alternative bases can be generated using the optional argument \var{legOrdering} is set to a permutation of the leg labels $[\var{n}]$, which permutes the ordering of legs accordingly.
}

\defnBox[31]{a1ClebschBasis}{\built{\{}\var{externalRepWeights}\patternTwo\built{Integer}\built{\}},\var{legOrdering}\pattern\built{List}\!\optArg{}}{returns a \built{List} of \mbox{\tensor{clebschTensor}\brace{}} objects which form a linearly-independent, colour-orthogonal \emph{basis} for colour tensors relevant for the scattering of $\var{n}$ particles transforming under the representations specified by \var{externalRepWeights} in \mbox{\algL{a}\brace{$1$}} gauge theory.\\[-10pt]

Each \mbox{\tensor{clebschTensor}\brace{}} object appearing encodes a tensor with slots of dimension corresponding to the ordered \var{externalRepWeights}. These representations may be arbitrarily arranged on the tree graph defining the basis of tensors, with different choices for this ordering resulting in different bases (which differ more substantially than merely a \built{Transpose} of their \built{Slot} \built{Sequence}s). To specify how these representations should be ordered around the graph defining the basis, the optional argument \var{legOrdering} may be used.\\[-10pt]

\mbox{}\hspace{-10pt}\textbf{Note}: the weights of representations specified by \var{externalRepWeights} may be taken as negative to represent raised/lowered tensor indices. 
}

\subsubsectionAppendix{Relations Among Colour Tensors in $\mathfrak{\algL{a}}$\brace{{{$1$}}} Gauge Theory}{a1_tensor_relations}

\defnBox{a1TraceTensorToClebschCoefficients}{\var{$n$}\pattern}{returns a \built{SparseArray} encoding the coefficients of the expansion of \tensor{trF}\brace{} objects of the list \funL{realMultiTraceTensors}\brace{\var{$n$}} into the basis of tensors \mbox{\funL[4]{a1ClebschBasis}\brace{\var{$n$}}} \emph{in the context of} \algL{a}\brace{$1$} gauge theory.%
}

\defnBox{a1TraceTensorEliminationRules}{\var{$n$}\pattern}{returns a \built{List} of linear relations (if any exist) among (real) multi-trace tensors \tensor{trF}\brace{} involving $\var{n}$ legs \emph{in the context of} \algL{a}\brace{$1$} gauge theory.%
}

\defnBox{a1ddmTensorToClebschCoefficients}{\var{$n$}\pattern}{returns a \built{SparseArray} encoding the coefficients of the expansion of \tensor{ddmTensor}\brace{} objects of the list \funL[4]{ddmTreeAmpTensors}\brace{\var{$n$}} into the basis of tensors \funL[4]{a1ClebschBasis}\brace{\var{$n$}} \emph{in the context of} \algL{a}\brace{$1$} gauge theory.%
}

\defnBox{a1ddmTensorEliminationRules}{\var{$n$}\pattern}{returns a \built{List} of linear relations (if any exist) among \tensor{ddmTensor}\brace{} objects of \funL{ddmTreeAmpTensors}\brace{\var{$n$}}  \emph{in the context of} \algL{a}\brace{$1$} gauge theory.%
}

\newpage
%================================================================================================================
\vspace{8pt}\sectionAppendix{Building Concrete Colour Tensors for Particular Theories}{appendix:buildiung_abstract_colour_tensors}\vspace{-0pt}
%================================================================================================================ 

\mbox{Appendix~\ref{naming_colour_tensors}} described a number of \emph{symbolic} tensors relevant to physics---namely, (multi-)trace tensors \tensor[4]{trR}, \tensor[4]{trF}, and \tensor[4]{trAd}; \tensor[4]{ddmTensor}; \tensor[4]{joTensor}; and---in the case of  \algL{a}\brace{$1$} gauge theory---\tensor[4]{clebschTensor}. All but the last of these only have \emph{concrete} meaning as tensors in the context of a particular Lie algebra or some representation $\mathbf{\b{R}}$. (\textbf{Note}: the specification of a Lie algebra automatically provides a meaning to any references to the fundamental (`$\mathbf{\b{F}}$') or adjoint (`$\mathbf{\r{ad}}$') representations.)\\[-10pt]

The package will convert these symbolic objects into concrete \built{SparseArray} objects using \funL[4]{builtTensors}\brace{\var{algebraOrRep}} and \mbox{\funL[4]{builtTensorsDual}\brace{\var{algebraOrRep}},} where \var{algebraOrRep} must be some simple Lie algebra (see \mbox{appendix~\ref{naming_of_lie_algebras}}), or a particular representation thereof---encoded as either a \built{SparseArray} or specified by \mbox{\repL{F}\brace{\var{algebra}}} or \mbox{\repL{ad}\brace{\var{algebra}}}.\\[-12pt]

\subsectionAppendix{Construction of Concrete Colour Tensors}{building_colour_tensors}

\defnBoxTwo{buildTensors}{\var{algebraOrRep}\pattern}{\var{expression}\pattern}{converts any symbolic tensors (see \mbox{appendix~\ref{naming_colour_tensors}}) appearing in \var{expression} to their concrete realizations as \built{SparseArray} objects, constructed from concrete representations.%
\mathematicaBox{
\mathematicaSequence{egTensor=\built{RandomChoice}\brace{\funL[1]{fermionicTreeAmpTensors}\brace{3,3}}}{\mbox{\fun{joTensor}\brace{\{\hspace{-1.4pt}\fun{f}\brace{\hspace{-1pt}1\hspace{-1pt}}\hspace{-1.7pt},\hspace{-1.1pt}\fun{g}\brace{\hspace{-1pt}5\hspace{-1pt}}\hspace{-1.7pt},\hspace{-1.1pt}\fun{f}\brace{\hspace{-1pt}2\hspace{-1pt}}\hspace{-1.7pt},\hspace{-1.1pt}\fun{g}\brace{\hspace{-1pt}6\hspace{-1pt}}\hspace{-1.7pt},\hspace{-1.1pt}\fun{fb}\brace{\hspace{-1pt}3\hspace{-1pt}}\hspace{-1.7pt},\hspace{-1.1pt}\fun{f}\brace{\hspace{-1pt}3\hspace{-1pt}}\hspace{-1.7pt},\hspace{-1.1pt}\fun{fb}\brace{\hspace{-1pt}1\hspace{-1pt}}\hspace{-1.7pt},\hspace{-1.1pt}\fun{g}\brace{\hspace{-1pt}4\hspace{-1pt}}\hspace{-1.7pt},\hspace{-1.1pt}\fun{fb}\brace{\hspace{-1pt}2\hspace{-1pt}}\}\hspace{-1.4pt},\hspace{-1.4pt}\!\{\!\{1\hspace{-1.7pt},\hspace{-1.4pt}9\}\hspace{-2.1pt},\hspace{-1.8pt}\{3\hspace{-1.7pt},\hspace{-1.4pt}5\}\hspace{-2.1pt},\hspace{-1.8pt}\{6\hspace{-1.7pt},\hspace{-1.4pt}7\}\!\}\hspace{-1pt}}}}
\mathematicaSequence{\funL[1]{nice}\brace{egTensor}\\[-10pt]}{\fig[-2pt]{1}{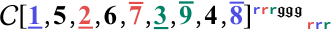}}
\mathematicaSequence{\funL[1]{drawTensors}\brace{egTensor}\\[-10pt]}{\fig[-2pt]{1}{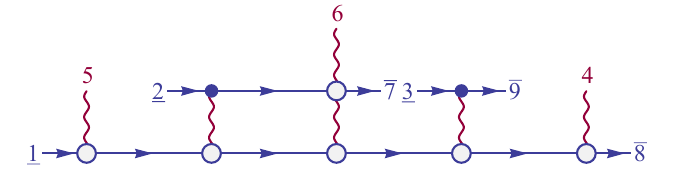}\vspace{-6pt}}
\mathematicaSequence[1]{t0=\built{AbsoluteTime}\brace{};\\
egBuilt=\funL[1]{buildTensors}\brace{\repL{F}\brace{\algL{g}\brace{2}}}\brace{egTensor}\\[-10pt]}{\fig[-2pt]{1}{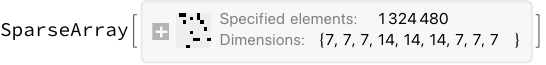}\vspace{-4pt}}
\mathematicaSequence{\funL[1]{niceTime}\brace{\built{AbsoluteTime}\brace{}-t0}}{\fig[-2pt]{1}{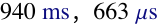}}
\mathematicaSequence{\{\funL[1]{nice}\brace{egBuilt},\funL[1]{byteCount}\brace{egBuilt}\}\\[-10pt]}{\{\fig[-2pt]{1}{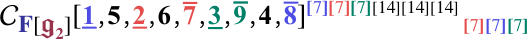},\fig[-2pt]{1}{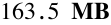}\}}
}
}

\defnBoxTwo{buildTensorsDual}{\var{algebraOrRep}\pattern}{\var{expression}\pattern}{converts any symbolic tensors (see \mbox{appendix~\ref{naming_colour_tensors}}) appearing in \var{expression} to their concrete \emph{duals}, given as \built{SparseArray} objects.
\mathematicaBox{
\mathematicaSequence{egTensor=\built{RandomChoice}\brace{\funL[1]{fermionicTreeAmpTensors}\brace{3,3}}}{\mbox{\fun{joTensor}\brace{\{\hspace{-1.4pt}\fun{f}\brace{\hspace{-1pt}1\hspace{-1pt}}\hspace{-1.7pt},\hspace{-1.1pt}\fun{g}\brace{\hspace{-1pt}6\hspace{-1pt}}\hspace{-1.7pt},\hspace{-1.1pt}\fun{g}\brace{\hspace{-1pt}5\hspace{-1pt}}\hspace{-1.7pt},\hspace{-1.1pt}\fun{f}\brace{\hspace{-1pt}2\hspace{-1pt}}\hspace{-1.7pt},\hspace{-1.1pt}\fun{f}\brace{\hspace{-1pt}3\hspace{-1pt}}\hspace{-1.7pt},\hspace{-1.1pt}\fun{fb}\brace{\hspace{-1pt}1\hspace{-1pt}}\hspace{-1.7pt},\hspace{-1.1pt}\fun{g}\brace{\hspace{-1pt}4\hspace{-1pt}}\hspace{-1.7pt},\hspace{-1.1pt}\fun{fb}\brace{\hspace{-1pt}3\hspace{-1pt}}\hspace{-1.7pt},\hspace{-1.1pt}\fun{fb}\brace{\hspace{-1pt}2\hspace{-1pt}}\}\hspace{-1.4pt},\hspace{-1.4pt}\!\{\!\{1\hspace{-1.7pt},\hspace{-1.4pt}9\}\hspace{-2.1pt},\hspace{-1.8pt}\{4\hspace{-1.7pt},\hspace{-1.4pt}8\}\hspace{-2.1pt},\hspace{-1.8pt}\{5\hspace{-1.7pt},\hspace{-1.4pt}6\}\!\}\hspace{-1pt}}}}
\mathematicaSequence{\funL[1]{nice}\brace{egTensor}\\[-10pt]}{\fig[-2pt]{1}{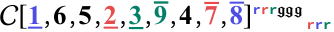}}
\mathematicaSequence{\funL[1]{drawTensors}\brace{egTensor}\\[-10pt]}{\fig[-2pt]{1}{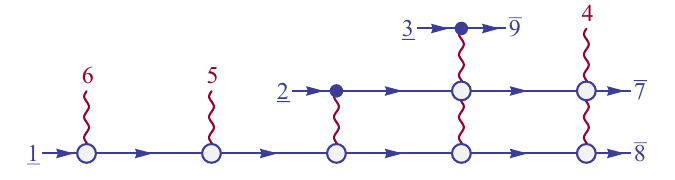}\vspace{-6pt}}
\mathematicaSequence[1]{t0=\built{AbsoluteTime}\brace{};\\
egBuilt=\funL[1]{buildTensorsDual}\brace{\repL{F}\brace{\algL{g}\brace{2}}}\brace{egTensor}\\[-10pt]}{\fig[-2pt]{1}{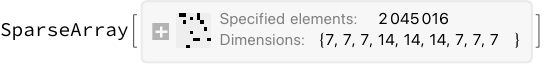}\vspace{-4pt}}
\mathematicaSequence{\funL[1]{niceTime}\brace{\built{AbsoluteTime}\brace{}-t0}}{\fig[-2pt]{1}{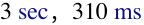}}
\mathematicaSequence{\{\funL[1]{nice}\brace{egBuilt},\funL[1]{byteCount}\brace{egBuilt}\}\\[-10pt]}{\{\fig[-2pt]{1}{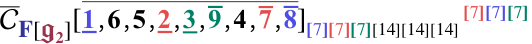},\fig[-2pt]{1}{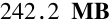}\}}
}
}

\subsectionAppendix{\emph{Direct} Analysis of Sets of Concrete Colour tensors}{analysis_of_colour_tensors}

\defnBoxTwo{contractTensors}{\var{algebraOrRep}\pattern}{\var{tensorsA}\pattern,\var{tensorB}\pattern}{computes the contraction between any symbolic tensors (see \mbox{appendix~\ref{naming_colour_tensors}}) appearing in \var{tensorsA} with those of \var{tensorsB}. If both arguments consist of a \built{List} of symbolic tensors, then the pairwise overlap matrix will be returned as a \built{SparseArray} object; if either argument consists of a single symbolic tensor, then an ordinary \built{List} is returned; and if both arguments are a symbolic tensor, then just the value of their contraction is returned.}

\defnBoxTwo{tensorOverlap}{\var{algebraOrRep}\pattern}{\var{symbolicTensors}\patternTwo\built{List}}{returns a \var{SparseArray} square matrix encoding the pairwise, complete contractions between tensors of \var{symbolicTensors} with their duals \emph{for the specified} \var{algebraOrRep}. Equivalent to \funL[4]{contractTensors}\brace{\var{algebraOrRep}}\brace{\var{\texttt{\#\hspace{-1pt}\,}},\var{\texttt{\#\hspace{-1pt}\,}}}\&@\var{symbolicTensors}.
}

\defnBoxTwo{independentTensorsQ}{\var{algebraOrRep}\pattern}{\var{symbolicTensors}\patternTwo\built{List}}{returns \built{True} if the \built{List} of \var{symbolicTensors} are linearly independent \emph{for the specified} \var{algebraOrRep}, and \built{False} if they are not independent. Importantly, this assessment is made \emph{entirely} from the concrete tensors constructed for the \var{algebraOrRep} specified. 
}

\defnBoxTwo{tensorRelations}{\var{algebraOrRep}\pattern}{\var{symbolicTensors}\patternTwo\built{List}}{returns a \built{List} of \built{Rules} encoding relations among the \built{List} of \var{symbolicTensors} \emph{for the specified} \var{algebraOrRep}.
}

\newpage
%================================================================================================================
\vspace{8pt}\sectionAppendix{General Utilities and Mathematical Functions}{appendix:general_utilities_and_amthematical_functions}\vspace{-0pt}
%================================================================================================================ 

\subsectionAppendix{Operations Built and Optimized for \built{SparseArray} Objects}{sparse_array_functions}

\defnBox{tensorContract}{\built{\{}\var{tensors}\patternTwo\built{\}},\built{\{}\var{indexPairs}\patternTwo\built{\}}}{an improved and optimized version of the built-in \textsc{Mathematica} function \built{TensorContract}}

\defnBox{dot}{\var{arraySequence}\patternTwo}{an improved and optimized version of \textsc{Mathematica}'s function \built{Dot}, making use of \funL[4]{tensorContract}.}

\defnBox{inverse}{\var{matrix}\pattern}{is equivalent to \built{SparseArray}@\built{Inverse}\brace{\var{matrix}}.}

\defnBox{nullSpace}{\var{arraySequence}\patternTwo}{analogous to the built-in \textsc{Mathematica} function \built{NullSpace}, but where linear relations are found among an \var{arraySequence} of \built{SparseArray} objects (of arbitrary---but identical---\built{Dimensions}). Specifically, the \built{Output} of \funL[4]{nullSpace} will return a \built{SparseArray} where each row encodes an independent linear relation satisfied by the \built{SparseArray} objects appearing in \var{arraySequence}.}

\defnBox{findSimilarity}{\var{initialRep}\pattern,\var{targetRep}\pattern}{returns a square \built{SparseArray} encoding the change of basis that takes \var{initialRep} into \var{targetRep}. That is, given two similar, concrete representations $\mathbf{\b{R}}\!\simeq\!\mathbf{\r{S}}$, \funL[4]{findSimilarity}\brace{$\mathbf{\b{R}},\mathbf{\r{S}}$} will return a matrix $\mathbf{M}$ such that
\eq{(\mathbf{M}^{{-}1}).\mathbf{\b{R}}.\mathbf{M}=\mathbf{\r{S}}.\label{similarity_matrix_defined}\vspace{-12pt}}
\mathematicaBox{
\mathematicaSequence[3]{egRep=\funL[1]{fundamentalRep}\brace{\algL{d}\brace{4}};\\
egM=\built{SparseArray}@\built{RandomInteger}\brace{\{1,50\},\{8,8\}};\\
imageRep=\funL[1]{dot}\brace{\funL[1]{inverse}\brace{egM},egRep,egM};\\
\funL[1]{showGenerators}\brace{\fun{h},imageRep}\brace{\brace{1}}\\[-8pt]}{\fig[-6pt]{0.7}{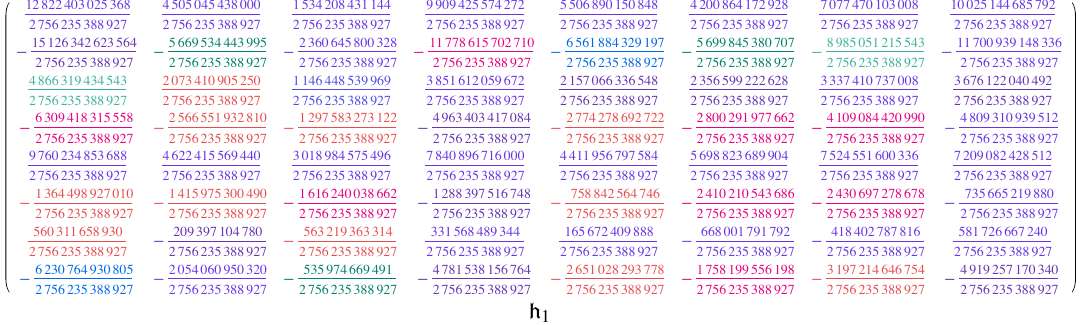}\vspace{-4pt}}
\mathematicaSequence{\funL[0]{findSimilarity}\brace{egRep,imageRep}\\[-8pt]}{\fig[-2pt]{1}{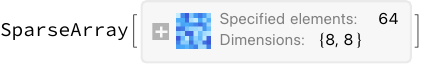}\vspace{-4pt}}
\mathematicaSequence{\funL[1]{nice}@\%\\[-8pt]}{\fig[-2pt]{1}{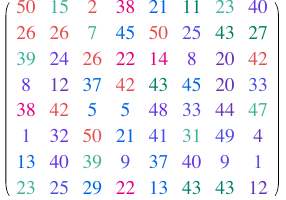}\vspace{-4pt}}
\mathematicaSequence{egM==\%\%}{\texttt{True}}
}

\mbox{}\hspace{-10pt}\textbf{Note}: it is often the case that $\mathbf{M}$ is non-unique; when this happens, the output of \funL[4]{findSimilarity} will include unspecified `free' parameters, for which (\ref{similarity_matrix_defined}) will be satisfied by any generic choice of their values. 
}

\subsectionAppendix{Mathematical and Miscellaneous Functions}{mathematical_functions}

\defnBox{randomPerm}{\var{$n$}\pattern}{returns a \built{List} of \built{Integers} chosen in $[\var{n}]$ encoding a random permutation---a randomly chosen element of $\mathfrak{S}([\var{n}])$.}

\defnBox{riordanNumber}{\var{$n$}\pattern}{returns the \var{$n$}th Riordan number \cite{oeis-riordan} given by 
\eq{r(\var{n})\equivR\sum_{\b{q}{=}0}^{\var{n}}(\text{-}1)^{\var{n}}\binom{\var{n}}{\b{q}}\binom{\b{q}}{\lfloor \b{q}/2\rfloor}\,.}
}

\defnBox{shuffle}{\var{listA}\pattern\built{List},\var{listB}\pattern\built{List}}{returns a \built{List} consisting of permutations of \mbox{\built{Join}\brace{\var{listA},\var{listB}}} for which the relative ordering of elements in each list is preserved.}

\subsectionAppendix{Installing the Package}{install_the_package}

\defnBox{installColourTensors}{}{copies the main source-code for the package, the file \rpackage\!\texttt{\textbf{.m}}, to the user's \built{\$Path}, so that it can be loaded in \built{Notebooks} \emph{not} located in the same directory as the package file downloaded from the \texttt{arXiv}.}

\newpage
\addtocontents{toc}{\protect~\\[-24pt]\mbox{\vspace{0pt}}\protect\hrulefill\par}
\addtocontents{toc}{\protect~\\[-36pt]}
\renewcommand{\indexname}{Alphabetic Glossary of Functions Defined by the Package}
\hypertarget{index}{}\printindex

\newpage
\providecommand{\href}[2]{#2}\begingroup\raggedright\endgroup
\end{document}